\documentclass[twocolumn, noshowpacs]{revtex4}
\usepackage{epstopdf}
\usepackage{amssymb}
\usepackage{amsmath}
\usepackage{amsfonts}
\usepackage{graphicx}
\begin{document}
\def\br{{\bf r}}
\def\vcmi{v_{cm,i}}

\title{Effective Soft-Core Potentials and Mesoscopic Simulations of Binary Polymer Mixtures}
\author{J. McCarty}
\author{I. Y. Lyubimov}
\author{M. G. Guenza
\footnote{To whom correspondence should be addressed. E-mail: mguenza@uoregon.edu}}
\affiliation{Department of Chemistry and Institute of Theoretical Science, University of Oregon, Eugene, Oregon 97403}
\date{\today}

\begin{abstract}
Mesoscopic molecular dynamics simulations are used to determine the 
large scale structure of several binary polymer mixtures of various 
chemical architecture, concentration, and thermodynamic conditions. 
By implementing an analytical formalism, which is based on the 
solution to the Ornstein-Zernike equation, each polymer chain is 
mapped onto the level of a single soft colloid. From the appropriate 
closure relation, the effective, soft-core potential between coarse-grained units is obtained and used as input to our mesoscale 
simulations. The potential derived in this manner is analytical and 
explicitly parameter dependent, making it general and transferable to 
numerous systems of interest. From computer simulations performed 
under various thermodynamic conditions the structure of the polymer 
mixture, through pair correlation functions, is determined over the 
entire miscible region of the phase diagram. 
In the athermal regime mesoscale simulations exhibit quantitative agreement with united atom 
simulations. Furthermore, they also provide information at larger 
scales than can be attained by united atom simulations and in the thermal regime 
approaching the phase transition. 
\end{abstract}

\maketitle

\section{Introduction}
\label{SX:INTR}
The mixing of two or more types of polymers is of great scientific and technological interest,
as it opens up the possibility of creating new materials emerging with specific physical and chemical properties.\cite{STRBL}  
However, although polymer blends have been very much a
part of everyday life, these continue to be a source of extensive
scientific inquiry.  The rich physics in polymer mixtures stems in
part from the fact that their structure and dynamics change as
thermodynamic conditions that cause phase separation (i.e.\ spinodal
decomposition) are approached.  Mixture stability is not only driven
by temperature and composition, but also by differences in chain
length and monomer architecture that may contribute substantial
entropic effects.  From physical and engineering standpoints, a goal
is to understand and predict changes that a polymer will undergo when
mixed with another polymer system \cite{BALSA,OLABI}.
Theoretical efforts have focused on this area, resulting in the development of 
several approaches connecting monomeric structure to static and dynamic properties 
as a function of thermodynamic
parameters
\cite{SWC94,SWC97,FRD02,Lodge,Lodge1,LIPSN}.

Molecular dynamics (MD) studies have aided our understanding of the
correlation between local (intra- and intermolecular) structure and
global fluid properties that govern polymer mixtures
\cite{MULER,BINDR,MATTI,HEINE,JARAM,Thomp,Bedrov}.  However, these
computational approaches have been limited to relatively small length
and time scales because present-day computing power limits the
attainment of extended scales.\cite{FRNKL}.  For polymer blends, this is particularly
problematic because as the spinodal decomposition is approached, a
divergent length scale in concentration fluctuations develops, thereby
readily exceeding box sizes commonly used to model polymeric
ensembles.  In this manner, the determination of large-scale
properties for polymer mixtures with atomic resolution requires increasingly larger
simulation boxes, and becomes rapidly unfeasable.  

One way to circumvent this difficulty is to introduce a coarse-grained description of the polymer mixture, where small-scale atomistic details are statistically averaged out.  The resulting structure allows for computer simulations of large
systems, including a high number of polymers, chains with large molecular weights, and for a long timescale.  Recent mesoscopic descriptions based on various coarse-grained approaches have been presented in the literature and effectively utilized to investigate the demixing of polymer solutions\cite{Yelash}, immiscible polymer blends\cite{Sun}, and star-linear polymer mixtures.\cite{Camargo}
By combining
information obtained from simulations of the coarse-grained system with information obtained
at smaller lengthscales and shorter times, in the context of a multiscale procedure, it is
possible to obtain a complete and exhaustive description of the polymer mixture at
all lengthscales of interest.\cite{Baeurle} We have previously shown that a procedure, which combines information at the two length scales of the effective segment length and the radius of gyration, accurately determines the structural properties of homopolymer melts.\cite{McCarty} This multiscale modeling procedure has been recently extended to treat polymer mixtures.\cite{McCarty2}
 
In this publication, we focus on performing simulations of polymer mixtures where the ensemble of polymers is mapped onto a system of interacting soft,
colloidal particles.  At this level of coarse-graining each polymer is represented as one soft colloidal
particle centered at the polymer center-of-mass. The concept of modeling flexible polymer chains as soft spheres was originally proposed by Flory and Krigbaum\cite{Flory} for dilute polymer solutions; however, the resulting effective potential exhibited an incorrect trend in repulsive interactions with increasing chain length or polymer density. For dilute to semidilute solutions, Louis and co-workers were able to obtain correct scaling behavior for polymer chains mapped onto soft spheres using liquid state theory in conjunction with Monte Carlo simulations.\cite{HNS00} This approach was later extended to poor solvent conditions where coils contract to avoid contact with the solvent.\cite{LOUIS2003} Importantly, this work shows that for a given set of state conditions, a simple effective pair potential, $v(r)$, is capable of reproducing the radial distribution function, $g(r)$, which depends on many body interactions.
Due to this reduction in interactions that need to be calculated at each computational step, mesoscale simulations of polymers represented as soft colloids, are useful in determining many structural as well as dynamical properties of polymer blends, including how morphologies develop depending on the thermodynamic parameters of the system\cite{Malescio}. The advantage of this extreme coarse-graining of the polymer is that it is possible to simulate very large ensembles of particles, i.e. adopt large simulation boxes, with a modest increase in the computational time. Such an extreme level of coarse-graining becomes important for simulations of systems where the relevant range of length scales is particularly large, for example in micellar aggregates of ionic surfactants.\cite{Baeurle2004}

The implementation of mesoscale simulations requires  attaining the effective
``bare'' potential, $v^{cc}(r)$, that characterizes pairwise
decomposable interactions between molecular center-of-mass (com)
sites. 
Since  $v^{cc}(r)$ corresponds to a free energy obtained from
the monomer frame of reference, it depends on the thermodynamic state
of the system, as specified by the density, temperature, polymer molecular
weight, and composition of the mixture.  

The determination of a reliable, fully transferable, coarse-grained potential is a highly desirable goal.\cite{Noid}
While phenomenological as
well as rigorously numerical approaches have been described in the
literature \cite{DHALL,KREMR,HNS00,HNS01} to determine effective
coarse-grained pair potentials for polymer systems, their reliance on
acquired microscopic simulation data partially defeats the gain in
computational time that is possible with a coarse-graining procedure.
Typically, mesoscopic potentials are optimized to full atomistic simuluations under a given set of thermodynamic conditions, limiting their transferability. Moreover, the optimization procedure generally starts from the mean-force potential, which is the effective potential between two particles \textit{in the field} of the surrounding particles, and is conceptually different from the bare potential,  $v^{cc}(r)$. 
 We have developed an analytical formalism to calculate the effective potential, $v^{cc}(r)$, starting from liquid state theory and solving the Ornstein-Zernike equation.\cite{YAPRL,BLNDS} The potential obtained in such a manner is explicitly related to the structural parameters of the polymer, making it specific to any system we desire to simulate, but also fully transferable to systems with different molecular structure and thermodynamic conditions.

More specifically, our non-phenomenological expressions for the com-com total pair
intermolecular correlation functions, $h^{cc}_{\alpha\beta}(r)$,
between self $(\alpha\alpha)$ and cross $(\alpha\beta)$ interactions,
are obtained from a generalized Ornstein-Zernike integral equation for
binary polymer mixtures where atomistic sites are accounted for as real sites, while coarse-grained sites are treated
as auxiliary sites.  The equation formally bridges information
from the microscopic (monomer) domain to mesoscopic (molecular)
scales.  The derived equations for $h^{cc}_{\alpha\beta}(r)$ reproduce
well and with no fitting parameters united atom (UA) molecular
dynamics (MD) simulation data in both real and reciprocal spaces.  Our
renormalized description,\cite{YAPRL,BLNDS} correctly recovers the known analytical expressions for
density and concentration fluctuations of mixtures of colloidal liquids, by Kirkwood and Buff,\cite{KB} and Bhatia and Thornton,
\cite{BT} slightly modified as our expressions account for soft potentials instead of hard-core ones.  Together, these tests provide a
benchmark which supports the foundation of our renormalization
procedure.  

In the current publication we extend our previous work to demonstrate that the derived coarse-grained effective potential, when input to mesoscale MD simulations of binary polymer blends, is capable of reproducing the large scale structure of the polymer, which depends on many body interactions at the monomer level. We demonstrate that our approach is useful to produce mesoscale simulations of binary homopolymer mixtures at various temperatures and concentrations approaching the spinodal. 
In this way, we are able to test the derived potential in terms of how well it reproduces the mesoscopic structure of the mixture over different regions of the phase diagram, where atomistic level simulations are computationally exhaustive.
We make use of the hypernetted-chain (HNC) closure to calculate the effective potential,
$v^{cc}_{\alpha\beta}(r)$, as a function of the total correlation function, $h^{cc}_{\alpha\beta}(r)$.  In turn, the pair potentials, $v^{cc}_{\alpha\beta}(r)$, are used as an input to a
coarse-grained simulation, where polymers interact as soft colloidal
particles.  

Systems investigated are blends of polyethylene,
polyisobutylene, as well as polypropylenes in their head-to-head,
istotactic, as well as syndiotactic forms. We show that our method is robust and allows for the equilibrium structural properties of the fluid mixture to be readily calculated under any thermodynamic parameters of interest. While the focus here is on static properties, the derived potential is widely applicable to non-equillibrium systems, and may be useful in other methods commonly employed, such as dissipative particle dynamics\cite{Groot}, which currently rely on numerical potentials. 

To demonstrate the reliability of our mesoscale simulations, we first consider systems which are far from the spinodal temperature, for which the liquid is well-mixed, and show that in this regime, our $v^{cc}_{\alpha\beta}(r)$ reproduces quantitatively the liquid structure at the level of com obtained from UA MD simulations\cite{HEINE,JARAM}, establishing consistency between the different levels of description.
We then extend our analysis to different thermodynamic conditions for which no UA data is available, focusing on mixtures of head-to-head polypropylene and polyethylene (hhPP/PE) in the miscible region of the phase diagram. The temperature dependence is expressed in terms of a single interaction parameter that enters our simulation through the analytical form of the potential and depends on the local interactions between monomers. From the simulation trajectories, we calculate pair distribution functions which depend on the parameters of the system and manifest the trends for demixing of the coarse-grained mixture.  From this treatment, and from the fact that our expressions recover known expressions for colloidal liquids, we calculate the concentration fluctuation contribution to the scattering function, $S(k)$, which is related to the scattered light intensity measured in experimental studies of critical binary polymer mixtures.  In this way, we are able to test the derived potential in terms of how well it reproduces the mesoscopic structure of the mixture over different regions of the phase diagram. Results show that the effective potential between center of masses of polymers in a mixture can be used to produce mesoscopic simulations under a variety of thermodynamic conditions, making the procedure useful to a number of application in polymer physics.
To test the versatility of our approach we show how the proposed theory may be applied to blends that present a lower critical solution temperature (LCST). We study the hhPP/PIB mixture using the temperature dependence of the $\chi$ parameter from the literature and run mesoscale simulations at various temperatures. The concentration fluctuation structure factor shows that fluctuations in concentration become smaller as the temperatures decreases, as it is expected.

The current publication is organized as follows.  In
Section \ref{SX:HOFR}, a review of our derivation for
$h^{cc}_{\alpha\beta}(r)$ is provided.  These results are then used in
Section \ref{SX:BACK} for the calculation of
$v^{cc}_{\alpha\beta}(r)$, obtained from the HNC closure.  In Section
\ref{SX:MESO}, our representations for $v^{cc}_{\alpha\beta}(r)$ are
implemented in mesoscopic simulations of binary polymer blends, where
the mixture is modeled as an assembly of soft, interacting colloidal
particles.  Section \ref{SX:H} compares predicted total correlation functions, analytical and from mesoscale simulations, with 
data from united atom molecular dynamic simulations.
In Section \ref{SX:SCAT} data obtained from mesoscopic simulations are used to calculated liquid partial structure factors which are used to determine the phase diagram of the mixture. Section \ref{SX:DEBYE} investigates how corrections to the Debye form factor affect the precision of the predicted total correlation functions. Finally, in  
Section \ref{SX:LCST} we show how our approach is effective in predicting high temperature demixing for the LCST blend hhPP/PIB.
\section{Mesoscopic Pair Correlation Functions for
Asymmetric Binary Polymer Blends}
\label{SX:HOFR}
Pair correlation functions (pcfs) comprise a standard approach in
liquid-state theory to describe the structure of liquids.  Since it is
generally sufficient to account for two-body correlations, pcfs can be
employed to determine all structural and thermodynamic properties of a
system \cite{HNSMC}.  Moreover, these are input to
the equation of motion describing cooperative dynamics, where the effect of the 
surrounding medium on single-chain dynamics is taken into
account \cite{MMACR,MPRLP}.  In the context of our
coarse-graining procedure, pcfs provide a convenient manner of
calculating, with the aid of an appropriate closure, the bare pair potential needed to perform 
mesoscopic simulations.  In this section,
a brief review is given of the theoretical formalism we previously developed to describe coarse-grained
 binary mixtures of asymmetric polymer blends \cite{YAPRL,BLNDS}.

Our model for a binary blend consists of $A$ and $B$ homopolymers,
having $N_A$ and $N_B$ monomer sites with segment lengths $\sigma_A$
and $\sigma_B$, respectively.  For simplicity, these monomer sites are
taken to span equivalent volumes so that the polymer volume fraction
is given by $\phi=n_AN_A/(n_AN_A+n_BN_B)$, where $n_\alpha$ is the
number of molecules of type $\alpha$ in the mixture with
$\alpha\in\{A,B\}$.  While $\rho=(n_AN_A+n_BN_B)/V$ quantifies the
total number of monomer sites contained in a region of space spanned
by $V$, the site and chain number densities for molecules of type
A are given by $\rho_{A}=n_AN_A/V=\phi\rho$ and
$\rho_{c,A}=n_A/V$, respectively.

The derivation of our analytical expressions for total intermolecular
com pcfs in a polymer mixture extends from a procedure outlined by
Krakoviack, \emph{et al.}\ \cite{KRAKO} for homopolymer solutions.  The
key step in this approach is to include molecular coms as
\textit{auxiliary} sites, along with monomer sites serving as
\textit{real} sites.  The generalized Ornstein-Zernike integral
equation is solved in reciprocal space and is given by
\begin{equation}
\label{EQ:OZGE}
\mathbf{H}(k) = \mathbf{\Omega}(k)\mathbf{C}(k)
\left[\mathbf{\Omega}(k)+\mathbf{H}(k)\right]\, ,
\end{equation}
where $\mathbf{H}(k)$ is the matrix of total intermolecular
pcfs, $\mathbf{C}(k)$ is the matrix of direct intermolecular pcfs, and
$\mathbf{\Omega}(k)$ represents the matrix of intramolecular pcfs.
Specializing to the case of a binary polymer mixture, each matrix in
Equation \ref{EQ:OZGE} is of rank four, composed of four $2\times2$
blocks that account for monomer-monomer ($mm$), com-com ($cc$), and
the corresponding cross ($cm$ and $mc$) interactions.  For instance,
\begin{equation}
\mathbf{H}(k)=
\left[
\begin{array}{ll}
\mathbf{H}^{mm}(k) & \mathbf{H}^{mc}(k) \\
\mathbf{H}^{cm}(k) & \mathbf{H}^{cc}(k)
\end{array}
\right]\, .
\label{EQ:HBLK}
\end{equation}
The remaining matrices in Equation \ref{EQ:OZGE} follow an
arrangement analogous to that of Equation \ref{EQ:HBLK}.  Each block in
Equation \ref{EQ:HBLK} contains self $(\alpha\alpha)$ interactions along
its diagonal, whereas cross $(\alpha\beta)$ interactions occupy
off-diagonal positions.

As a next step, the individual block elements that define the matrices
in Equation \ref{EQ:OZGE} are defined.  The intermolecular total pcf
matrix $\mathbf{H}(k)$ is composed of the chain-averaged
monomer-monomer pcfs ${H}_{\alpha\beta}^{mm}(k) =
\rho_{\alpha}\rho_{\beta} h_{\alpha\beta}^{mm}(k)$, the
com-monomer pcfs $\mathit{H}_{\alpha\beta}^{cm}(k) = \rho_{c,\alpha}
\rho_{\beta}h^{cm}_{\alpha\beta}(k) =
\mathit{H}^{mc}_{\beta\alpha}(k)$, and com-com pcf
$\mathit{H}_{\alpha\beta}^{cc}(k) =
\rho_{c,\alpha}\rho_{c,\beta}h_{\alpha\beta}(k)$.  Note that in
general, $h_{\alpha \beta}^{cm}(k) = h_{\beta \alpha}^{mc}(k)$ while
$h_{\alpha \beta}^{cm}(k) \neq h_{\alpha \beta}^{mc}(k)$ when $\alpha
\neq \beta$.  The intramolecular pcf matrix $\mathbf{\Omega}(k)$ is
similarly composed of $\mathit{\Omega}^{mm}_{\alpha\beta}(k) =
\rho_{\alpha}\rho_{\beta} \omega_{\alpha\beta}^{mm}(k)
\delta_{\alpha\beta} z_{\alpha\beta}$,
$\mathit{\Omega}^{cm}_{\alpha\beta}(k) = \rho_{c,\alpha}\rho_{\beta}
\omega_{\alpha\beta}^{cm}(k) \delta_{\alpha\beta} z_{\alpha\beta} =
\mathit{\Omega}^{mc}_{\beta\alpha}(k)$, and
$\mathit{\Omega}^{cc}_{\alpha\beta}(k) =
\rho_{c,\alpha}N_\beta\rho_{c,\beta} \delta_{\alpha\beta}
z_{\alpha\beta}$, where $z_{\alpha\beta} =
[\phi_{\beta}\rho(2-\delta_{\alpha\beta})]^{-1}$.  Finally, the
intermolecular direct total pcf matrix $\mathbf{C}(k)$ acquires a much
simpler structure when including the approximation that any
interaction involving auxiliary sites is negligible.\cite{SAM2006}  Under these conditions, the only non-zero block in
$\mathbf{C}(k)$ involves monomer-monomer pcfs, defined as
$C^{mm}_{\alpha\beta}(k)=c^{mm}_{\alpha\beta}(k)$.

Using the matrix definitions described above, Equation \ref{EQ:OZGE} is
solved to obtain the intermolecular mesoscopic total pcfs, which are
given by the relation
\begin{eqnarray}
\label{EQ:HCCK}
h_{\alpha \beta}^{cc}(k) =
\left[\frac{\omega_{\alpha\alpha}^{cm}(k)\omega_{\beta\beta}^{cm}(k)}
{\omega_{\alpha\alpha}^{mm}(k)\omega_{\beta\beta}^{mm}(k)}\right]
h_{\alpha \beta}^{mm}(k)\, .
\end{eqnarray}
Upon inspection, it is readily seen that Equation \ref{EQ:HCCK} formally connects com distribution functions to
monomer-monomer intra- and intermolecular distribution functions.  In
this manner, one calculates mesoscale properties from information on
the local polymer scale.  As mentioned before, this feature is relevant 
because properties on the mesoscale ultimately depend
on small-scale interactions.

To obtain analytical solutions for $h^{cc}_{\alpha\beta}(k)$, a brief
description is given for each of the correlation functions entering
into Equation \ref{EQ:HCCK}.  The com-monomer intramolecular pcf can be
approximated in reciprocal space with a Gaussian distribution as
\begin{equation}
\omega^{cm}_{\alpha\alpha}(k)=N_\alpha
e^{-k^2R_{g\alpha}^2/6} \, ,
\label{EQ:WCMK}
\end{equation}
with the molecular radius
of gyration defined as $R_{g\alpha}=(N/6)^{1/2}\sigma_{\alpha}$.  On
the other hand, the monomer-monomer intramolecular pcf is given by the
Debye formula,
\begin{equation}
\label{EQ:DBYE}
\omega^{mm}_{\alpha\alpha}(k) =
\frac{2N_\alpha\left[e^{-k^2R_{g\alpha}^2}-1+k^2R_{g\alpha}^2\right]}
{k^4R_{g\alpha}^4} \, .
\end{equation}
For analytical convenience, however, it is costumary to
approximate Equation \ref{EQ:DBYE} with its Pad\'e approximant given by\cite{DOI}
\begin{equation}
\label{EQ:PADE}
\omega^{mm}_{\alpha\alpha}(k) \approx
\frac{N_\alpha}{1+k^2R_{g\alpha}^2/2} \, .
\end{equation}
Although approximated, inclusion of Equation \ref{EQ:PADE} allows for a convenient analytic expression for $h^{cc}(r)$ given below by Equation \ref{EQ:HCCR}, which has been shown to give good agreement with simulations for the total pair correlation function in both real and reciprocal spaces\cite{YAPRL}. We discuss the implication for this approximation in detail below. In the current publication, we use both Equation \ref{EQ:DBYE} and Equation \ref{EQ:PADE} for $\omega^{mm}(k)$, and compare the resulting mesoscopic $h^{cc}(k)$  from Equation \ref{EQ:HCCK}.

The respective monomer-monomer intermolecular total pcfs used are
taken from the thread limit of the Polymer Reference Interaction Site
Model \cite{SWC94,SWC97} (PRISM).  The initial analytical
treatment in the context of PRISM for polymer mixtures \cite{TNGSW}
has been extended by Yatsenko et al.\ \cite{BLNDS} to account for
chain asymmetry effects in the system. In this approach, a new parameter enters the formalism, $\gamma=\sigma_B/\sigma_A$, which defines the monomer asymmetry.

While the thread model for
polymer chains coarsely describes the liquid structure on local
scales, it accurately captures the onset of the ``correlation hole''
effect at a length scale of $R_g$.  Given that the spatial dimension
of interest in our description is $R_g$, the thread limit of PRISM is
an adequate representation for the intended purpose of the present
work.  The solutions are given by \cite{YAPRL,BLNDS}
\begin{eqnarray}
\label{EQ:HMMR}
h_{AA}^{mm}(r)
&= \frac{3}{{\pi\rho r \sigma_{AB}^2}}
\bigg[\frac{{1-\phi}}{\phi} e^{-r/\xi_{\phi}}
+ \gamma^2 e^{-r/\xi_{\rho_{AA}}}
-\frac{1}{{\phi}} \frac{\sigma_{AB}^2}{\sigma_A^2}
e^{-r/\xi_{cA}} \bigg] \nonumber,   \\
&h_{BB}^{mm}(r) =  \frac{3}{{\pi\rho r \sigma_{AB}^2}}
\bigg[\frac{\phi}{1-\phi}
e^{-r/\xi_{\phi}}
+\gamma^{-2}
e^{-r/\xi_{\rho_{BB}}} \nonumber, \\
&-\frac{1}{1-\phi} \frac{\sigma_{AB}^2}{\sigma_B^2}
e^{-r/\xi_{cB}}\bigg] \ , \\
&h_{AB}^{mm}(r)  =  \frac{3}{{\pi\rho r \sigma_{AB}^2}}
\bigg[-e^{-r/\xi_{\phi}}+e^{-r/\xi_{\rho_{AB}}}\bigg] \ ,
\nonumber
\end{eqnarray}
where
\begin{eqnarray}
\xi_{\phi}=\frac{\sigma_{AB}}{\sqrt{24\phi(1-\phi)\chi_s(1-\chi/\chi_s)}}
\ ,
\label{EQ:XI}
\end{eqnarray}
is the length scale governing concentration fluctuations,
which diverges at the spinodal temperature.  Here, $\chi$ is a single interaction parameter that depends on the specific nearest neighbor pair energies between two AA, AB, or BB monomers, and is given by
$\chi=\epsilon_{AB}-(\epsilon_{AA}+\epsilon_{BB})/2$.  In a mesoscopic treatment which averages out the specific monomer interactions, $\chi$ is an input parameter corresponding to the temperature dependance of a specific polymer architecture.  From our definitions it clear that the quantity  $\chi/\rho$ is the analog of the Flory-Huggins interaction parameter, and at the spinodal
temperature $\chi \rightarrow \chi_s$, where $\chi_s=[2 N_A
\phi]^{-1}+ [2 N_B (1-\phi)]^{-1}$.  The quantity, $(1-\chi/\chi_s)$ can be seen as a reduced temperature that indicates how far the system is from its spinodal temperature. Also in Equation \ref{EQ:HMMR},
$\xi_{c\alpha}=R_{g\alpha}/2^{1/2}$ is the length scale of the
correlation hole while $\xi_{\rho \alpha \beta}^{-1}= \pi\rho
\sigma_{\alpha\beta}^2/3+\xi_{c \alpha \beta}^{-1}$ is the density
correlation length scale with $\sigma_{\alpha\beta}^2 =
\phi_{\beta}\sigma_{\alpha}^2 + \phi_{\alpha}\sigma_{\beta}^2$.  This
latter definition reintroduces finite-size effects, local
semiflexibility, and branching that pertain to each component through
a melt-like description.  The effective segment length scales are
determined from the radius of gyration of each component polymer,
through the relation $\sigma_{\alpha}=(6/N_{\alpha})^{1/2}R_g$.

Inserting the definitions from Eqs.\ (\ref{EQ:WCMK}), (\ref{EQ:PADE}),
and (\ref{EQ:HMMR}) into (\ref{EQ:HCCK}), the intermolecular total
pcfs at the com level read
\begin{eqnarray}
\label{EQ:HCCR}
h^{cc}_{AA}(r)& = & \frac{1-\phi}{\phi}I^{\phi}_{AA}(r)+
\gamma^{2}I^{\rho}_{AA}(r) \ ,
\nonumber \\
h^{cc}_{BB}(r) & = & \frac{\phi}{1-\phi}I^{\phi}_{BB}(r)+
\gamma^{-2}I^{\rho}_{BB}(r) \ , \\
h^{cc}_{AB}(r) & = & -I^{\phi}_{AB}(r)+
I^{\rho}_{AB}(r) \ ,
\nonumber
\end{eqnarray}
where $I^{\phi}_{\alpha\beta}(r)$ and
$I^{\rho}_{\alpha\beta}(r)$ identify the concentration and density
fluctuation contributions, respectively.  We introduce here a compact
notation with the function $I^{\lambda}_{\alpha\beta}(r)$ defined as
\begin{eqnarray}
&I^{\lambda}_{\alpha\beta}(r)= \frac{3}{4}\sqrt{\frac{3}{\pi}}\frac{\xi_{\rho}^\prime}
{R_{g\alpha\beta}}\vartheta_{\alpha\beta1}
\left(1-\frac{\xi_{c\alpha\beta}^2}{\xi_{\lambda}^2}\right) e^{
  -3r^2/(4R_{g\alpha\beta}^2)} \nonumber  \\
&-\frac{1}{2}\frac{\xi_{\rho}^\prime}{r}\vartheta_{\alpha\beta2}
\left(1-\frac{\xi_{c\alpha\beta}^2}{\xi_{\lambda}^2}
  \right)^2 e^{
  R_{g\alpha\beta}^2/(3\xi_{\lambda}^2)} \nonumber \\
&  \times \left[ e^{r/\xi_{\lambda}}\mbox{erfc} 
\left(\frac{R_{g\alpha\beta}}{\xi_{\lambda} \sqrt{3}}+
 \frac{r\sqrt{3}}{2R_{g\alpha\beta}}\right) \right. \nonumber \\ 
&  \left. -e^{ -r/\xi_{\lambda}}\mbox{erfc}\left(\frac{R_{g\alpha\beta}}{
\xi_{\lambda}\sqrt{3}}-
\frac{r\sqrt{3}}{2R_{g\alpha\beta}} \right) \right]
\end{eqnarray}
and
\begin{eqnarray}
&\vartheta_{\alpha\beta1}&=
\frac{\left(1-\frac{\xi_{c\alpha\alpha}^2\xi_{c\beta\beta}^2}
{\xi_{c\alpha\beta}^2\xi_{\lambda}^2}\right)}
{\left(1-\frac{\xi_{c\alpha\beta}^2}{\xi_{\lambda}^2}\right)},\\
&\vartheta_{\alpha\beta2}&= 
\frac{\left(1-\frac{\xi_{c\alpha\alpha}^{2}}{\xi_{\lambda}^2}\right)
\left(1-\frac{\xi_{c\beta\beta}^2}{\xi_{\lambda}^2}\right)}
{\left(1-\frac{\xi_{c\alpha\beta}^2}{\xi_{\lambda}^2}\right)^2},
\end{eqnarray}
where $\xi_{\lambda}\in \{\xi_{\phi},\xi_{\rho}\}$ and
$\xi_{\rho}'=3/(\pi\rho\sigma_{AB}^2)$.  Radii of gyration in the
blend are defined such that $2R_{g \alpha \beta}^2 = R_{g\alpha}^2 +
R_{g\beta}^2=4\xi_{c\alpha \beta}^2$, with $\xi_{c\alpha\alpha} \equiv
\xi_{c\alpha}$.

The development presented here is the required input to the derivation
of the effective pair interaction potentials, a topic that will be
addressed in the following section.

\section{The Effective Soft Core Potential}
\label{SX:BACK}
Our theoretical approach is based on an integral equation description of intermolecular pair correlation functions.  A closure approximation is needed to connect these liquid structure functions to the effective pair interaction potentials which are required to perform, in our case, mesoscopic simulations of polymer mixtures mapped onto ensembles of soft colloidal particles.  Since the fundamental units in our description interact through a soft-core potential, use is made of the hypernetted-chain (HNC) closure, which is known to be accurate for such systems.\cite{McQuarrie}  The relationship connecting the effective pair interaction potential $v^{cc}_{\alpha\beta}(r)$ with pcfs is given by the HNC closure as
\begin{equation}
(k_BT)^{-1} v^{cc}_{\alpha\beta}(r) = h^{cc}_{\alpha\beta}(r)
- \ln\left[1+h^{cc}_{\alpha\beta}(r)\right]
- c^{cc}_{\alpha\beta}(r)\, ,
\label{EQ:VOFR}
\end{equation}
where $c^{cc}_{\alpha\beta}(r)$ is the direct pcf.  Following the matrix definitions presented in Section
\ref{SX:HOFR}, and taking our system to be a simple liquid comprised
by soft colloidal particles, the direct pair correlation functions are
defined by
\begin{eqnarray}
c^{cc}_{\alpha\alpha}(k) & = & \frac{1}{\rho_{c,\alpha}}
- \frac{S^{cc}_{\beta\beta}(k)}
{\left(\rho_{c,\alpha}+\rho_{c,\beta}\right)
|\mathbf{S}_{cc}(k)|} \nonumber  \ , \\
c^{cc}_{\alpha\beta}(k) & = & \frac{S^{cc}_{\alpha\beta}(k)}
{\left(\rho_{c,\alpha}+\rho_{c,\beta}\right)|\mathbf{S}_{cc}(k)|} \ ,
\label{EQ:COFR}
\end{eqnarray}
where $S^{cc}_{\beta \beta}$ and $S^{cc}_{\alpha \beta}$ are the static structure factors for a binary mixture, and  $|\mathbf{S}_{cc}(k)| = S^{cc}_{AA}(k)S^{cc}_{BB}(k) -
\left[S^{cc}_{AB}(k)\right]^2$ is the determinant of the mesoscopic
static structure factor matrix.  For a binary mixture these static structure factors are given by 
\begin{eqnarray}
\label{EQ:XEFF3}
S_{AA}(k) &=& \phi +\phi^2 \rho_{ch} h^{cc}_{AA}(k) \nonumber \ , \\
S_{BB}(k) &=& 1- \phi +(1- \phi )^2 \rho_{ch} h^{cc}_{BB}(k) \ , \\
S_{AB}(k) &=& \phi (1-\phi ) \rho_{ch} h^{cc}_{AB}(k) \nonumber \ ,
\end{eqnarray}
where the total chain density, $\rho_{ch}=\rho/N$. By inserting Eqs.\ (\ref{EQ:HCCR})
and (\ref{EQ:COFR}) into (\ref{EQ:VOFR}), the
$v^{cc}_{\alpha\beta}(r)$ are obtained.

Since the potential is obtained from an inversion procedure utilizing the HNC closure, the  adequacy of this method is limited to systems for which the HNC is valid, which includes the weakly interacting soft particles modeled here. However, for systems of hard spheres or for low density ionic fluid models, such as the restricted primitive model (RPM), the HNC has been shown to be inaccurate\cite{HOYE}, and more sophisticated closures are required.\cite{HNS01, ZERAH}

The analytical solution of our mesoscopic approach represents an advantage to previous work, where effective pair potentials are derived from simulations performed on a microscopic level.  Such a requirement partially defeats the computational time gains behind a coarse-graining procedure.  Overall, our approach bypasses the need to perform atomistic simulations for each thermodynamic state point of interest, which is necessary in numerical implementations since the effective pair interaction potentials depend on the state of the system.  This can be readily appreciated from the pcfs that enter into the HNC closure, which are themselves state-dependent.

We investigated the effect that the use of the Debye formalism, Equation \ref{EQ:DBYE}, or of its Pad\'e approximant, Equation \ref{EQ:PADE}, for the monomer form factor in the denominator of Equation \ref{EQ:HCCK} has on the calculation of the potential. The  Pad\'e approximant is less precise than the Debye equation, but it allows for the analytical solution of the total correlation functions, Equation \ref{EQ:HCCR}.
We observe that when Equation \ref{EQ:PADE} is used, singular points arise in the low k regime in the solution of Equation \ref{EQ:COFR} for $c^{cc}(k)$ as the determinant of the mesoscopic static structure factor, $|\bold{S}_{cc}(k)|=S^{cc}_{AA}(k)S^{cc}_{BB}(k)-[S^{cc}_{AB}(k)]^2$, passes through zero. This corresponds to an unphysical region of negative compressibility. 
When Equation \ref{EQ:DBYE} is used instead, such singular points do not arise. 

For homopolymer melts, it has previously been determined that the singularities in $c^{cc}(k)$ occur as a result of the intrinsic error introduced in Equation \ref{EQ:PADE} by the Pad\'e approximation.\cite{SAM2006} In order to obtain a usable form of the effective potential from Equation \ref{EQ:VOFR}, we tested two schemes: in scheme 1, we enforced the condition that $c^{cc}(k=0) \leq c^{cc}(k) \leq 0$ for low k, which effectively eliminates any singularities from the direct correlation function; in scheme 2, we enforced the isothermal compressibility limit, such that for regions where  $|\bold{S}_{cc}(k)| \leq |\bold{S}_{cc}(0)| $, we truncated $h^{cc}(k)$ so that $h^{cc}_{\alpha \beta}(k)=h^{cc}_{\alpha \beta}(k=0)$. The two schemes are equivalent and give identical results within the precision of our calculation. This is so, because polymer liquids are almost incompressible.

\begin{figure}[!ht]
\centering
\begin{tabular}{c}
\includegraphics[scale=0.8]{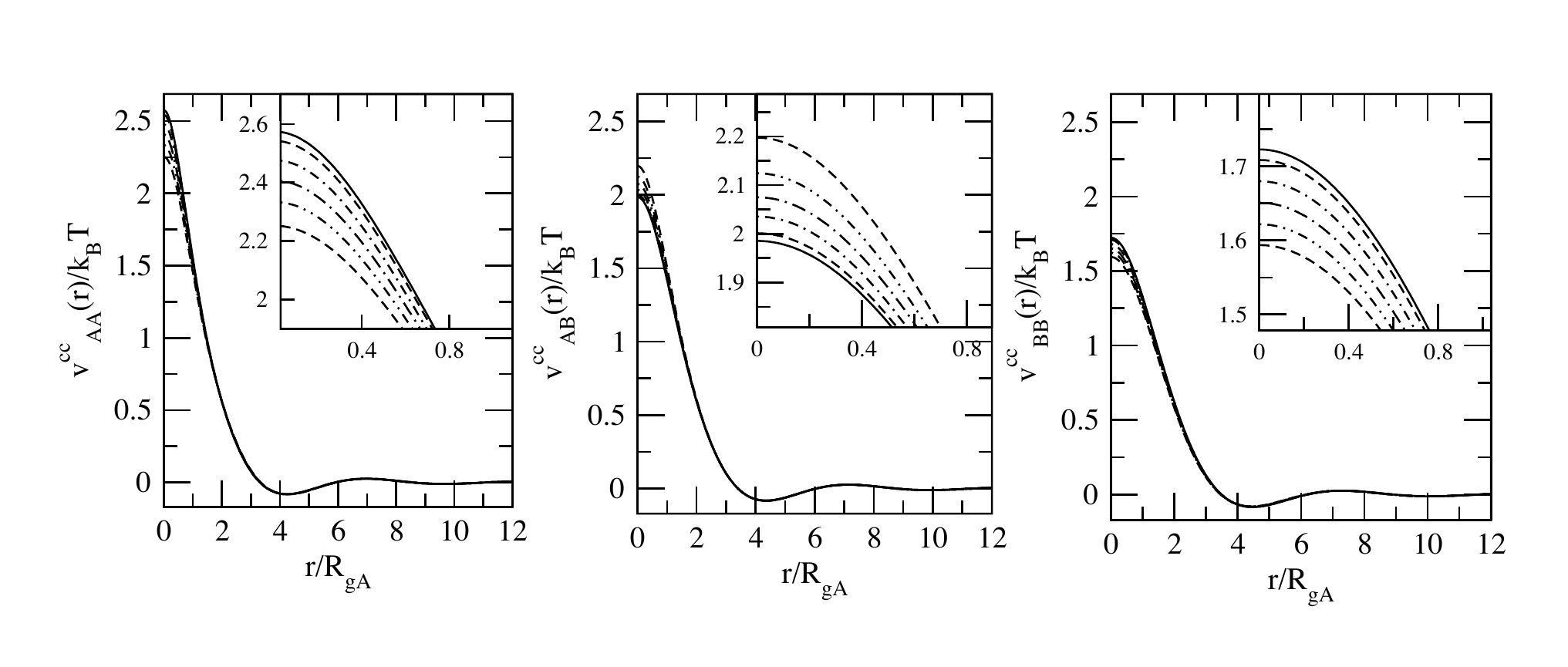}
\\
\includegraphics[scale=0.8]{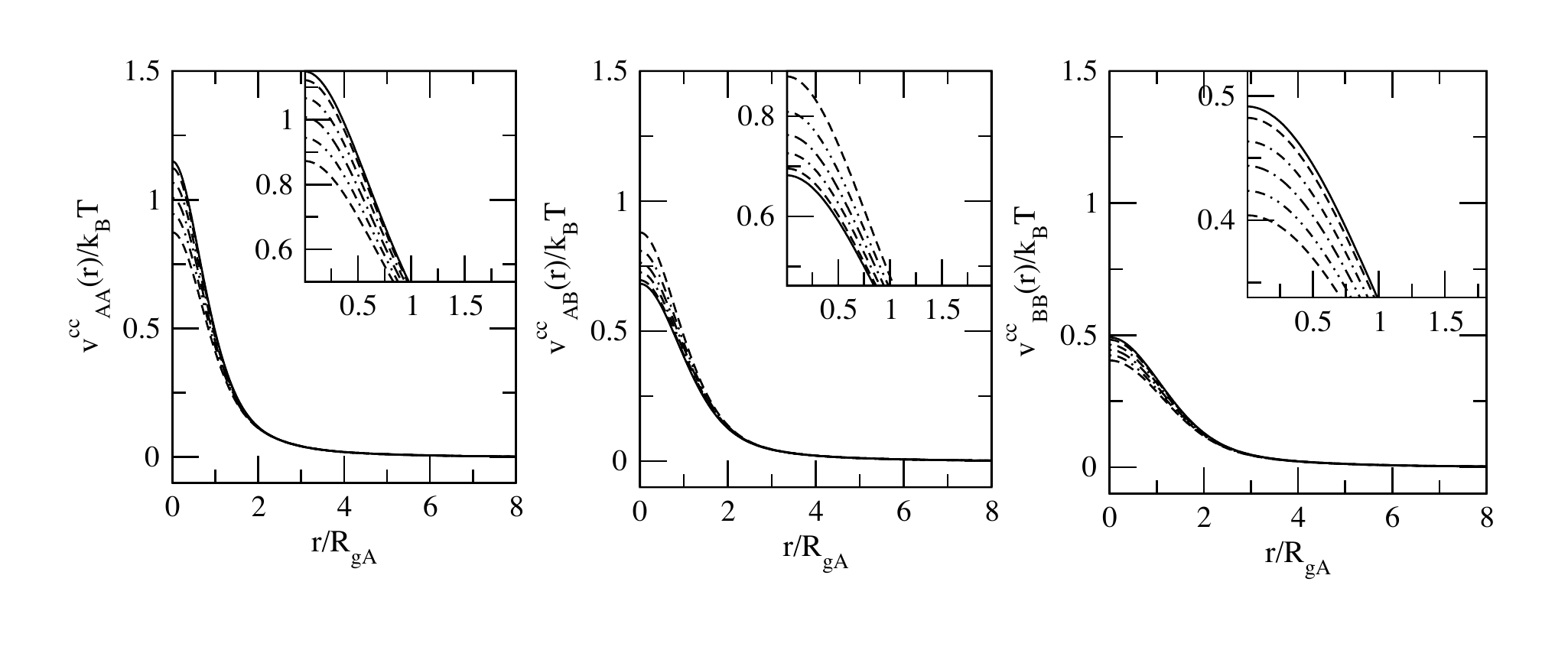}
\end{tabular}
\caption{\small{Comparison of the effective pair interaction potential
$v_{\alpha\beta}(r)$ derived from the HNC closure for the hhPP/PE
blend, $\phi=0.5$, with $\chi/\chi_s \in \{0.0, 0.1, 0.3, 0.5, 0.7, 0.9\}$. The upper panels show $v^{cc}(r)$ obtained via the Pad\'e approximation and the lower panels show $v^{cc}(r)$ from the Debye form. The inset highlights the change in the repulsive part of the potential as the reduced temperature is changed. The solid line represents the athermal regime ($\chi/\chi_s =0.0$). In both the AA and BB curves, the repulsive component decreases as the system approaches the spinodal ($\chi/\chi_s =1$), whereas the AB curve increases.}}
\label{FG:VOFR}
\end{figure}

In this work, we study polymer blends of polyethylene (PE), polyisobutylene (PIB), and polypropylenes in their head-to-head (hhPP), isotactic (iPP), and syndiotactic (sPP) forms.  
The effective pair potential, $v^{cc}(r)$ for interactions of type AA, BB and AB is calculated for the different binary polymer mixtures and for hhPP:PE under different values of $\phi$ and $\chi$ using both the Debye form and Pad\'e form of the intramolecular distribution function (Equation \ref{EQ:DBYE} and \ref{EQ:PADE}).  As a model calculation of the potential, we present the results for the prototypical hhPP/PE polymer blend in Figure \ref{FG:VOFR}, which shows how the potential depends on the reduced temperature $(1-\chi/\chi_s)$.  Input parameters to our theoretical calculations are reported in Table \ref{Table1} as data for the UA simulations against which we test our approach. \cite{HEINE,JARAM}  Although there is a noticeable difference in the potential obtained using either Equation \ref{EQ:DBYE} or \ref{EQ:PADE}, they are qualitatively similar in many respects. For example, under athermal conditions, the mixture is random and the number of $AB$ contacts is in between those of the self terms, $AA$ and $BB$.  Correspondingly, pair interactions accounting for $AB$ contacts must be intermediately repulsive.  This effect is reflected in the plot of $v^{cc}_{\alpha\beta}(r)$.  The $A$-type (flexible hhPP) particles display the highest repulsive response as a consequence of their stronger correlation hole effect.  The inset of Figure \ref{FG:VOFR} highlights the change in the repulsive component in the potential, as the ratio, $\chi/\chi_s$, is varied. 

\begin{table}[h!]
\caption{Polyolefin blends ($T=453$ K, $N_A=N_B=96$)}
\renewcommand{\thefootnote}{\alph{footnote}}
\begin{tabular}{cccccc} \hline \hline
Blend[$A/B$] & $\phi$ & $\rho$ [sites/\AA$^3$] &
$R_{gA}$ [\AA] & $\gamma$ & $\chi$\\ \hline
hhPP/PE  & 0.50 & 0.0332 & 12.32 & 1.34 &
$-0.0294+17.58/T$\footnotemark[1]$^,$\footnotemark[2] \\
PIB/PE   & 0.50 & 0.0343 &  9.76 & 1.68 &
$0.00257+4.99/T$\footnotemark[2] \\
PIB/sPP  & 0.50 & 0.0343 &  9.76 & 1.41 &
$\cdots$ \\
sPP/PE   & 0.50 & 0.0328 & 13.89 & 1.19 &
$\cdots$ \\
iPP/PE   & 0.25 & 0.0328 & 11.35 & 1.47 &
$0.005$\footnotemark[3] \\
iPP/PE   & 0.75 & 0.0328 & 11.33 & 1.48 &
$0.01$ \footnotemark[3] \\
hhPP/PIB & 0.50 & 0.0343 & 12.4 & 1.28& $0.027 -11.4/T$\footnotemark[2]$^,$\footnotemark[4]$^,$\footnotemark[5] \\
\hline \hline
\end{tabular}
\footnotetext{$^{\text{a}}$Reference \cite{JEONH}.}
\footnotetext{$^{\text{b}}$Reference \cite{LEEJH}.}
\footnotetext{$^{\text{c}}$Reference \cite{HEINE}.}
\footnotetext{$^{\text{d}}$Reference \cite{JARAM}.}
\footnotetext{$^{\text{e}}$Reference \cite{KRISH}.}
\label{Table1} 
\end{table}

\onecolumngrid

\twocolumngrid
While the full Debye form (Equation \ref{EQ:DBYE}) for the monomer form factor prevents an explicit analytic expression for $h^{cc}(r)$ in the form of Equation \ref{EQ:HCCR}, which was the motivation for adopting the Pad\'e approximation, a numerically obtained $h^{cc}(r)$ still can be readily obtained for any given system and so does not represent a limitation to our approach and avoids any singularities in the low k region for $c^{cc}(r)$. For this reason, in our following calculations the Debye approximation will be preferentially used. 

The potentials  $v^{cc}_{\alpha\beta}(r)$, calculated following the procedures discussed here, are required to carry out the simulations of the polymer liquid on a mesoscopic level.  In the next section, we discuss the implementation of the $v^{cc}_{\alpha\beta}(r)$ to our mesoscopic simulations and in the following sections we compare mesoscale simulation results with UA MD simulations and theoretical predictions. 

\section{Mesoscopic Simulations of Binary Mixtures}
\label{SX:MESO}
In this section, we implement molecular dynamics simulations for the systems presented in Table \ref{Table1}, and we describe our methodology for carrying out mesoscopic simulations. 

In our mesoscopic modeling approach, we implement classical MD simulations, where the ensemble is evolved in the microcanonical $(N,V,E)$ ensemble.  In the initialization step, all particles are placed on a cubic lattice with periodic boundary conditions.  We use reduced units such that all length units are scaled by $R_g$ $(r^*=r/R_g)$ and energies scaled by $k_BT$. The type of particle that occupies a particular site is determined at random.  Even though the molecular center-of-mass coordinates from a UA MD simulation can be used as an initial point, our calculations were started afresh as a more stringent test of our procedure and to allow us to increase the number of particles in the system (or equivalently, the spatial dimension) at will, capturing relevant features of the effective pair interaction potential and to improve the statistical sampling of the ensemble. The potential and its corresponding derivative are entered as tabulated inputs to the program, and linear interpolation was used to determine values not found in the supplied numerical data as the algorithm proceeds. Each site is given an initial velocity pooled from a Mersenne Twister random number generator \cite{MERSN}.  Subsequently, the system is evolved using a velocity Verlet integrator.  Equilibrium is induced in the ensemble by rescaling the velocity at regular intervals, and is manifested in the system when observing a Maxwell-Boltzmann distribution of velocities, a steady temperature, a stabilized Boltzmann $H$-theorem function, and a decayed translational order parameter.

Once the equilibration step is completed,  velocity rescaling is discontinued and trajectories are collected over a traversal of $\sim8R_g$.  The fastest benchmark in our studies is for a duration of $\sim4$ hours for a system consisting of $\sim6000$ particles, performed on a single-CPU workstation.  This compares extraordinarily well with a UA implementation that requires $\sim24$ hours for a system with 1600 particles performed in parallel on a 64-node cluster \cite{JARAM} for an equivalent trajectory.  We stress that our benchmarks for mesoscopic simulations represent an underestimate in the computational time since these have not been subjected to a parallelized algorithm.
\onecolumngrid
\begin{table*}[t!] 
\caption{Mesoscale Simulation Parameters for Blends of hhPP/PE}
\begin{tabular}{lcccc}
  \hline \hline
Form Factor & Interaction Parameter & Particles  & $\phi$ & $L/2$ [$R_g^{-1}$]\\
\hline
Pad\'e &$\chi/\chi_s =  \{0.1, 0.3, 0.5, 0.7\}$ & 5324 & 0.5 &  8.549 \\ 
Debye & $\chi/\chi_s =  \{0.1, 0.3, 0.5\}$ & 5324 & 0.5 &  8.549 \\ 
Debye & $\chi/\chi_s =  0.7$ & 10,648 & 0.5 &  10.771 \\ 
Debye & $\chi = \{0.008, 0.012, 0.016, 0.019\}$ & 10,648 & $\{0.5, 0.7, 0.9\}$ & 10.771 \\
\hline \hline
\end{tabular}
\label{TB:BLNDTHERM}
\end{table*}
\twocolumngrid 
Extensive mesoscale simulations were performed on a typical system, the  hhPP/PE mixture, to investigate the consistency of our approach. Simulations were performed for compositionally symmetric mixtures, but also while approaching the spinodal, $\chi = \{0.008, 0.012, 0.016, 0.019\}$,  while changing  the fraction of A and B species in the melt such that $\phi = \{0.5, 0.7, 0.9\}$.  
Mesoscale simulation parameters for all of the hhPP/PE systems are presented in Table \ref{TB:BLNDTHERM}. For systems with $\chi$ approaching $\chi_s$,  larger simulation boxes, with 10,648 particles, were used to properly account for the increase in the lengthscale of concentration fluctuations. Those systems also required longer equilibration.
These simulations were run using the LONI TeraGrid system\cite{TeraGrid} to facilitate performing numerous simulations at a time.

\section{Total Pair Correlation Functions of the Polymer Mixture from Mesoscale Simulations} 
\label{SX:H}

\begin{figure*}[t!]
\centering
\begin{tabular}{cc}
\begin{minipage}{2.8in}
\includegraphics[scale=0.28]{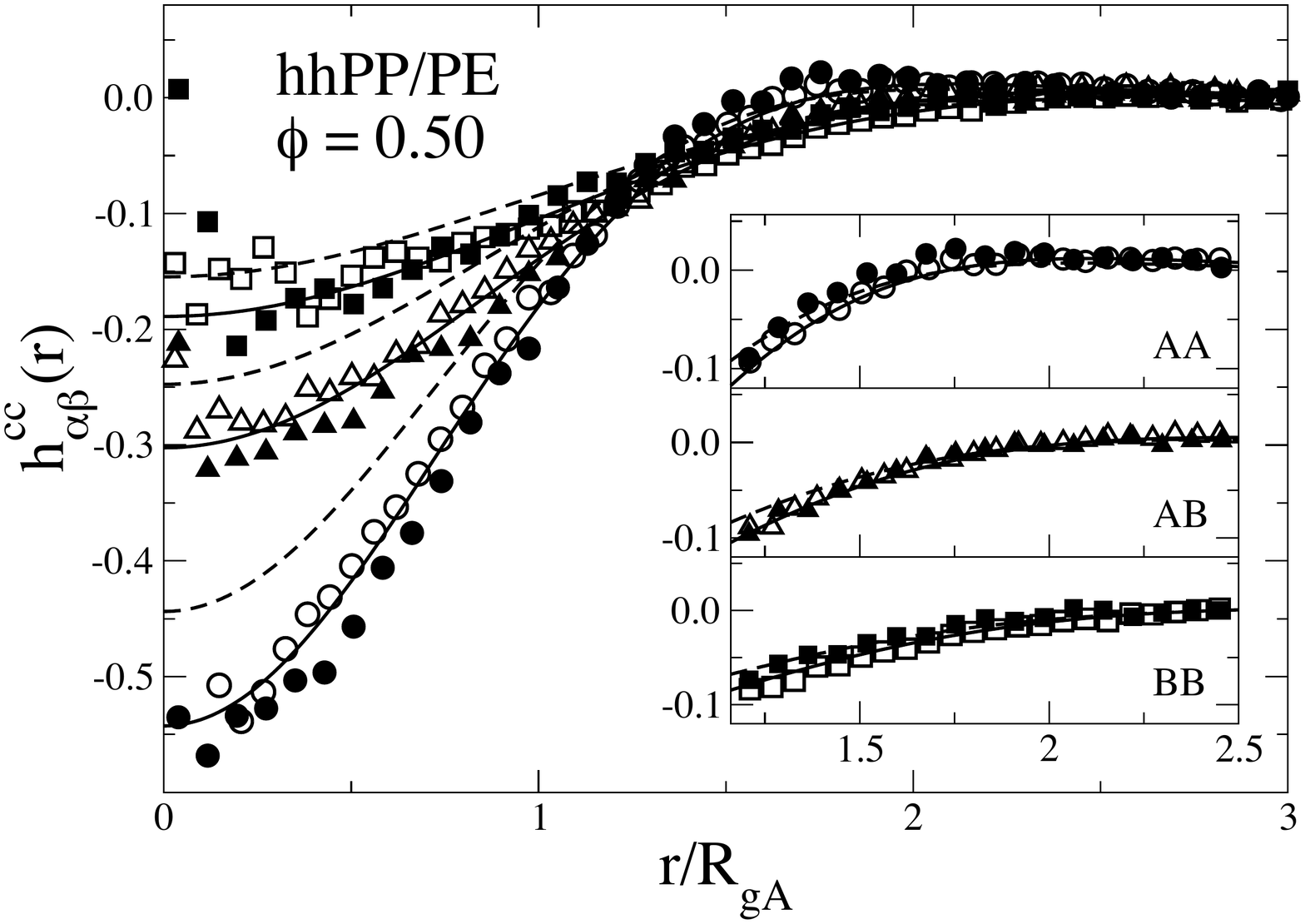}
\end{minipage}
&
\begin{minipage}{3in}
\includegraphics[scale=0.28]{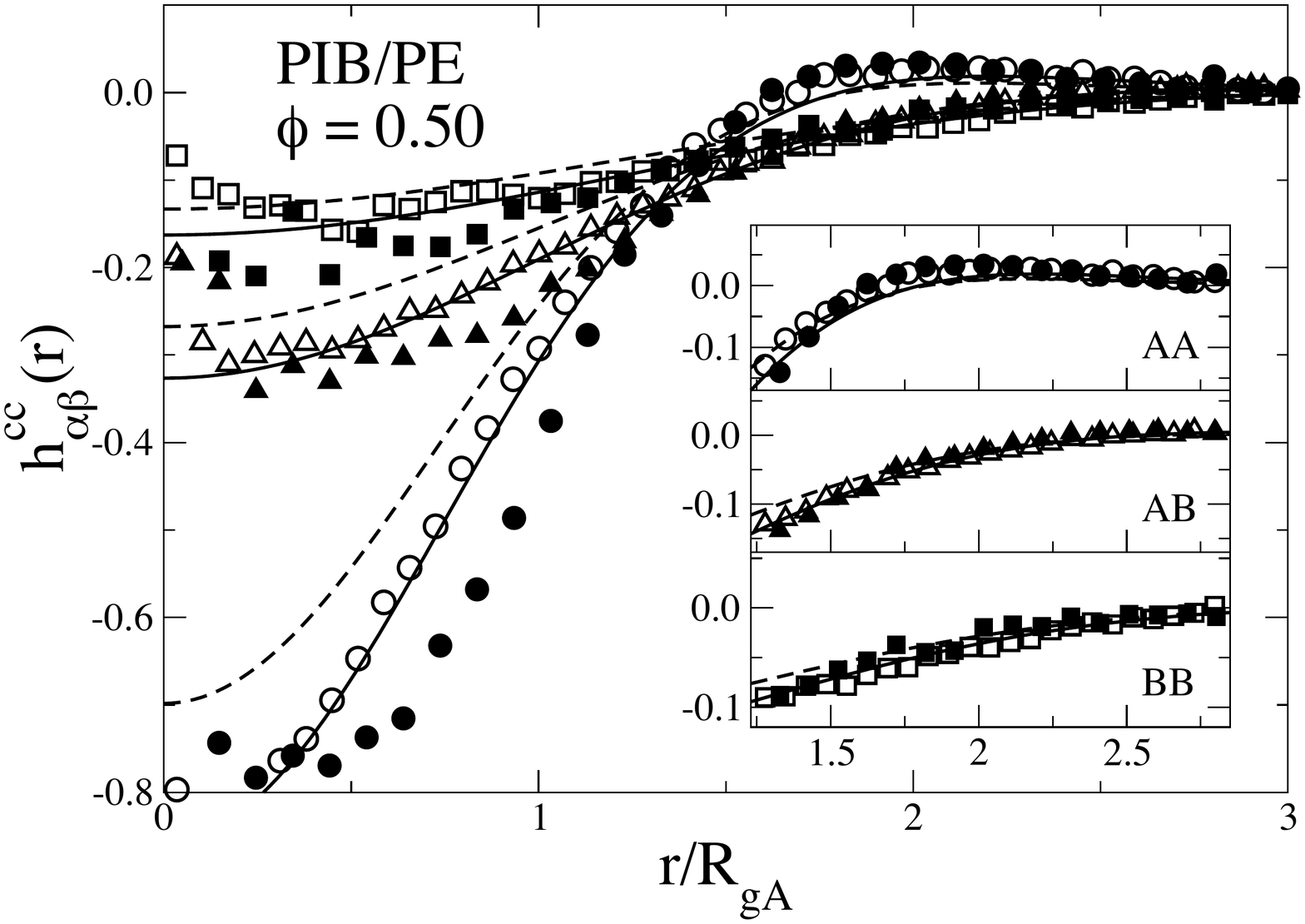}
\end{minipage}
\\
\begin{minipage}{2.8in}
\includegraphics[scale=0.28]{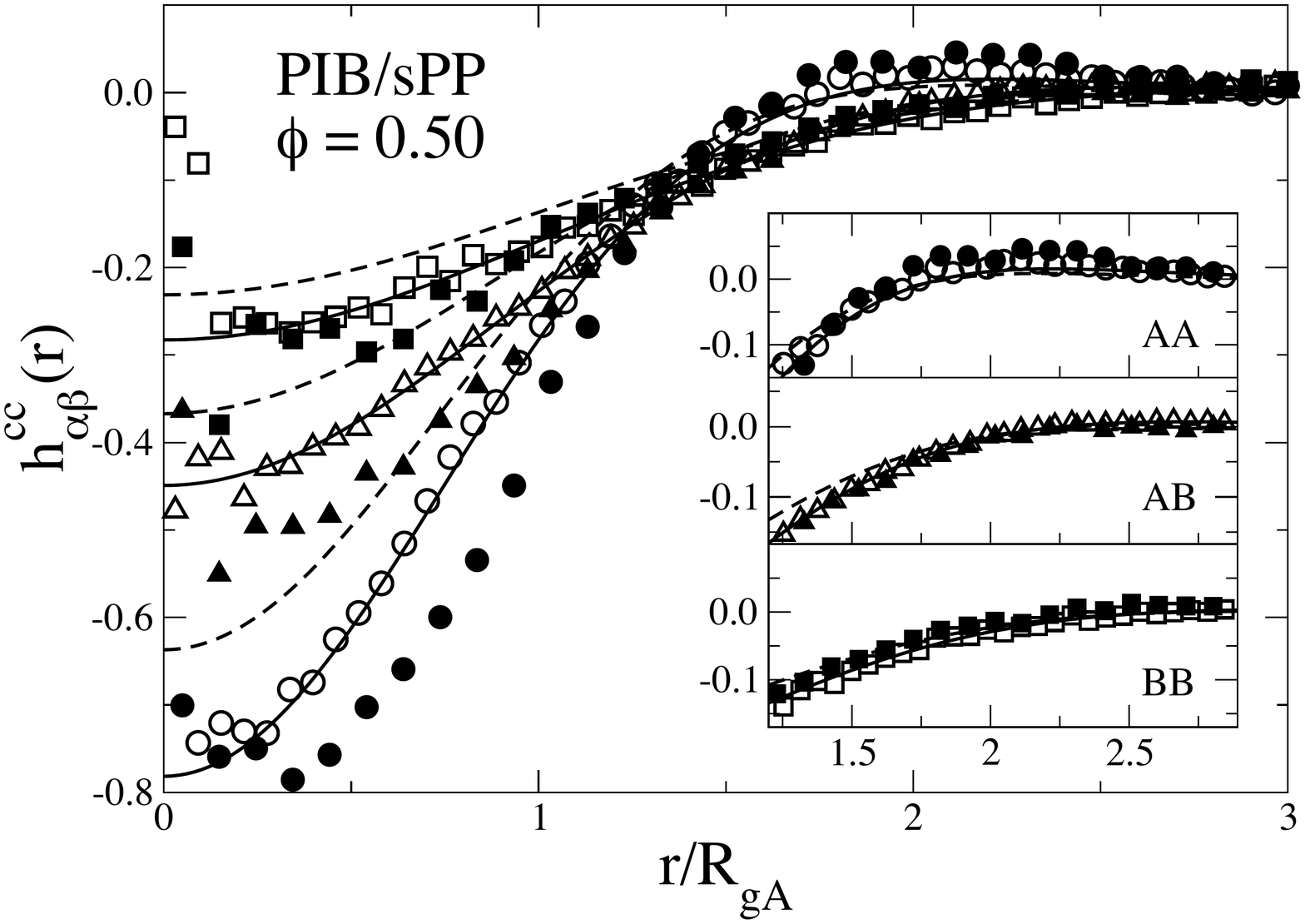}
\end{minipage}
&
\begin{minipage}{2.8in}
\includegraphics[scale=0.28]{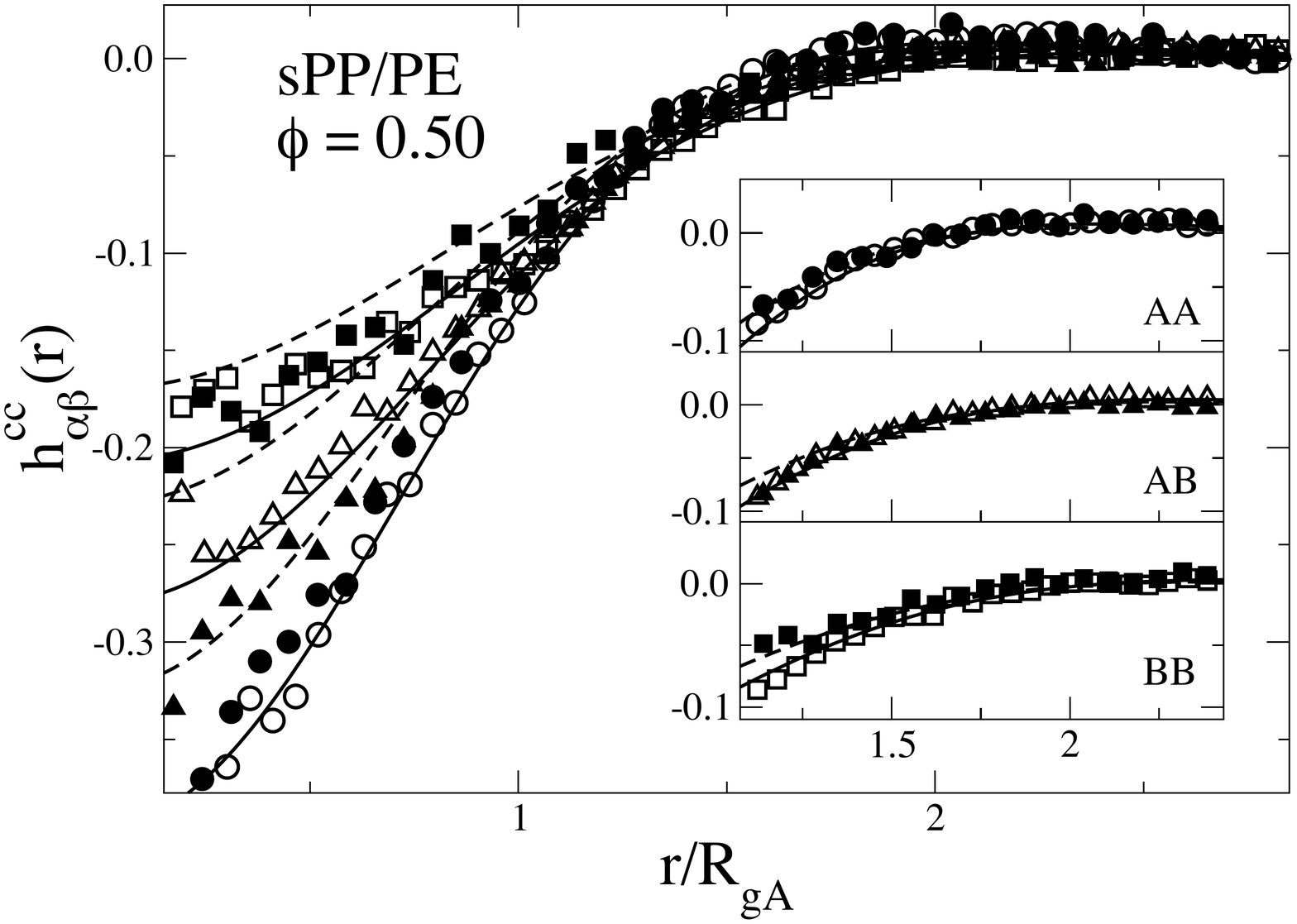}
\end{minipage}
\\
\begin{minipage}{2.8in}
\includegraphics[scale=0.28]{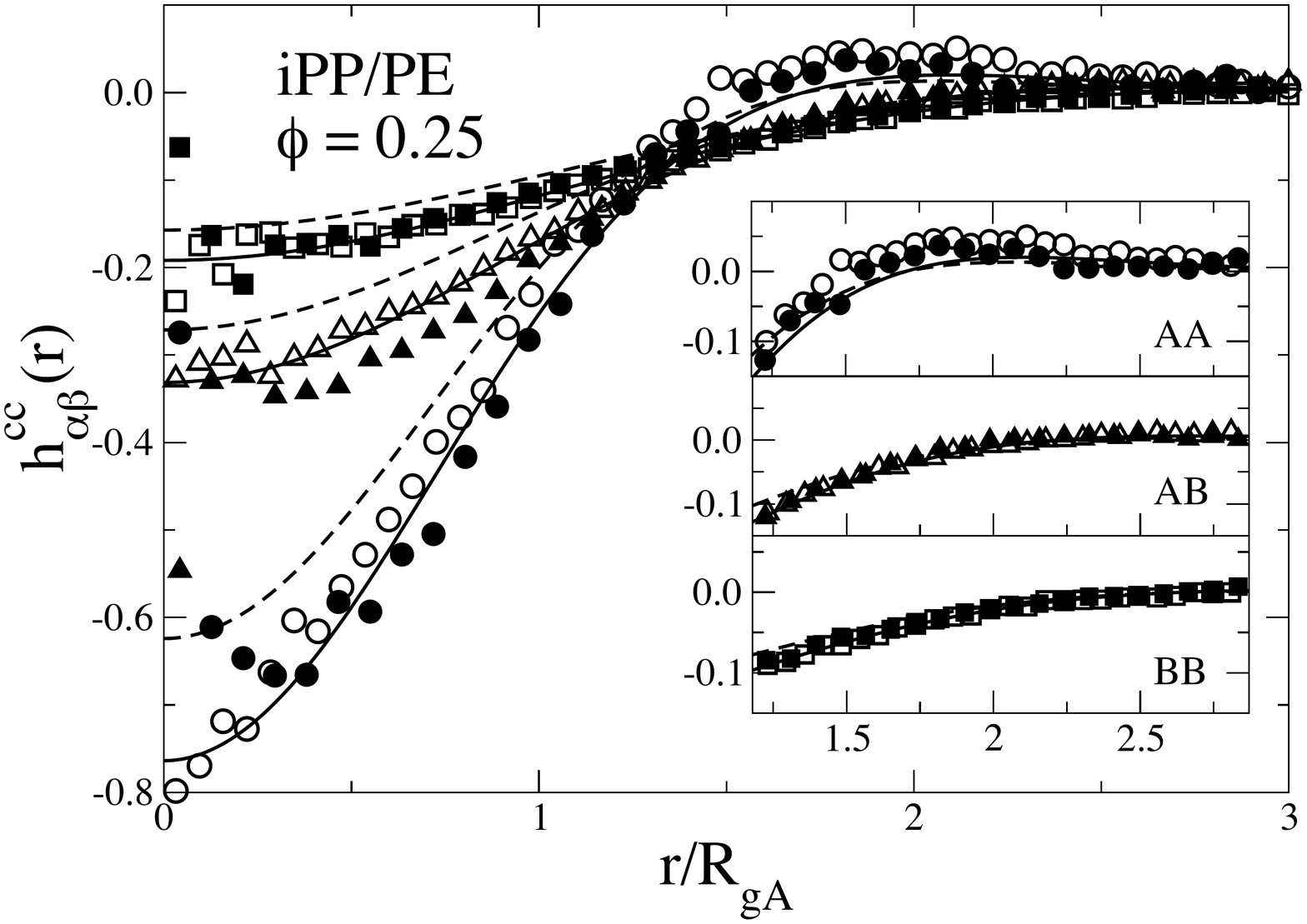}
\end{minipage}
&
\begin{minipage}{2.8in}
\includegraphics[scale=0.28]{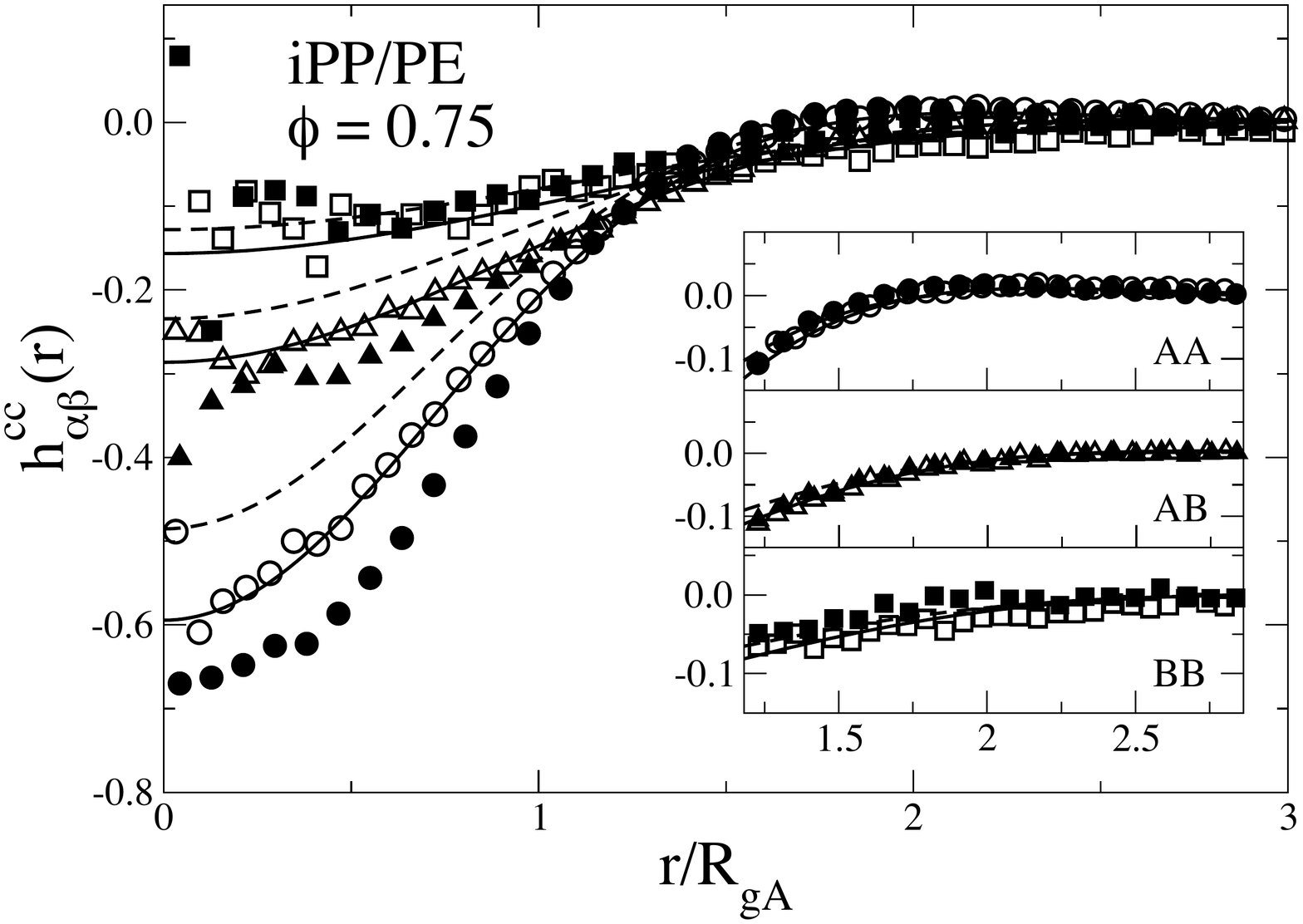}
\end{minipage}
\end{tabular}
\caption{\small{Comparison of mesoscopic simulations [open symbols] with UA
MD simulations [filled symbols] for the $h_{\alpha\beta}(r)$ of polymer mixtures under athermal
conditions.  Also shown are theoretical predictions [solid curves] based on our analytic expression, Equation \ref{EQ:HCCR}.
Presented are data from $AA$ [circles], $AB$ [triangles], and
$BB$ [squares] contributions for compositionally symmetric and
asymmetric systems. For comparison, numerical predictions obtained from Equation \ref{EQ:HCCK} using the Debye form are shown [dashed curves]. For clarity the inset highlights the peak region for each separate contribution.}}
\label{FG:BLNS}
\end{figure*}

From the trajectories of our mesoscopic simulations, the intermolecular total pcfs are computed.  Initially, we set $\chi=0$ to determine the liquid structure far from the spinodal temperature, \emph{i.e.} under athermal conditions, $(1-\chi/\chi_s) = 1$. Mesoscale simulation parameters for these blends are presented in Table \ref{TB:MSBLND}. For these simulations we compare the resulting pcfs to UA MD simulations. To obtain center of mass distribution functions from UA simulations, the center of mass coordinates for each chain are evaluated at each time step as the averaged sum of the position coordinates of each united atom. The radial distribution of center of mass coordinates is evaluated in the usual method employed for liquids.\cite{ALLEN} The resulting pcfs are shown in Figure\ \ref{FG:BLNS} for the systems listed in Table \ref{Table1}. Mesoscopic simulations are found to yield a coarse-grained liquid structure in agreement with our theoretical predictions from the analytical expression of Equation \ref{EQ:HCCR}, serving as a self-consistent check of our determination of the effective pair potential through the HNC closure.  
\begin{table}[h!] 
\caption{Mesoscale Simulation Parameters for Athermal Blends}
\begin{tabular}{lccc}
  \hline \hline
System & Particles  & $\phi$ & $L/2$ [$R_g^{-1}$]\\
\hline
hhPP/PE & 5324 & 0.5 &  8.549 \\ 
PIB/PE & 4096 & 0.5 & 8.365 \\
PIB/sPP & 5488 &  0.5 & 10.416  \\
sPP/PE & 1728 &  0.5 & 5.635  \\
iPP/PE  & 4913 & 0.25 & 8.482 \\
iPP/PE & 1728 & 0.75 & 6.016 \\
\hline \hline
\end{tabular}
\label{TB:MSBLND}
\end{table}
The results presented in Figure \ref{FG:BLNS} were obtained using the Pad\'e approximation (Equation \ref{EQ:PADE}) which works sufficiently well under athermal conditions where the low k behavior is less important since critical fluctuations are assumed to be small. 

The liquid structure from mesoscopic simulations are in general consistent with data obtained from UA MD simulations, with the exception of blends containing iPP and PIB for which theoretical predictions and mesoscopic MD predict a less pronounced correlation hole than UA MD simulations.  These observations are not surprising since these systems tend to possess very efficient intramolecular packing, leading to smaller isothermal compressibilities and thermal expansion coefficients when compared to other polyolefin blends \cite{JARAM,KRISH}.  The effective intramolecular packing arises from the attractive interactions between methyl moieties induced by their geometrical arrangement.  However, the theory and mesoscopic simulations do exhibit good agreement for $r\approx R_g$.
\begin{figure*}[t!]
\centering
\begin{tabular}{cc}
\begin{minipage}{2.8in}
\includegraphics[scale=0.26]{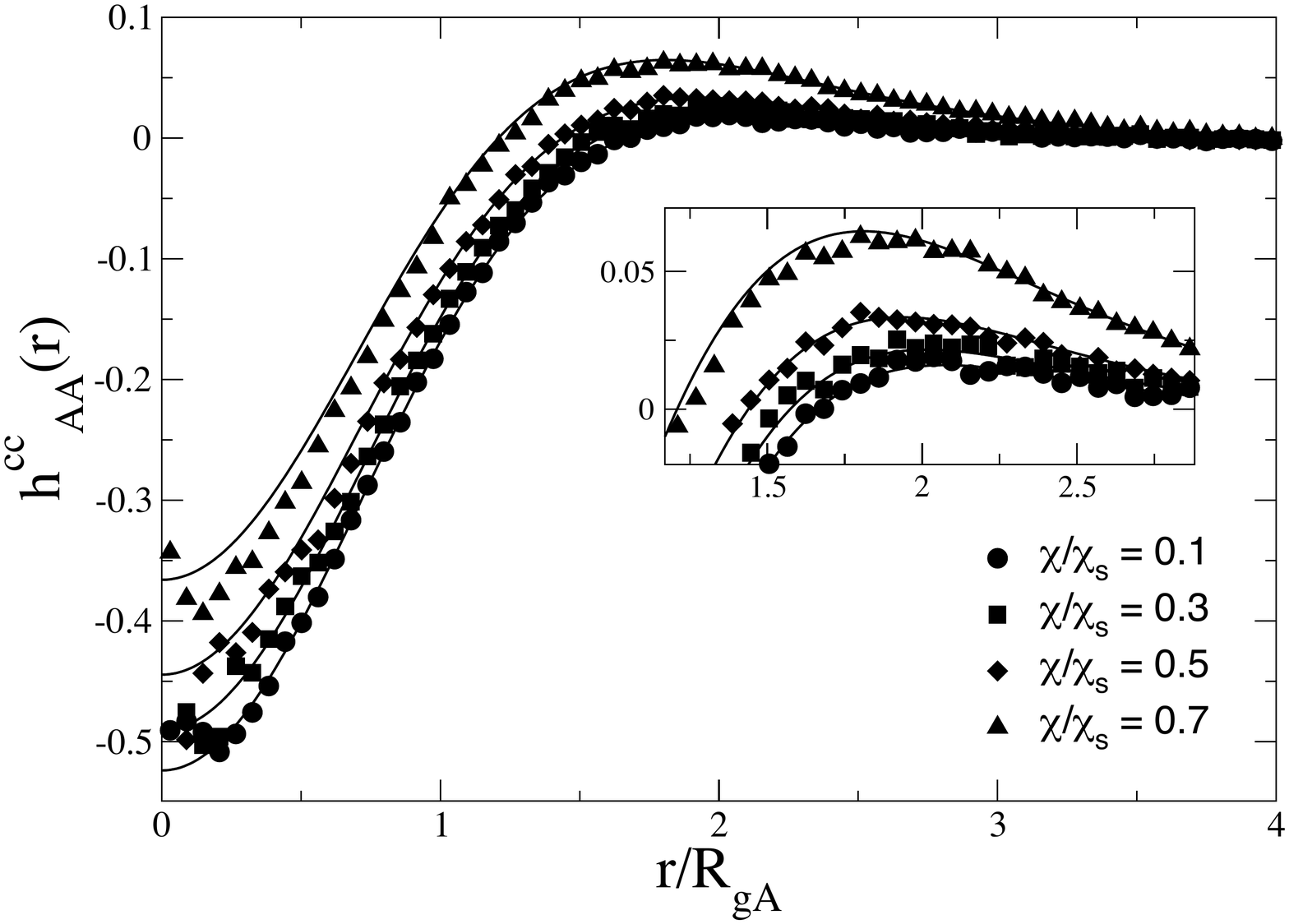}
\end{minipage}
&
\begin{minipage}{2.8in}
\includegraphics[scale=0.26]{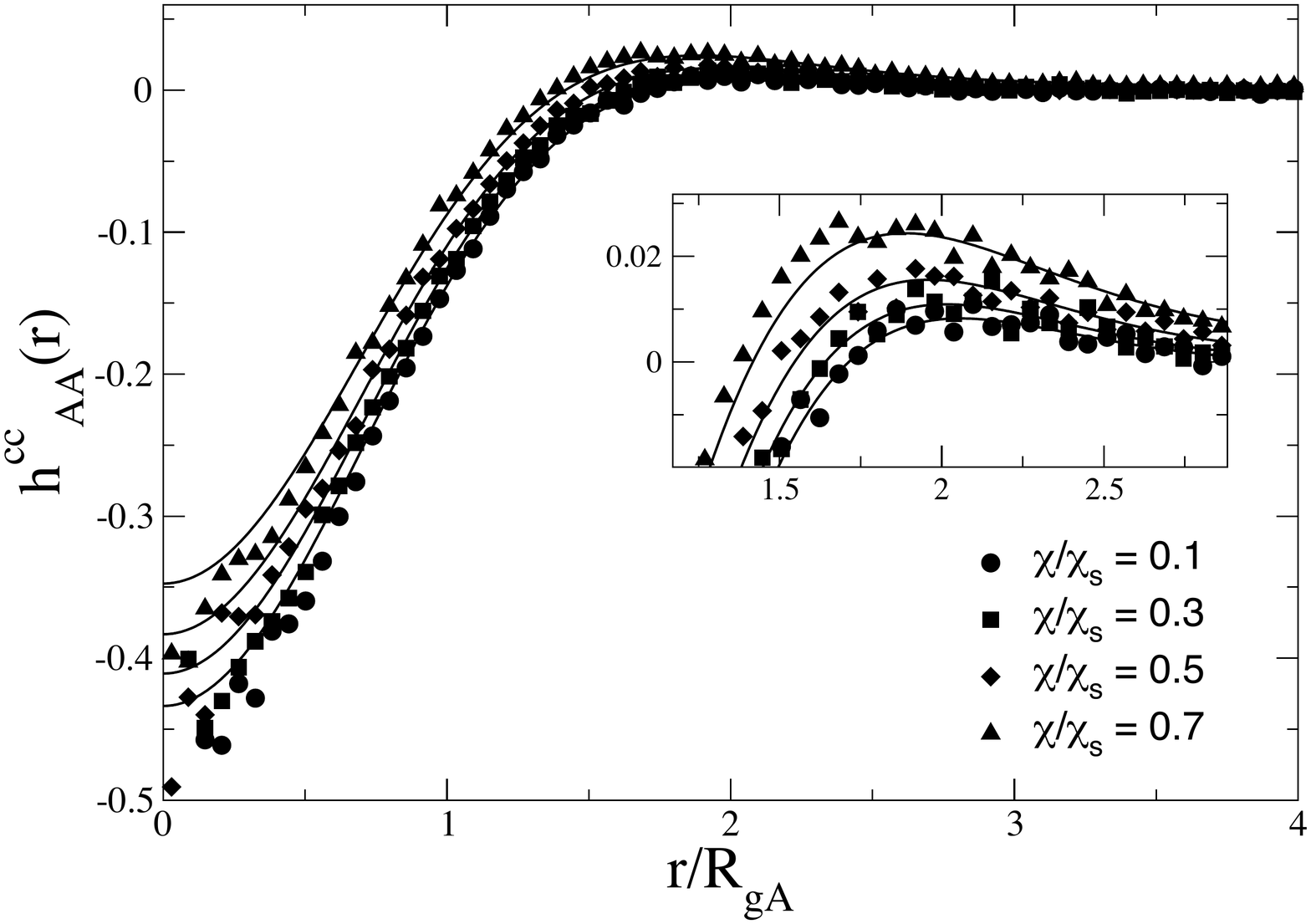}
\end{minipage}
\\
\begin{minipage}{2.8in}
\includegraphics[scale=0.26]{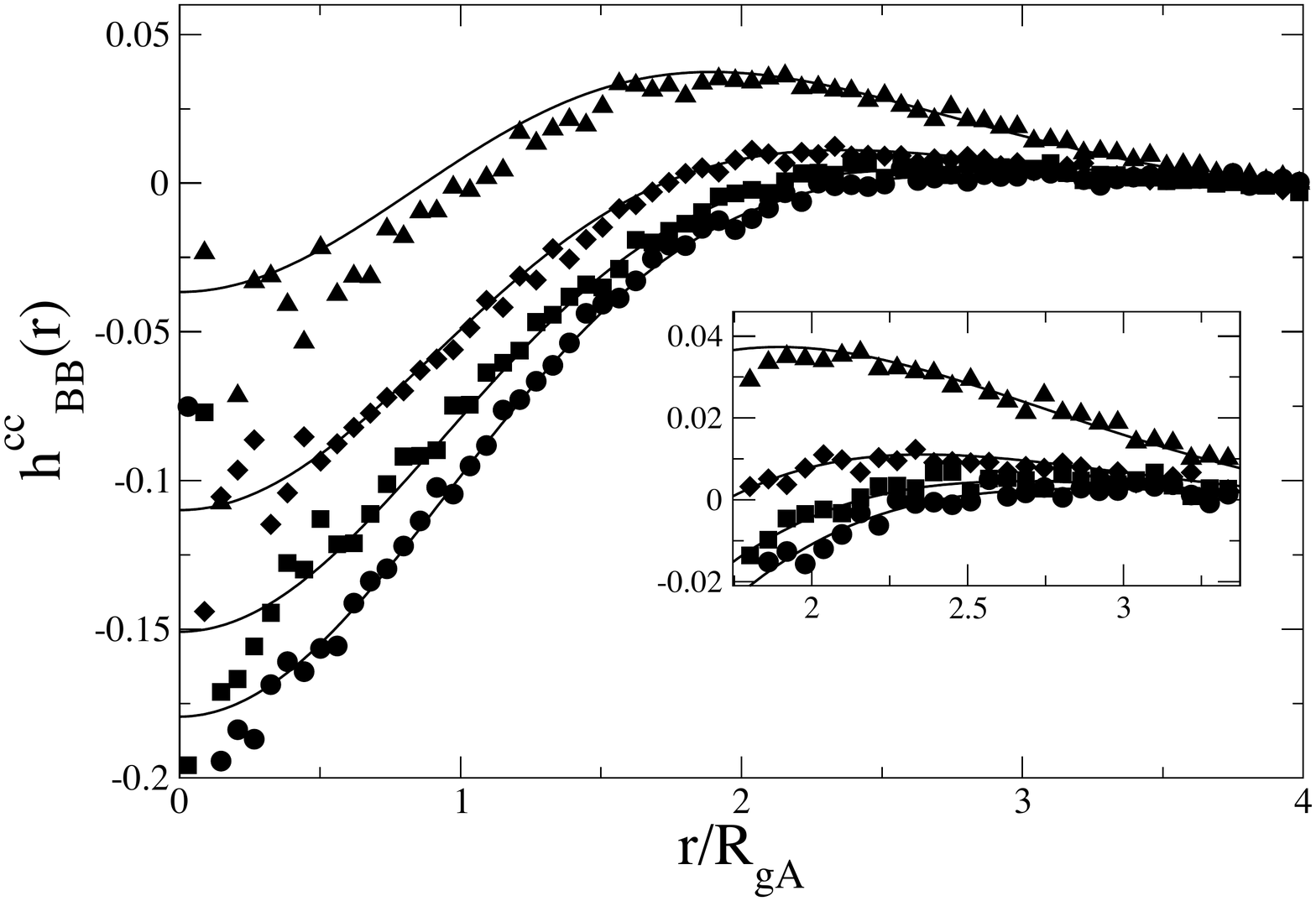}
\end{minipage}
&
\begin{minipage}{2.8in}
\includegraphics[scale=0.26]{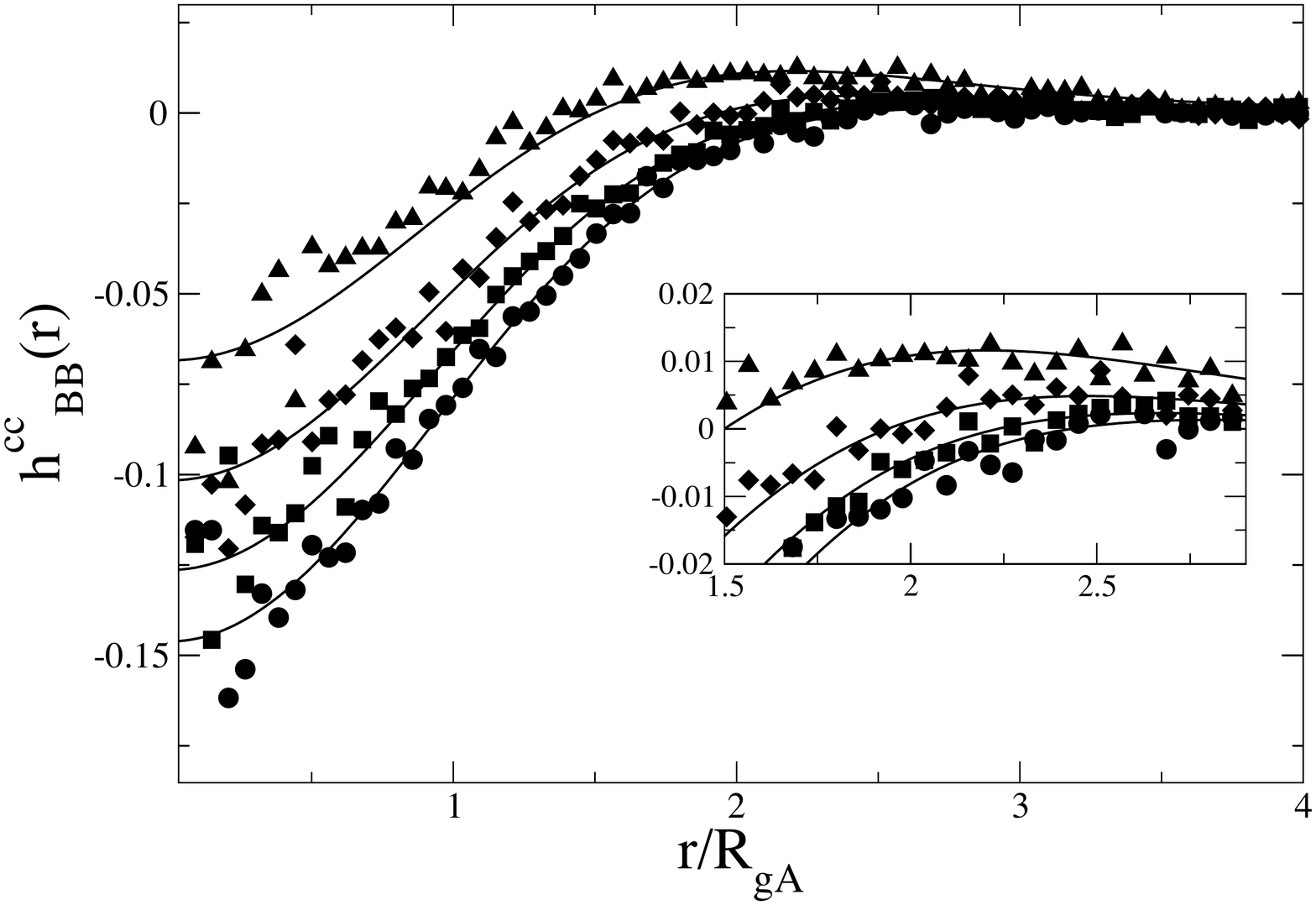}
\end{minipage}
\\
\begin{minipage}{2.8in}
\includegraphics[scale=0.26]{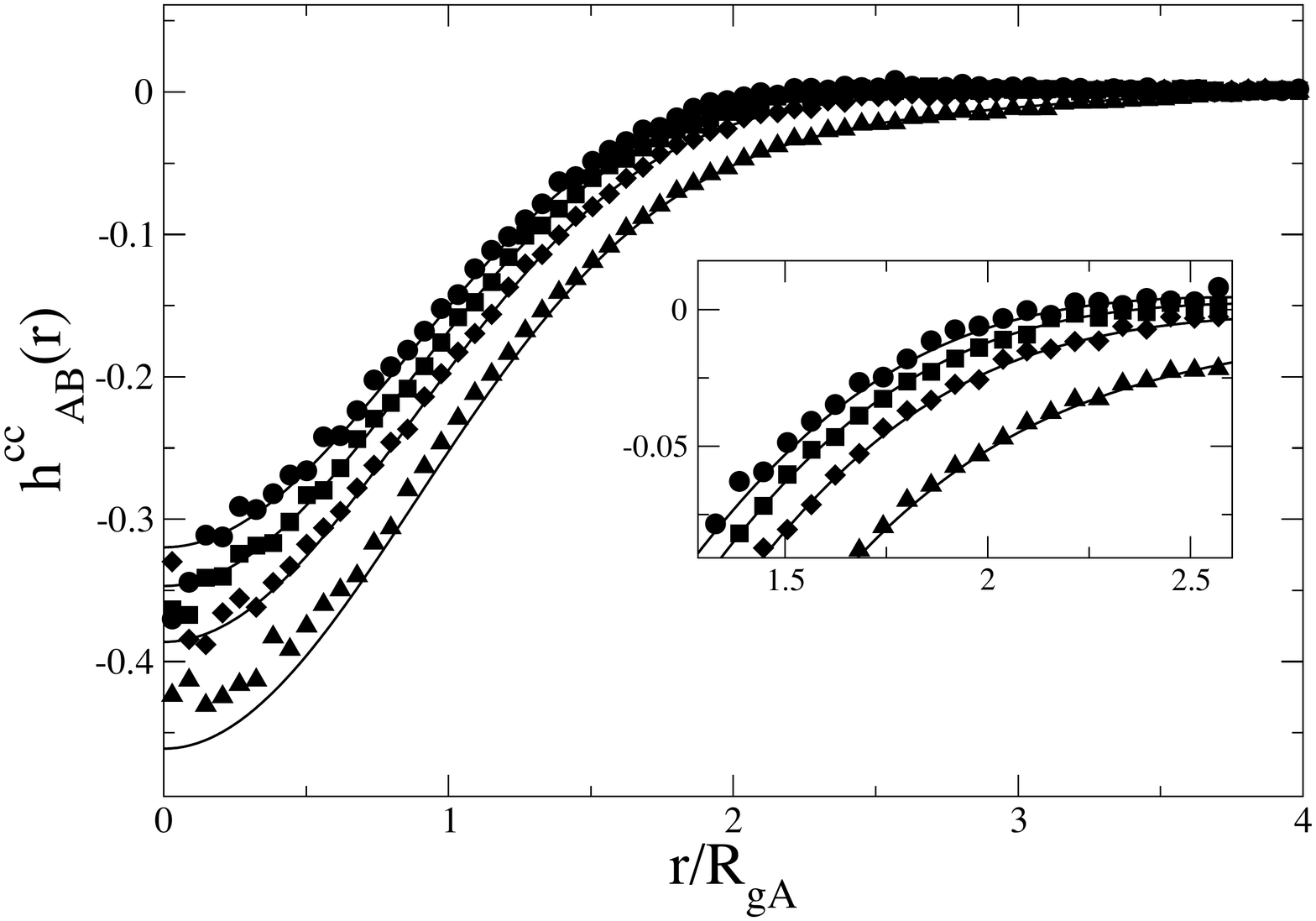}
\end{minipage}
&
\begin{minipage}{2.8in}
\includegraphics[scale=0.26]{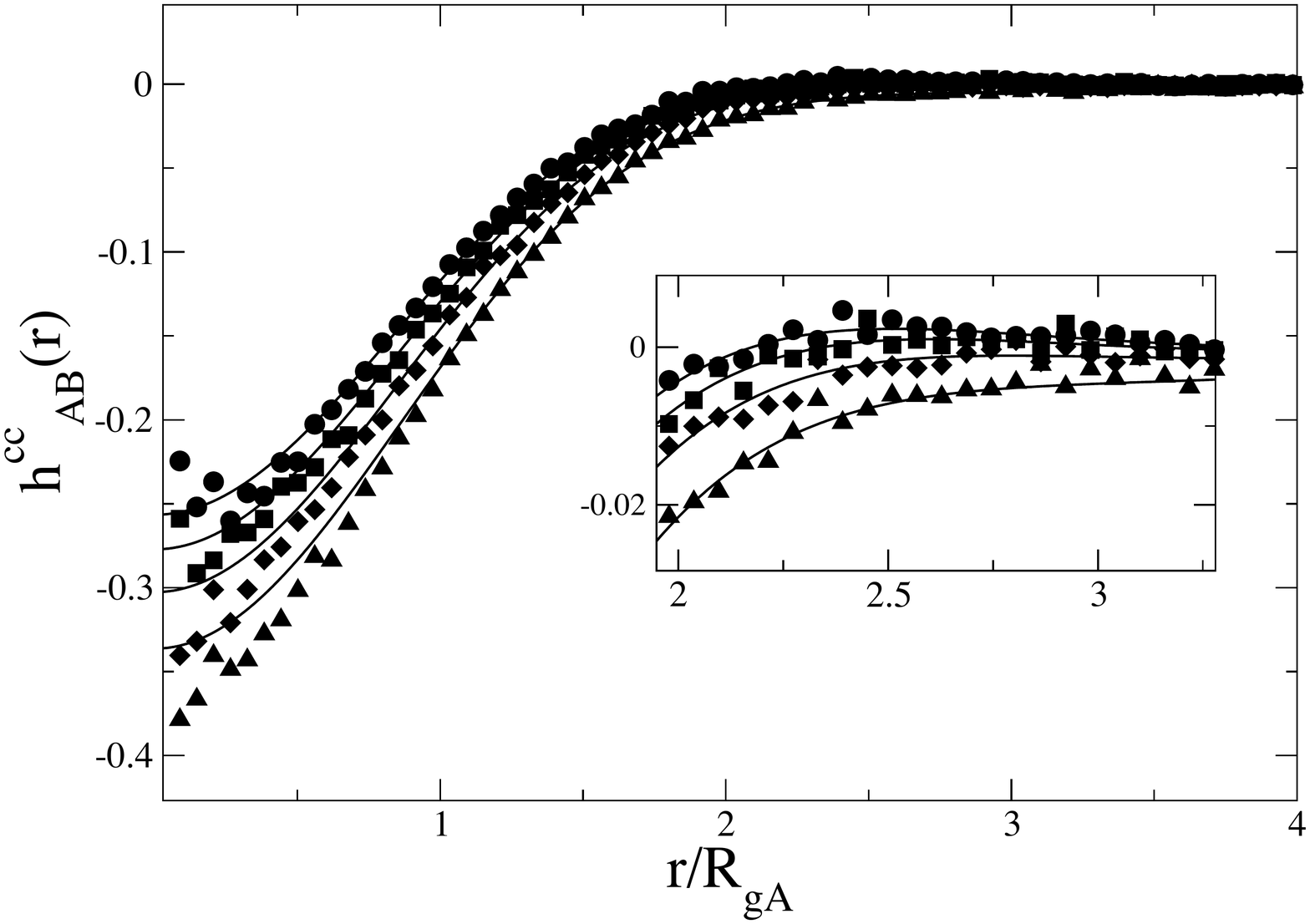}
\end{minipage}
\end{tabular}
\caption{\small{Comparison of mesoscopic simulations [symbols] with numerical predictions [curves] for the $h_{\alpha\beta}(r)$ of a 50:50 mixture of hhPP/PE for different values of the ratio, $\chi/\chi_s$. The left panel shows results obtained using the Pad\'e approximation with our truncation scheme. The right panel depicts the results when the Debye form is used. Mesoscale simulations are shown to capture the structural changes that occur as the system approaches the spinodal. The inset highlights the peak region of $h(r)$}}
\label{FG:THERM}
\end{figure*}

Moving to the thermal regime, where large scale fluctuations in the local concentration develop as the system approaches a second order phase transition, we present results for the typical 50:50 mixture of hhPP/PE, although the theory and methods employed are ubiquitous and generally applicable to a wide range of systems.  For these simulations the value of the $\chi$ parameter was varied such that $\chi/\chi_s = \left\{ 0.1, 0.3, 0.5, 0.7\right\}$, in order to see the changes in the pcfs as the system approaches the spinodal. Figure \ref{FG:THERM} shows the dependence of the partial correlation functions on the interaction parameter, $\chi$. The left side of Figure \ref{FG:THERM} shows the resulting correlation function from mesoscale simulations with the potential obtained via the Pad\'e approximation, after using our truncation scheme for $c^{cc}(r)$ in the HNC (see upper panels Fig \ref{FG:VOFR}). Use of the Pad\'e approximation has the advantage of allowing a fully analytic solution for $h^{cc}(r)$, as shown in Equation \ref{EQ:HCCR}, which shows quantitative agreement with UA simulations in the athermal limit (as shown in Figure \ref{FG:BLNS}). 

The right panels of Figure \ref{FG:THERM} shows the correlation function obtained using the potential derived from the Debye form (see lower panels of Fig \ref{FG:VOFR}). Here, comparison is again made to numerical predictions based on Equation \ref{EQ:HCCK}, since an analytic solution, such as that of Equation \ref{EQ:HCCR}, is not possible when the Debye form is used. In both the right and left panels of Figure \ref{FG:THERM} mesoscopic simulations show quantitative agreement with our theoretical predictions, indicating the self-consistency of our approach. Furthermore, despite the differences in the potential used in the simulation, Figure \ref{FG:THERM} shows that the resulting pcfs from either the Pad\'e or Debye form are qualitatively similar. Lastly, we note that despite the approximations made in obtaining the analytical form of Equation \ref{EQ:HCCR}, our analytical expressions recover the correct k=0 limit.\cite{YAPRL,BLNDS}  In fact, all of the forms for $h^{cc}(k)$ exhibit the same $k=0$ behavior.

The standard approach to describe the mixing behavior of polymers is the Flory-Huggins model. Under Flory-Huggins treatment, the phenomenon of demixing is understood in terms of contributions to the free energy of mixing.  Generally, at low enough temperatures the translational entropy, which is associated with the center of mass motion of the molecules and always favors mixing, is outweighed by local monomer-monomer interactions.  In most cases, van-der-Waals interactions are stronger between like pairs than those between unlike pairs, resulting in a positive free energy of mixing.  As a result, lower temperature favors spontaneous demixing due to changes in the local free energy of the system.\cite{STRBL}  In an empirical manner, the Florry-Huggins parameter, $\chi$, is used to describe these changes in local free energy.  At the limit of the spinodal temperature, $\chi \rightarrow \chi_s$, and  since $\chi \propto \frac{1}{T}$, positive values of $\chi$ always lead to incompatibility of the mixture.\cite{STRBL}  

In real systems, the simple Flory-Huggins model does not hold, and the $\chi$ parameter may be a complicated function of $N$, $\phi$, and $T$, leading to the variety of phase behaviors observed in polymer blends. For example, some blends phase separate upon cooling, while others show an opposite trend in demixing and phase separate upon heating. It is customary to fit the experimental temperature dependence of a mixture to the form $\chi=a+b/T$ where $a$ and $b$ may be either positive or negative depending on the system. Table \ref{Table1} shows the experimentally determined $a$ and $b$ parameters for a few of the systems investigated in this paper. It should be noted that  when applying an equation for the $\chi$ parameter from the literature, the $\chi$ value must be normalized by the average number of UA sites per monomer\cite{JARAM} to be consistent with the site-basis description adopted here.

In our present treatment, the interaction parameter, $\chi$ is treated as an adjustable parameter,
which describes the interactions that drive phase separation. It is analogous to the Flory-Huggins parameter; however, since in our model it represents a system specific parameter, it may be given any value positive or negative depending on the behavior of the system of interest. 
The advantage of a mesoscale approach is that once the system specific parameters are defined, the trends in phase behavior can be readily calculated without requiring restrictively large MD simulations. 

\begin{figure*}[t!]
\centering
\begin{tabular}{cc}
\begin{minipage}{3in}
\includegraphics[scale=0.38]{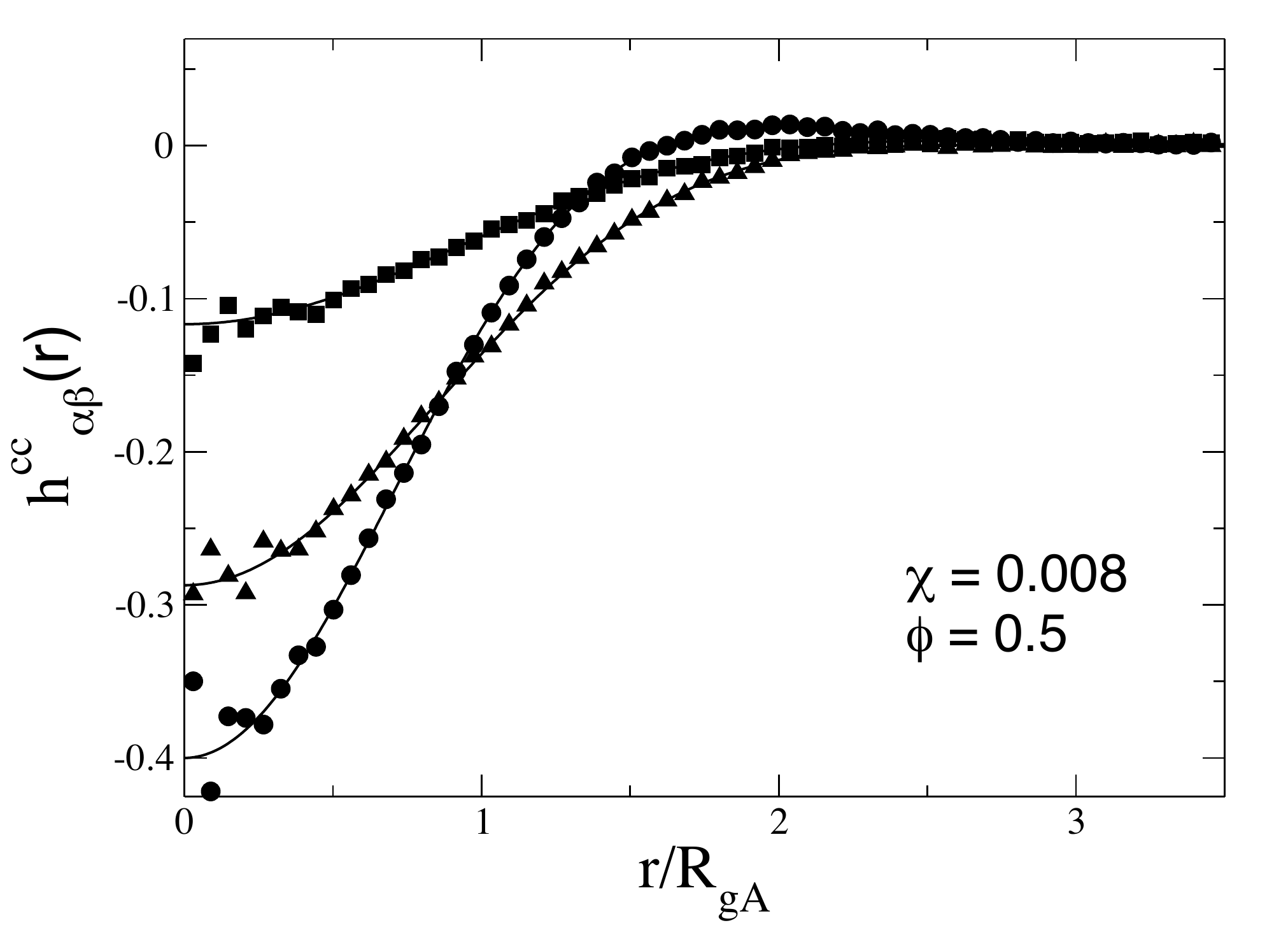}
\end{minipage}
&
\begin{minipage}{3in}
\includegraphics[scale=0.38]{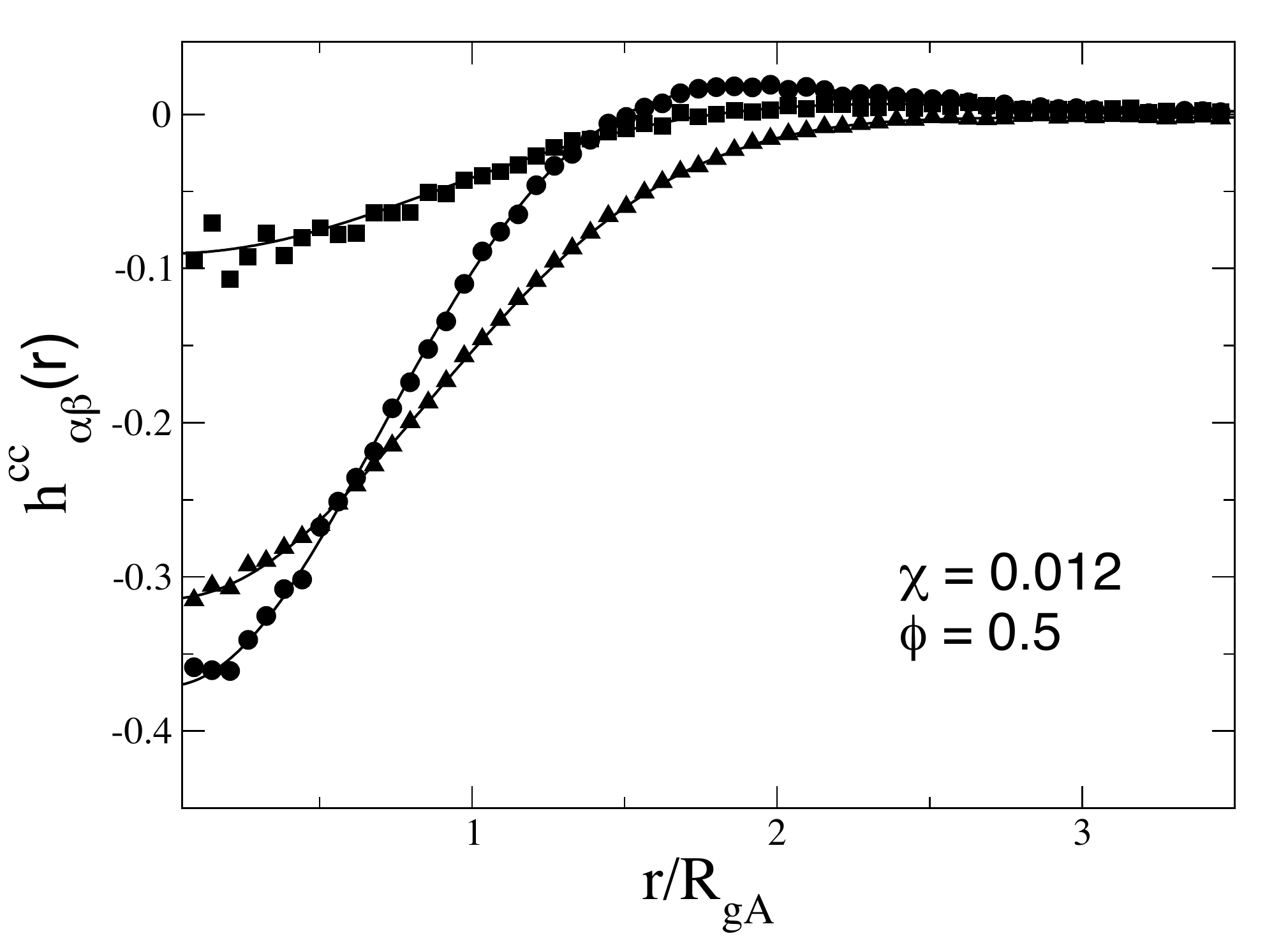}
\end{minipage}
\\
\begin{minipage}{3in}
\includegraphics[scale=0.38]{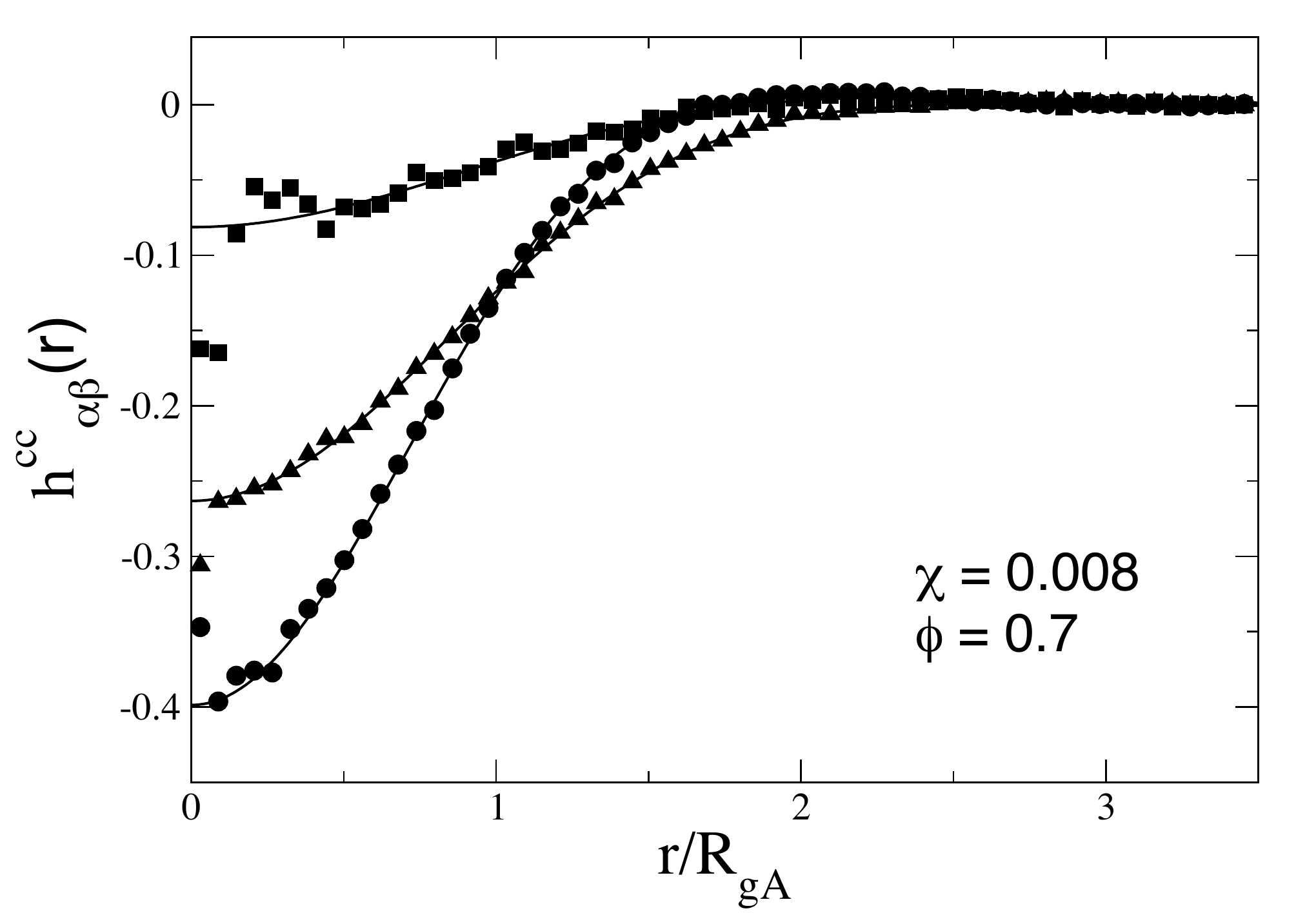}
\end{minipage}
&
\begin{minipage}{3in}
\includegraphics[scale=0.38]{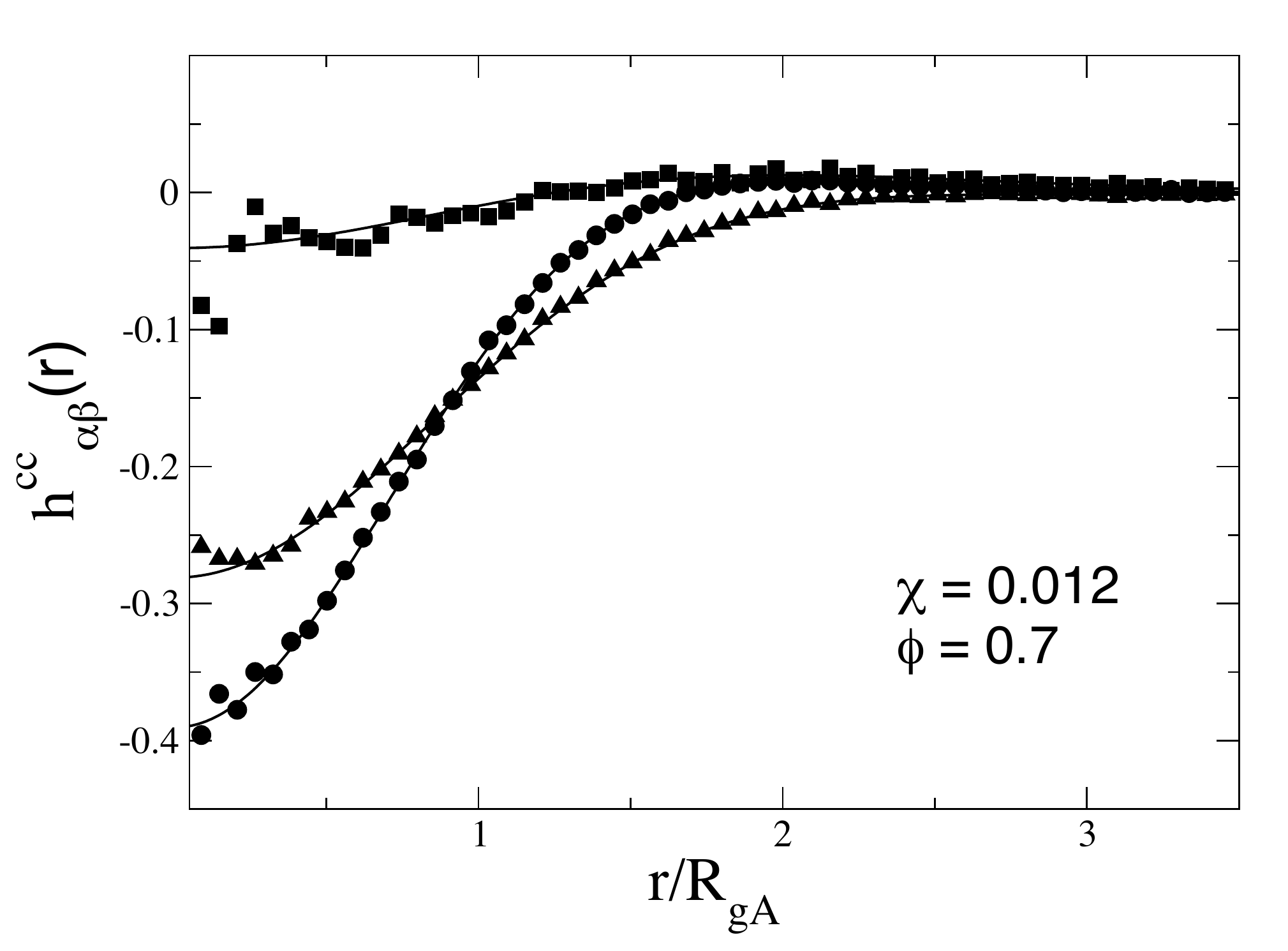}
\end{minipage}
\\
\begin{minipage}{3in}
\includegraphics[scale=0.38]{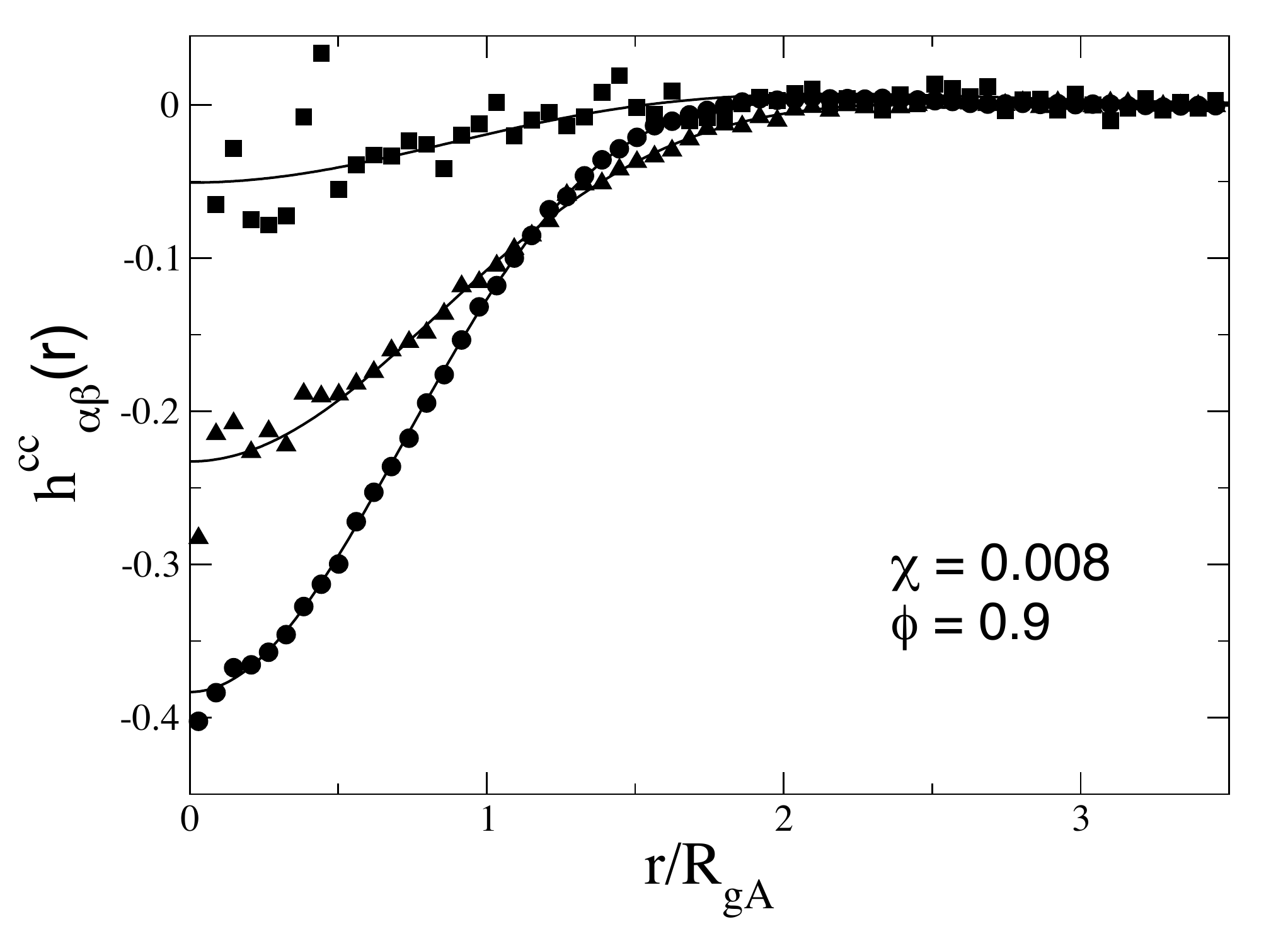}
\end{minipage}
&
\begin{minipage}{3in}
\includegraphics[scale=0.38]{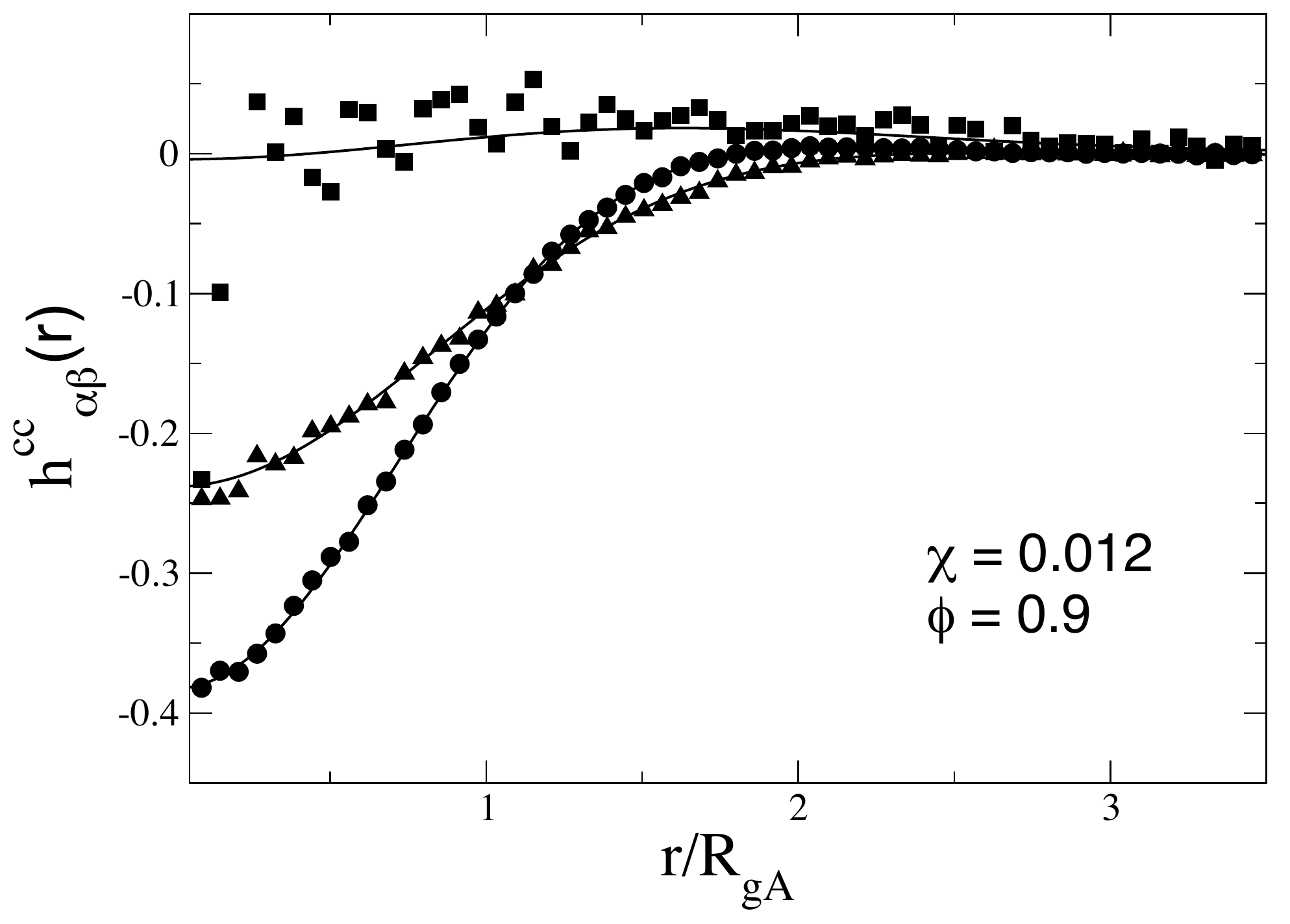}
\end{minipage}
\end{tabular}
\caption{\small{Comparison of mesoscopic simulations [symbols] with numerical predictions [curves] for the $h_{\alpha\beta}(r)$ of hhPP/PE for different values of $\phi$. Left panels show data when $\chi = 0.008$. Right panels show data for $\chi = 0.012$. Shown are the separate contributions for $AA$ [circles], $AB$ [triangles], and
$BB$ [squares] interactions. As $\phi$ increases, the fraction of species B in the simulation box decreases, and thus, the statistics become poorer for BB interactions.}}
\label{FG:PHIA}
\end{figure*}

\begin{figure*}[t!]
\centering
\begin{tabular}{cc}
\begin{minipage}{3in}
\includegraphics[scale=0.38]{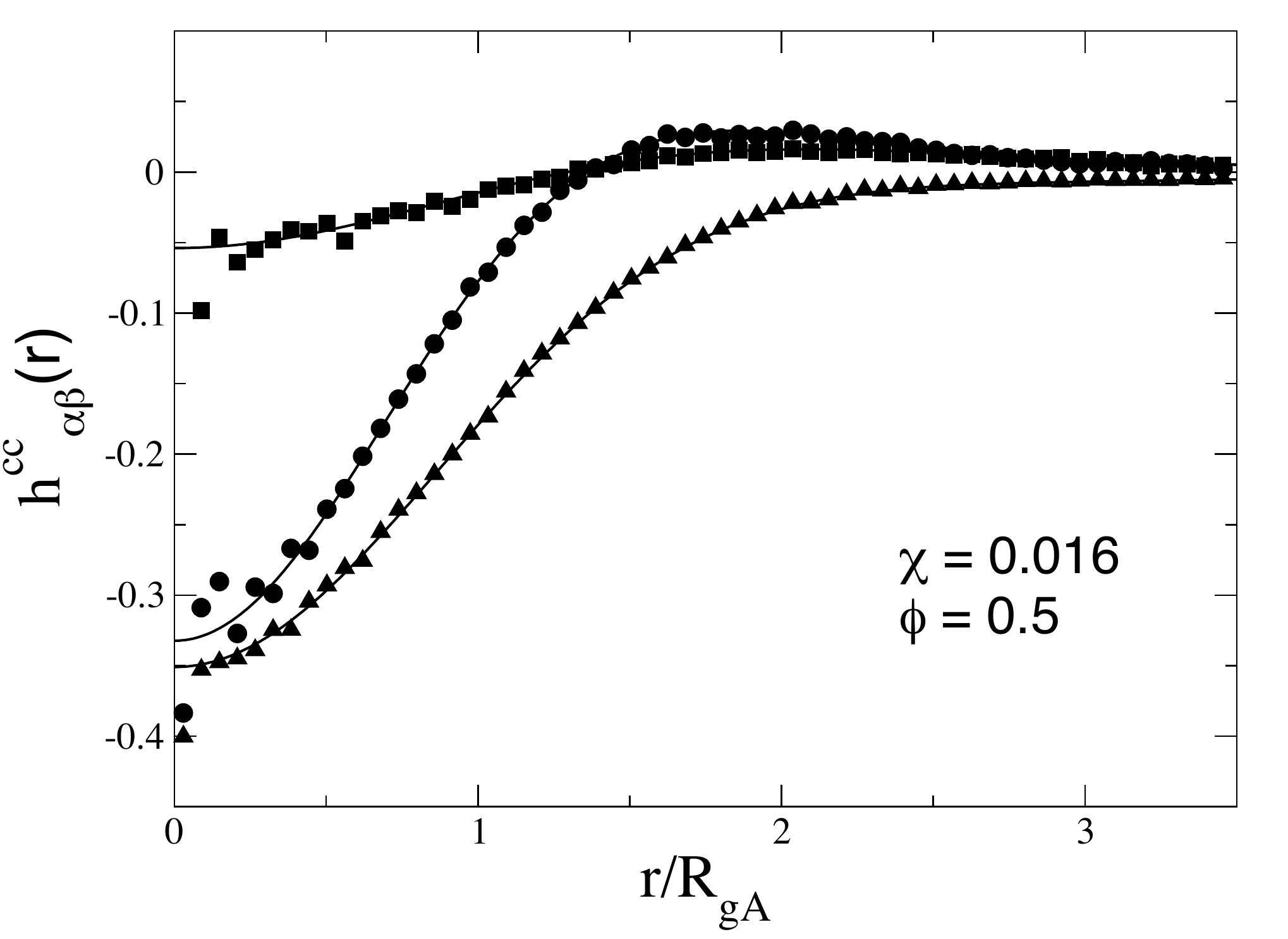}
\end{minipage}
&
\begin{minipage}{3in}
\includegraphics[scale=0.38]{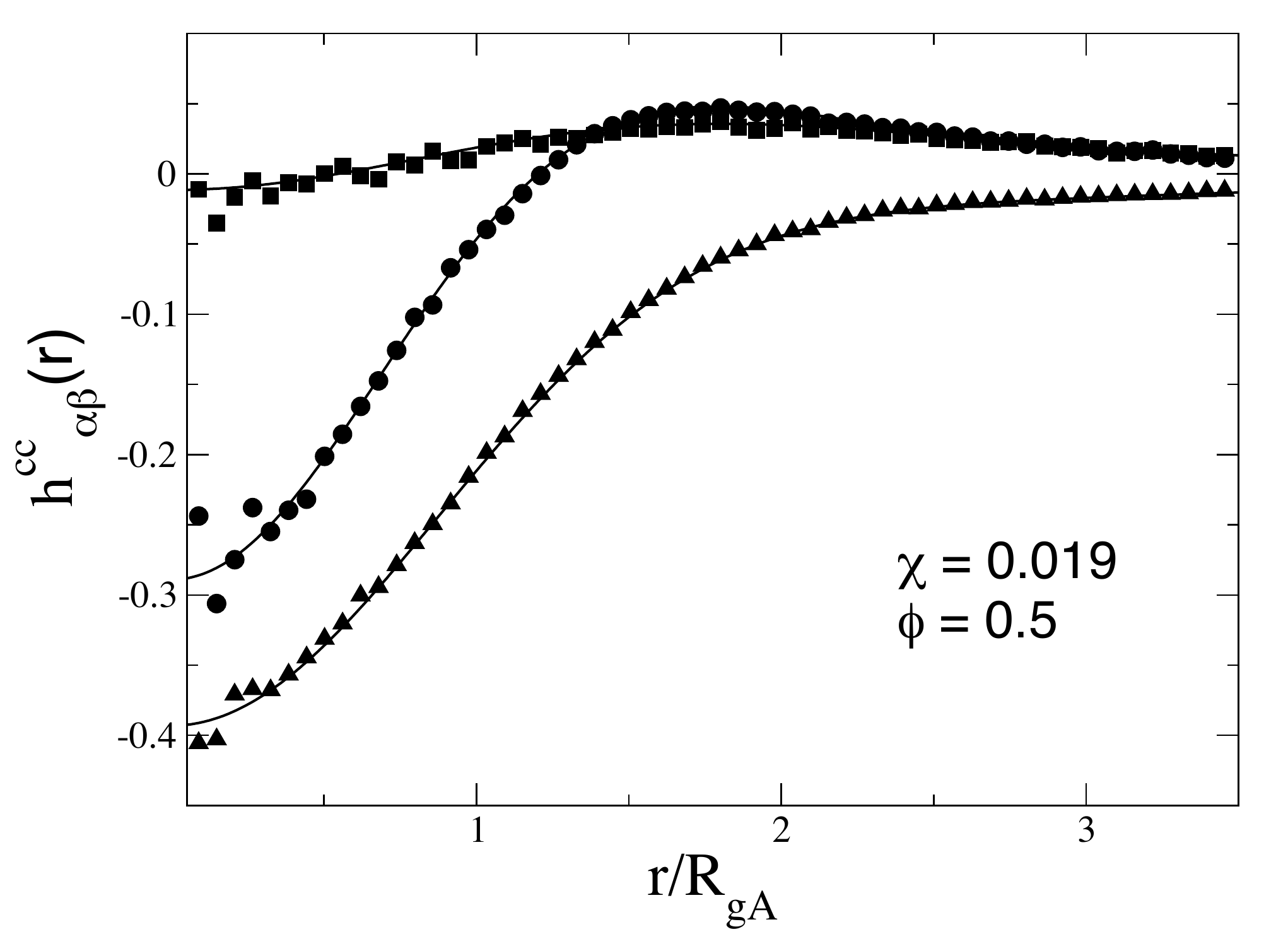}
\end{minipage}
\\
\begin{minipage}{3.1in}
\includegraphics[scale=0.38]{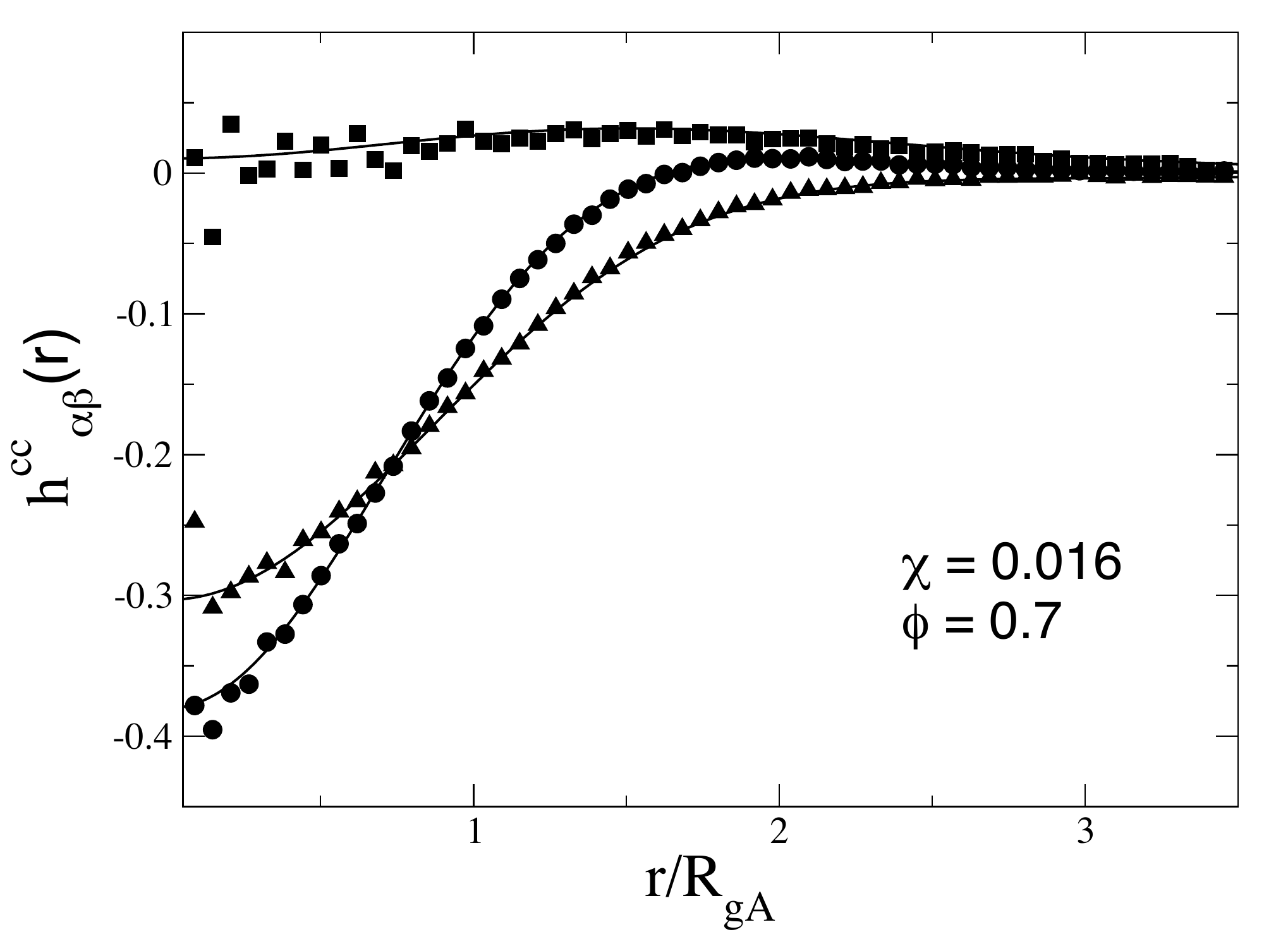}
\end{minipage}
&
\begin{minipage}{3in}
\includegraphics[scale=0.38]{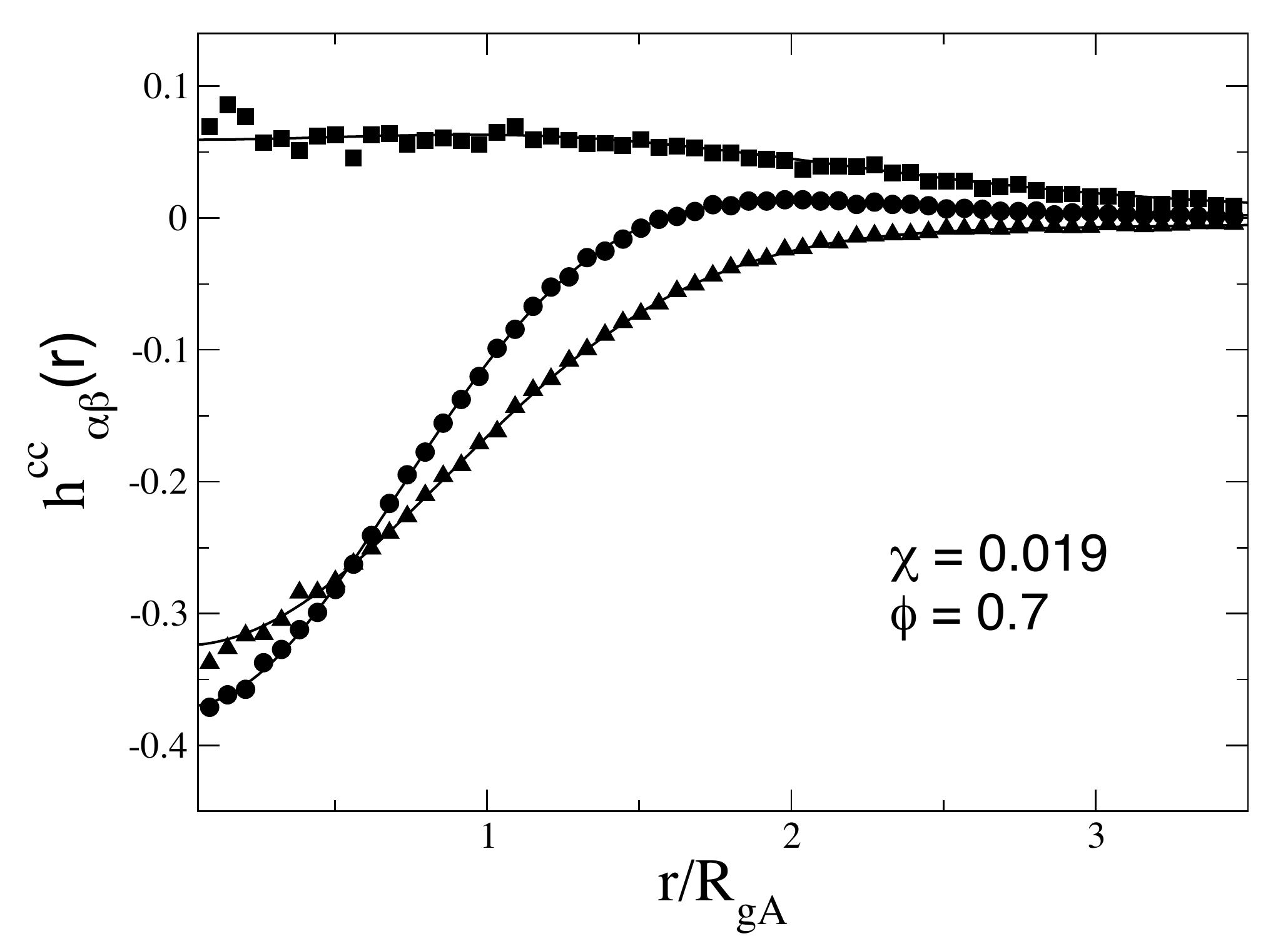}
\end{minipage}
\\
\begin{minipage}{3.1in}
\includegraphics[scale=0.38]{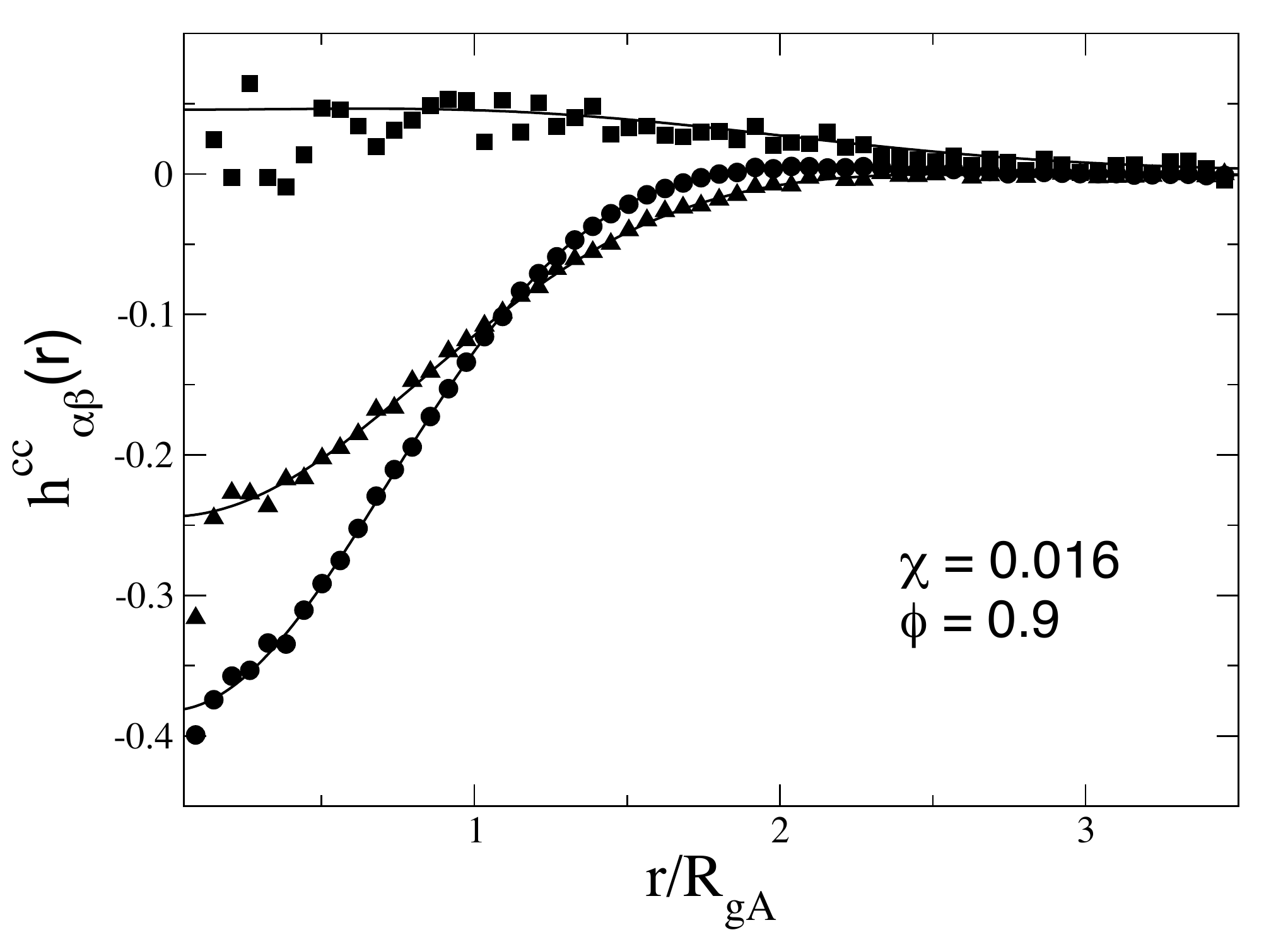}
\end{minipage}
&
\begin{minipage}{3in}
\includegraphics[scale=0.38]{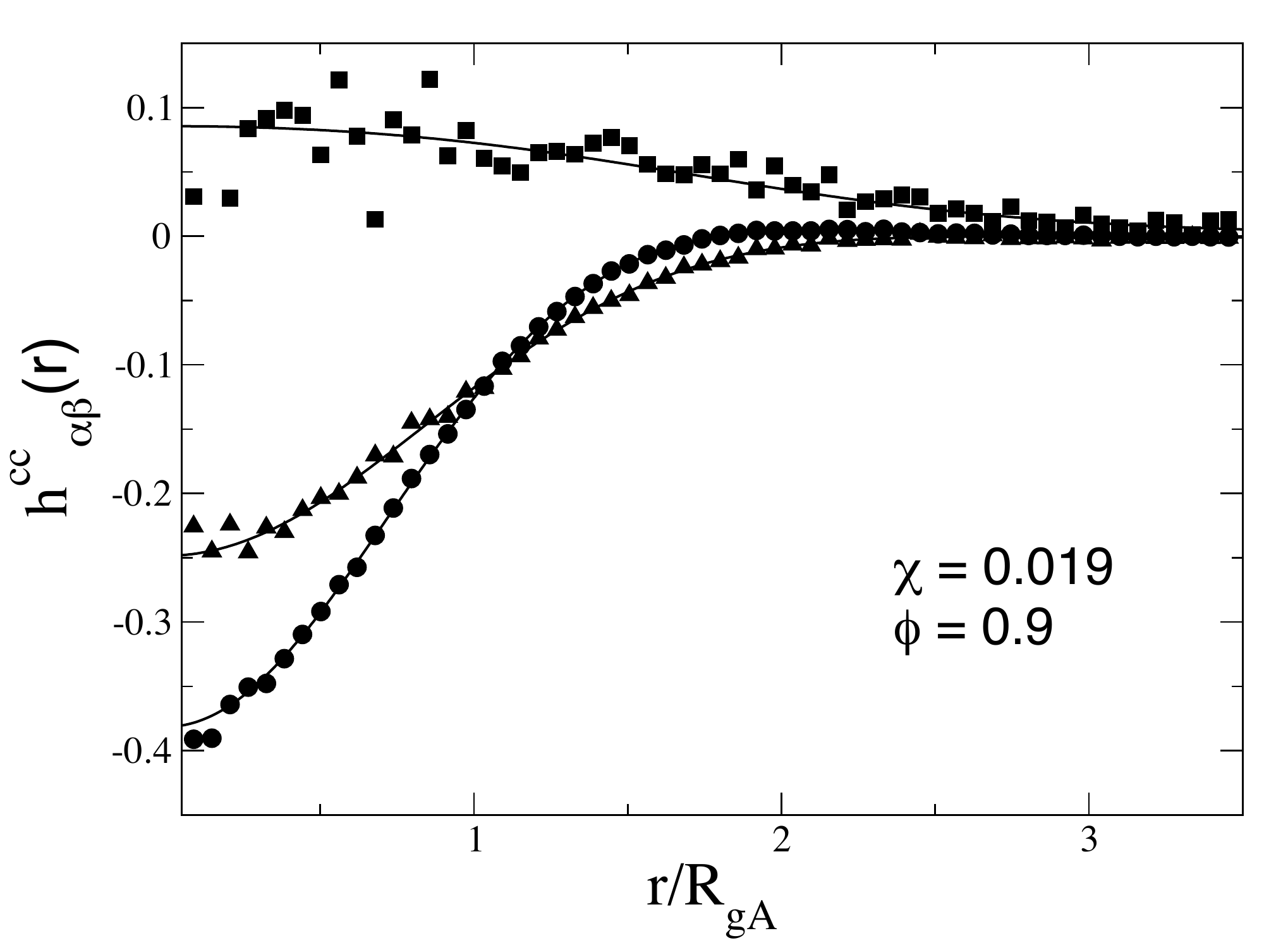}
\end{minipage}
\end{tabular}
\caption{\small{Same as Figure \ref{FG:PHIA} except that left panels show data when $\chi = 0.016$, and right panels show data for $\chi = 0.019$.}}
\label{FG:PHIB}
\end{figure*}

As a further implementation of our theory, we perform mesoscale simulations at several fixed values of $\chi$ for which the fraction of A and B species in the melt is varied. For these simulations, we again use hhPP/PE as a typical system and vary the volume fraction such that $\phi = \{0.5, 0.7, 0.9\}$.  In order to better capture the large scale structural changes, simulations were performed in a large box with 10,648 particles.  
Figure \ref{FG:PHIA} shows the resulting pair correlation functions for mesoscale simulations run with $\chi = 0.008$ and $\chi = 0.012$, and Figure \ref{FG:PHIB} shows the case where $\chi = 0.016$ and $\chi = 0.019$. In all cases, mesoscale simulations  correctly capture the structural changes that depend on the concentrations of the species in the mixture when comparison is made with our theoretical predictions. For these simulations we limit our consideration to using only the Debye form in Equation \ref{EQ:HCCK} to avoid any effects due to the truncation scheme in the low $k$ region. Once more, theory and mesoscale simulations appear to be fully consistent in predicting the structural information of the mixture in the lengthscales larger or equal to the polymer radius-of-gyration.

\section{Scattering Functions and Concentration Fluctuations}
\label{SX:SCAT}
The mesoscale pair correlation functions effectively describe the polymer fluid as a liquid of soft colloidal particles. Once these pcfs are obtained from simulation, any property of the liquid can be calculated, including the equation of state, internal energy, compressibility, and others.\cite{McQuarrie}
In this section, we examine the extent to which our classical MD simulations of soft colloidal particles reproduce the structural changes which occur as the system approaches the spinodal. Due to the increasing length scale of fluctuations as the system approaches the critical temperature for demixing, UA simulations can only reach a very limited region of the phase diagram. An advantage of using a procedure that captures the structure at the mesoscopic scale is that the relevant length scale of the simulation can increase considerably with respect to UA MD, and simulations can describe the increasing lengthscale of the fluctuations. Thus a mesoscopic picture greatly facilitates the ability to capture this phenomenon, since we are able to simulate many thousands of chains represented as soft spheres. Models using Monte Carlo methods with phenomenological potentials have been previously performed at the level of soft colloids, demonstrating the valuable information that may be gained about phase transitions.\cite{KREMR, SARIBAN} The advantage of the procedure presented here is that the potentials used to simulate the system are explicitly parameter dependent, being related to the system-specific molecular parameters, such as $R_g$. The potentials obtained in this manner allow for mesoscale simulations to be performed on any number of different, but specific, systems under different thermodynamic conditions, mapping them as soft colloids.

\begin{figure*}[t!]
   \centering
   \begin{tabular}{cc}
\begin{minipage}{3in}
\includegraphics[scale=0.38]{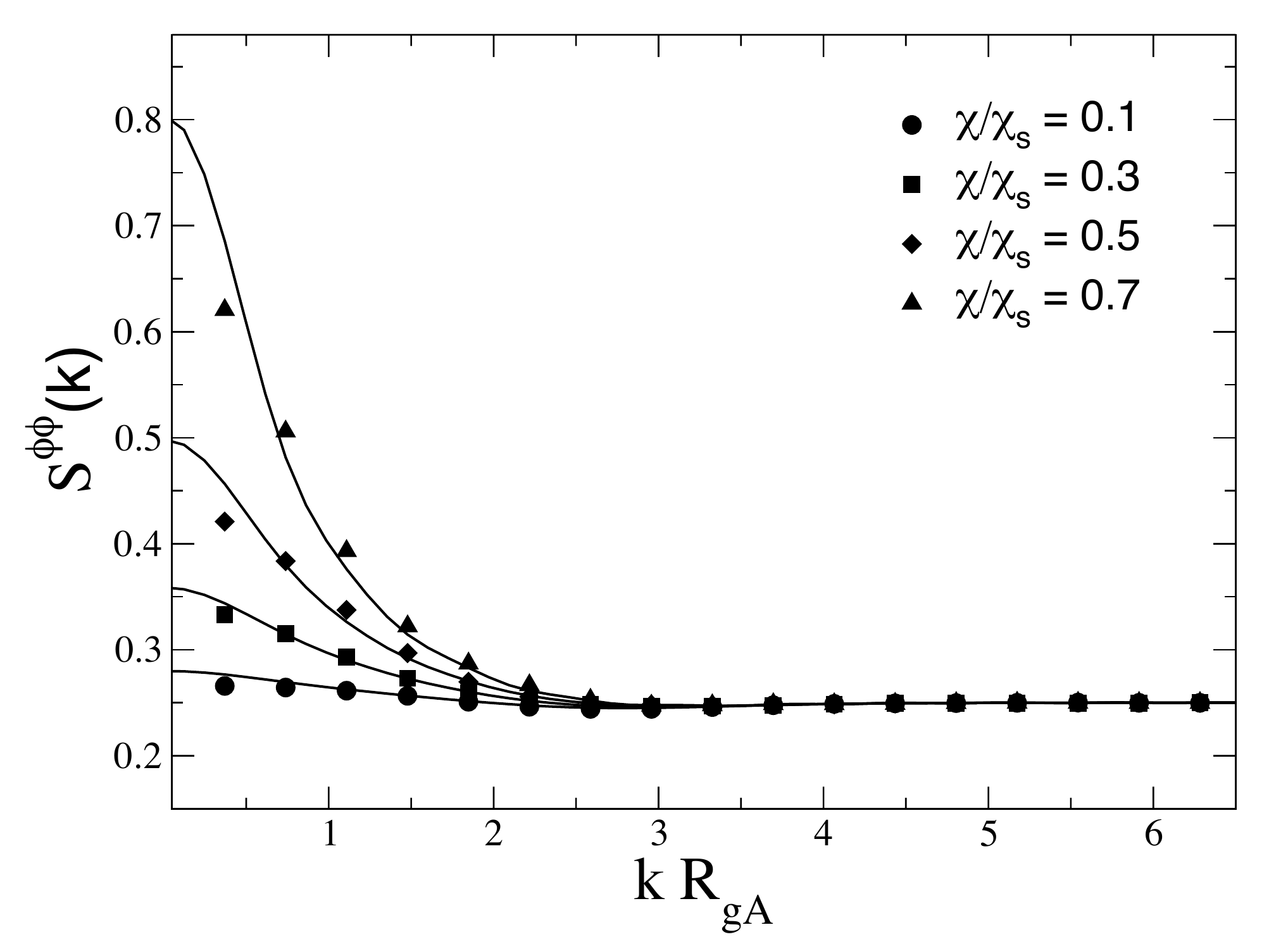}
\end{minipage}
&
\begin{minipage}{3in}
\includegraphics[scale=0.38]{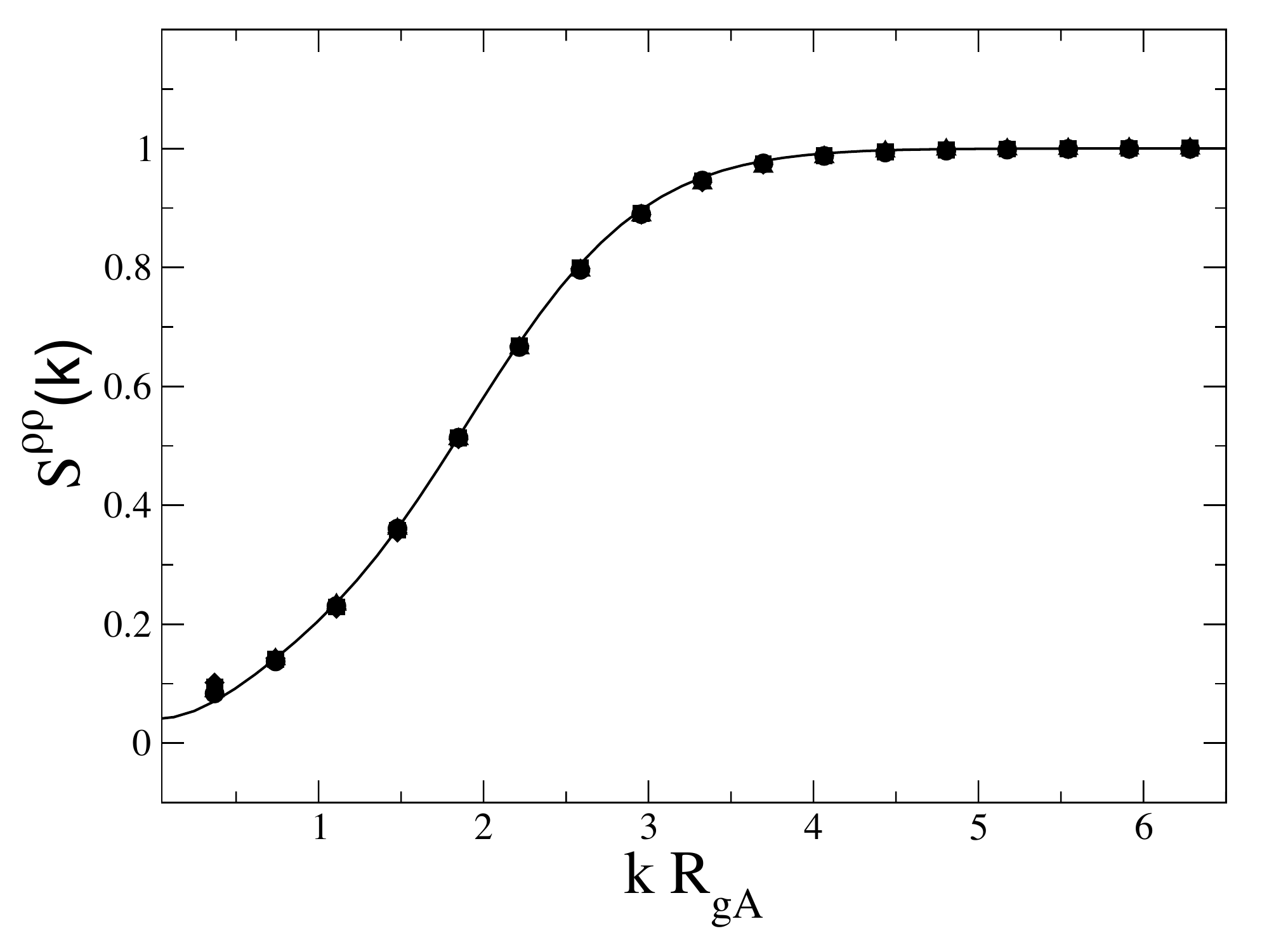}
\end{minipage}
\\
\begin{minipage}{3in}
\includegraphics[scale=0.38]{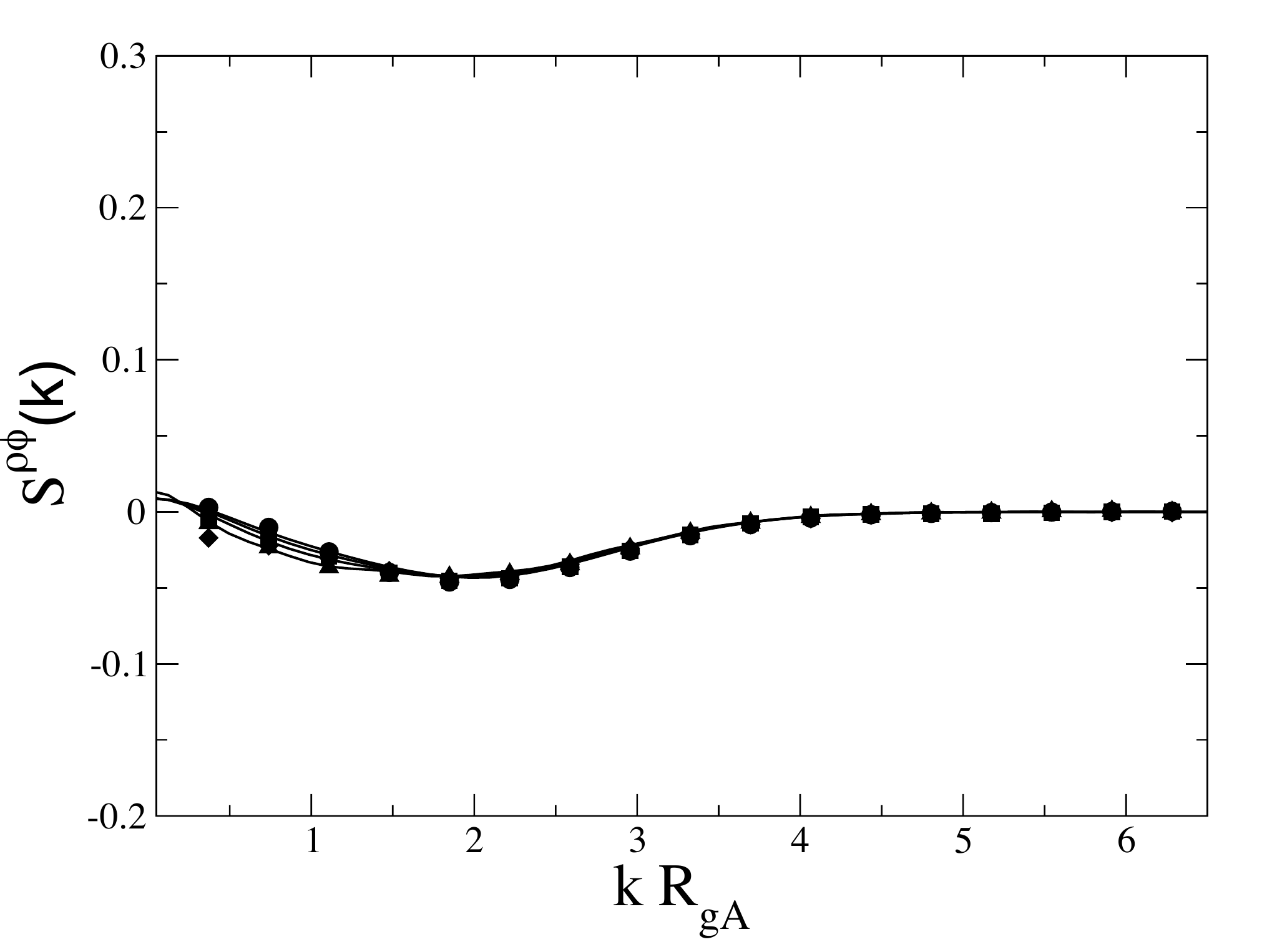}
\end{minipage}
&
\begin{minipage}{3in}
\includegraphics[scale=0.38]{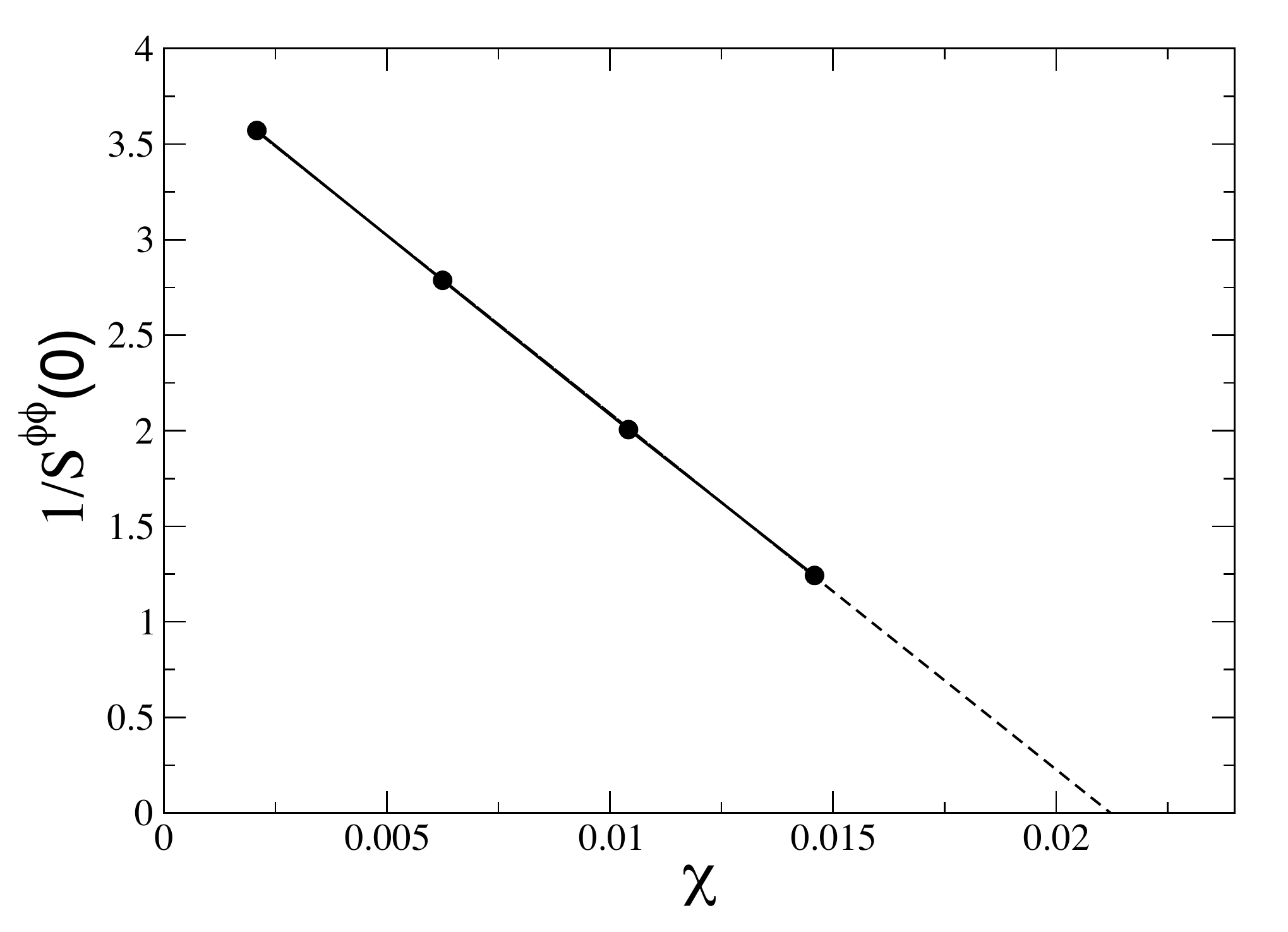}
\end{minipage}
\end{tabular}
\caption{\small{Top Left: Partial structure factor, $S^{\phi \phi}(k)$, obtained from mesoscopic simulations [symbols] of the coarse-grained mixture of 50:50 hhPP/PE with $\chi/\chi_s \in \{0.0, 0.1, 0.3, 0.5, 0.7\}$. The curves represent theoretical values obtained using the Debye function. Top Right: Partial structure factor, $S^{\rho \rho}(k)$ and Bottom Left: $S^{\rho \phi}(k)$ are also shown for different values of $\chi/\chi_s$. $S^{\rho \rho}(k)$ does not change noticeably with $\chi/\chi_s$ but $S^{\rho \phi}(k)$ has a slight $\chi/\chi_s$ dependence at low k. Bottom Right: Extrapolated $1/S^{\phi \phi}(0)$  values vs. $\chi$ [symbols]. The line represents a linear fit to the data and is extrapolated to the spinodal, $\chi_s$ (dashed line).} }
\label{FG:Scck}
\end{figure*}

The static structure factors for each component are calculated from our simulations by Fourier transform of the total correlation function,
\begin{eqnarray}
S_{AA}(k) = \phi +4 \pi \phi^2 \rho_{ch} \int_0^\infty r^2 \frac{\sin kr}{kr} h_{AA}(r) dr \nonumber \ , \\
S_{BB}(k) = 1- \phi +4 \pi (1- \phi )^2 \rho_{ch} \int_0^\infty r^2 \frac{\sin kr}{kr} h_{BB}(r) dr \nonumber ,\\
S_{AB}(k) = 4 \pi \phi (1-\phi ) \rho_{ch} \int_0^\infty r^2 \frac{\sin kr}{kr} h_{AB}(r) dr \ .
\label{EQ:Fourier}
\end{eqnarray}
Density and concentration fluctuation contributions can be written as linear combinations of the static structure factors according to the formalism of Bhatia and Thornton\cite{BT}.  Here, the density fluctuation, $S^{\rho \rho}(k)$ is given by 
\begin{equation}
S^{\rho \rho}(k) = S_{AA}(k) + S_{BB}(k) + 2S_{AB}(k) \ .
\label{EQ:XEFF4}
\end{equation}
The concentration fluctuation contribution,  $S^{\phi \phi}(k)$ may be expressed as
\begin{equation}
S^{\phi \phi}(k) = (1-\phi)^2S_{AA}(k) + \phi^2S_{BB}(k) -2 \phi (1-\phi)S_{AB}(k) \ ,
\label{EQ:XEFF5}
\end{equation}
and is particularly important since it provides information about the stability of the binary mixture against demixing. The coupling term, $S^{\rho \phi}(k)$, is given by 
\begin{equation}
S^{\rho \phi}(k) = (1-\phi)S_{AA}(k) - \phi S_{BB}(k) + (1-2\phi)S_{AB}(k).
\label{EQ:XEFF6}
\end{equation}
Figure \ref{FG:Scck} shows the colloidal partial structure factors,  $S^{\rho \rho}(k)$,  $S^{\phi \phi}(k)$, $S^{\rho \phi}(k)$, calculated from pcfs obtained from mesoscopic simulations shown in the right panel of Figure \ref{FG:THERM} using Equations \ref{EQ:Fourier}-\ref{EQ:XEFF6}. The data from the simulation is compared to predictions based on our numerical values for $h^{cc}(k)$, obtained from Equation \ref{EQ:HCCK} using the Debye function. Since it is particularly pertinent to capture the low $k$ behavior where concentration fluctuations will diverge as the spinodal is approached, we use the results for the Debye form since the Pad\'e approximation introduces unphysical effects in this regime, typically for $k R_g < 2$. 
As seen in Figure \ref{FG:Scck}, the curves of the density fluctuation contribution, $S^{\rho \rho}(k)$, which behaves similarly to the static structure factor for a single-component liquid,\cite{BLNDS} are indistinguishable over the range of $\chi$ investigated. The function $S^{\rho \phi}(k)$ exhibits a slight dependence on the ratio $\chi/\chi_s$ in which the minimum at low k becomes slightly more pronounced. The minimum in $S^{\rho \phi}(k)$ represents the length scale for asymmetry in the mixture arising from the difference in particle size.\cite{BLNDS} The partial structure factor,  $S^{\phi \phi}(k)$, exhibits a characteristic diverging behavior as the spinodal is approached, indicating an increase in the length scale of concentration fluctuations. 

As illustrated in the upper left of Figure \ref{FG:Scck}, $S^{\phi \phi}(0)$ increases as the ratio $\chi/\chi_s \rightarrow 1$.  As the system nears the phase transition, the divergence of $S^{\phi \phi}(k)$ is indicative of the concentration fluctuations becoming increasingly macroscopic.  Since concentration fluctuations occur on an increasingly large scale, the relevant region of the $S^{\phi \phi}(k)$  curve occurs in the low-k region; however, due to periodic boundary conditions, simulation data is only reliable at a distance less than half the length of the simulation box. This makes extrapolation of the k=0 limit from mesoscopic simulations still difficult, as seen in Figure \ref{FG:Scck}, even though thousands of particles were represented. In this respect, our numerical predictions may serve as a guide for extending $S(k)$ to the k=0 limit. Furthermore, we have previously shown that Equation \ref{EQ:HCCR}, also gives an estimate for $S^{\phi \phi}(0)$ given by\cite{BLNDS}
\begin{equation}
S^{\phi \phi}(0) = \frac{\phi (1-\phi)}{1-\chi / \chi_s} + \frac{\phi^2(1-\phi)^2(\gamma^2-1)^2}{(\phi \gamma^2 + 1 - \phi) \gamma^2} \frac{\xi^2_\rho}{\xi^2_{cA}} \ .
\label{EQ:XEFF7}
\end{equation}
Even though it is based on the Pad\'e approximation, Equation \ref{EQ:XEFF7} may be used to estimate $S^{\phi \phi}(0)$ since $h^{cc}(k)$ calculated from the Pad\'e approximation has the same $k=0$ limit as $h^{cc}(k)$ from the Debye form. The lower right of Figure \ref{FG:Scck} shows a linear plot of $1/S^{\phi \phi}(0)$ vs. $\chi$ for which the $k=0$ limit was determined by our theoretical predictions.

\begin{figure*}[t!]
\centering
\begin{tabular}{cc}
\begin{minipage}{3in}
\includegraphics[scale=0.38]{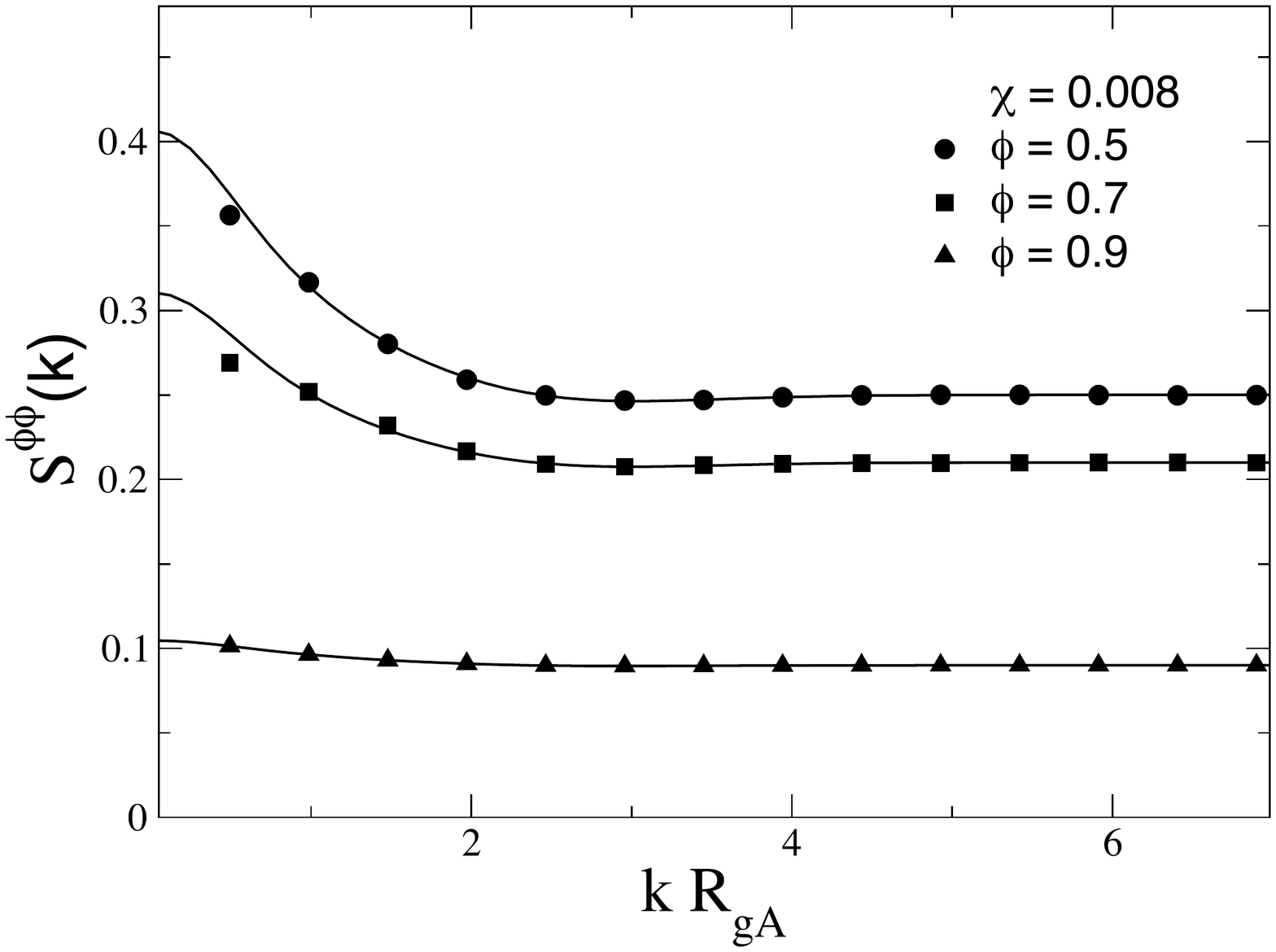}
\end{minipage}
&
\begin{minipage}{3in}
\includegraphics[scale=0.38]{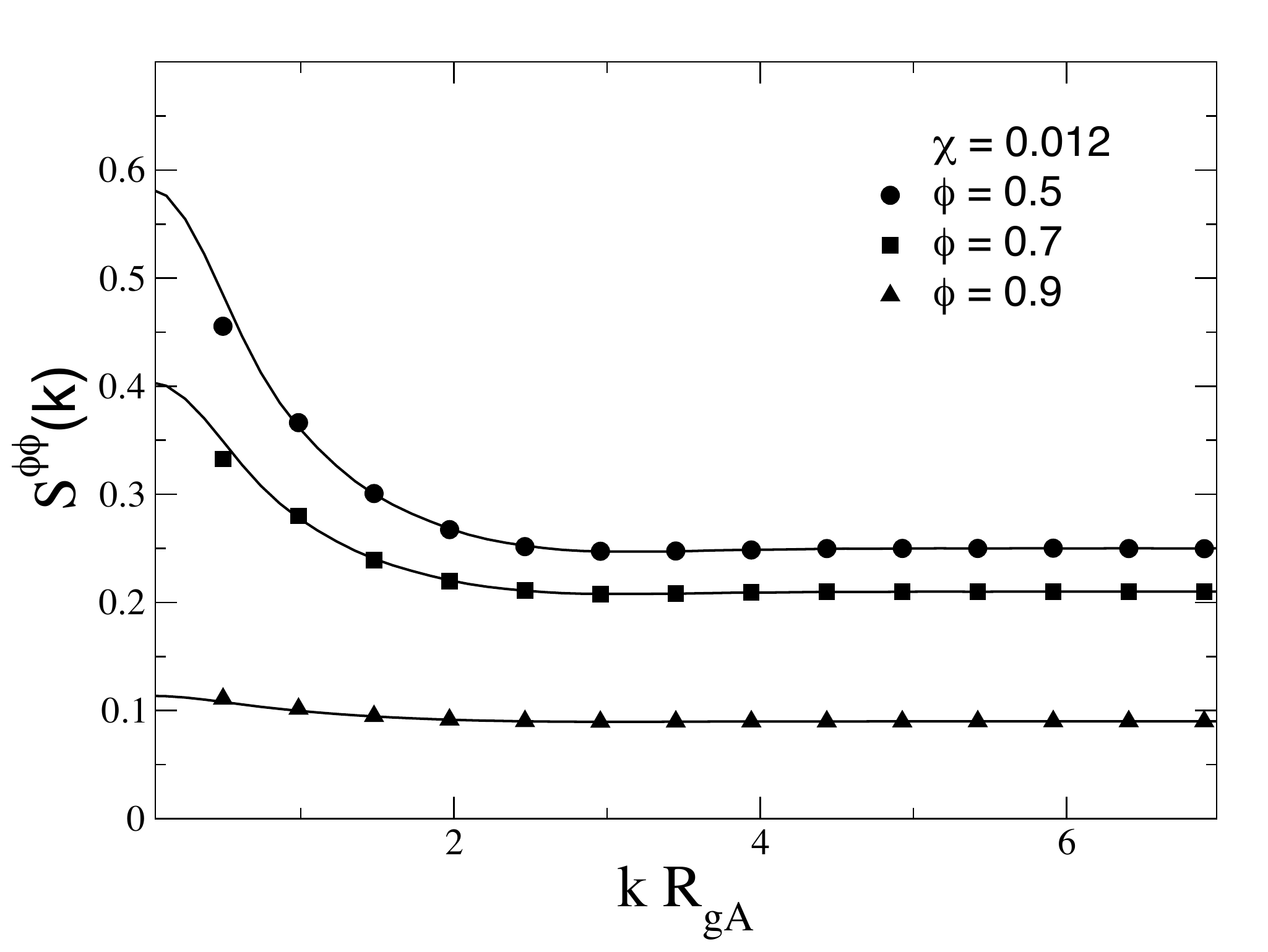}
\end{minipage} \\
\begin{minipage}{3in}
\includegraphics[scale=0.38]{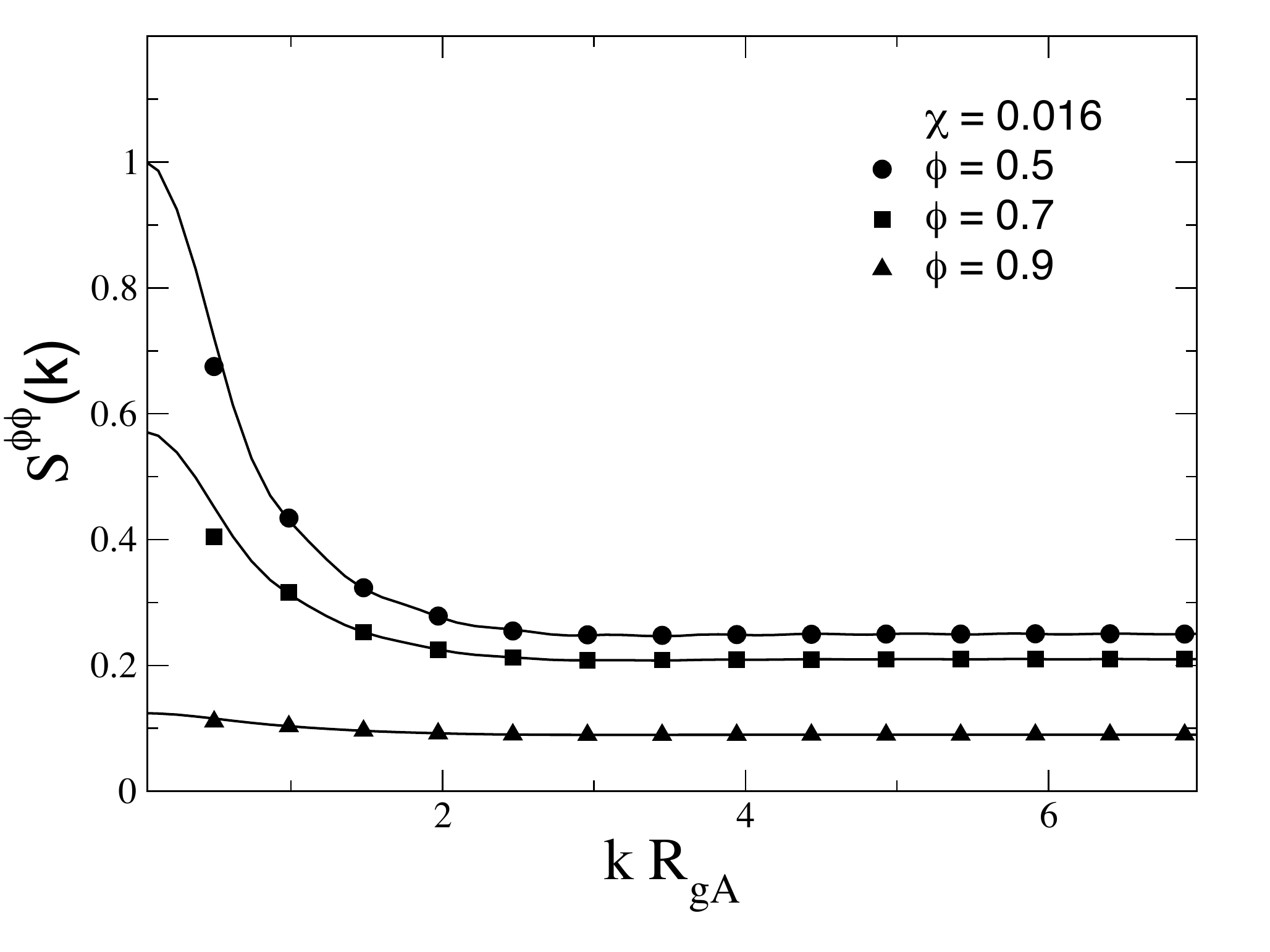}
\end{minipage}
&
\begin{minipage}{3in}
\includegraphics[scale=0.38]{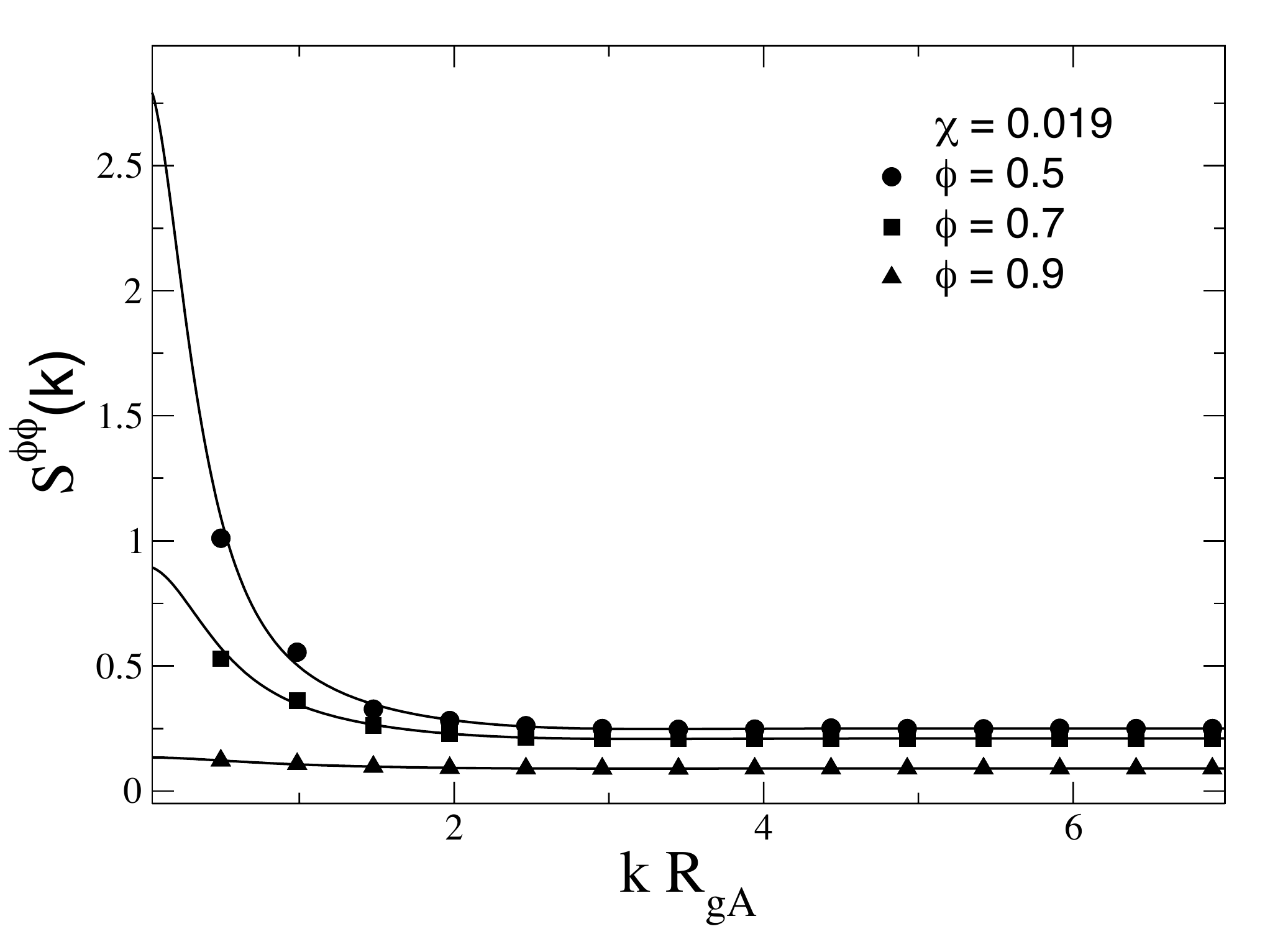}
\end{minipage}
\end{tabular}
\caption{The concentration fluctuation partial structure factor, $S^{\phi \phi}(k)$, calculated from mesoscale simulations [filled symbols] at different values of $\phi$ for the mixture hhPP/PE. The curves represent theoretical predictions.}
\label{FG:SPHIK}
\end{figure*}

Following Equations \ref{EQ:Fourier}-\ref{EQ:XEFF6}, the concentration fluctuation partial structure factor, $S^{\phi \phi}(k)$, was calculated from the mesoscale simulations presented in Figures \ref{FG:PHIA} and \ref{FG:PHIB}, where the volume fraction, $\phi$, was changed. The resulting $S^{\phi \phi}(k)$ is presented in Figure \ref{FG:SPHIK} along with theoretically predicted values using the Debye formula. Once again, mesoscale simulations show an increase in concentration fluctuations as the thermodynamic conditions are changed, and $\chi \rightarrow \chi_s$ or $\phi \rightarrow 0.5$. In general, mesoscale simulations are consistent with our theoretical predictions based on Equation \ref{EQ:HCCK} up to the limit set by the finite box size. As seen in Figure \ref{FG:SPHIK}, when $\chi$ is low or the polymer volume fraction of one species is large, the system is well mixed and the extrapolation to low k is straightforward.  However, for the case when $\phi=0.5$ and $\chi = 0.019$, as depicted in the lower right panel of Figure \ref{FG:SPHIK}, it becomes more difficult to reach the $k=0$ limit from mesoscale simulation, even if the precision is higher than for atomistic simulations for the reasons previously discussed. Since our simulations are consistent with our theoretical predictions as shown in Figures \ref{FG:BLNS} - \ref{FG:PHIB}, we estimate the extrapolated $S(k=0)$ limit based on these predictions.

\begin{figure}[t!]
\includegraphics[scale=0.43]{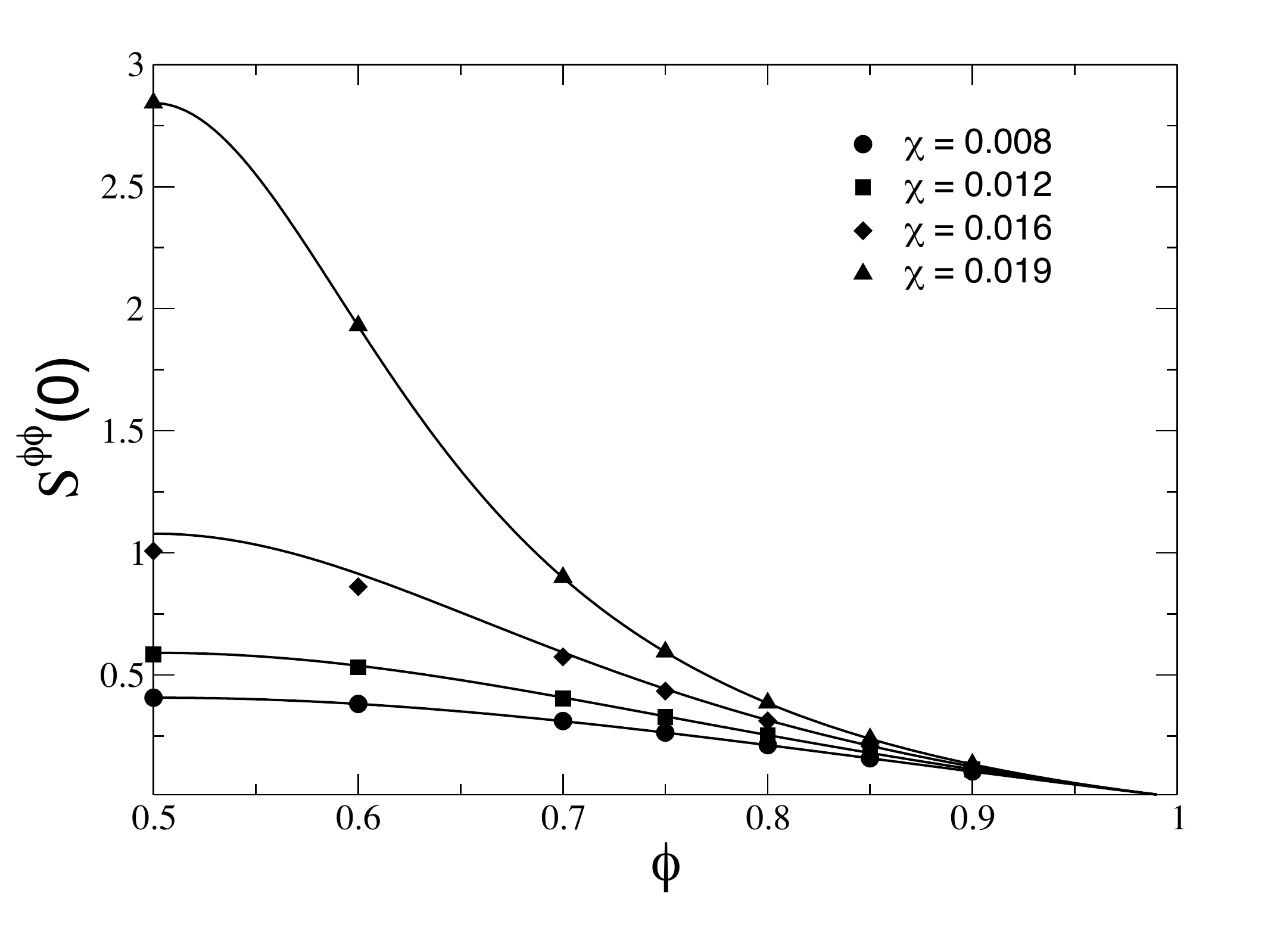}
\caption{The extrapolated $k=0$ limit of $S^{\phi \phi}(k)$ based on our numerical predictions [data points] and from our analytical expression, Equation \ref{EQ:XEFF7}, [curves] as a function of $\phi$ for different fixed values of $\chi$, for the mixture hhPP/PE.}
\label{FG:SKPHI}
\end{figure}

Once this method is employed, it is possible to discern the phase behavior of the mixture from the extrapolated $k=0$ limit. In order to include more data points, we calculate $S^{\phi \phi}(0)$ for a range of $\chi$ and $\phi$ values, based on our solution to Equation \ref{EQ:HCCK}. These are presented in Figure \ref{FG:SKPHI} which shows the structure factor as a function of the volume fraction for several fixed values of the $\chi$ parameter. The interpolation between the points is given by Equation \ref{EQ:XEFF7}, which demonstrates that our analytical expression is useful in determining the phase behavior.

\begin{figure}[h!]
\centering
\begin{tabular}{c}
\includegraphics[scale=0.43]{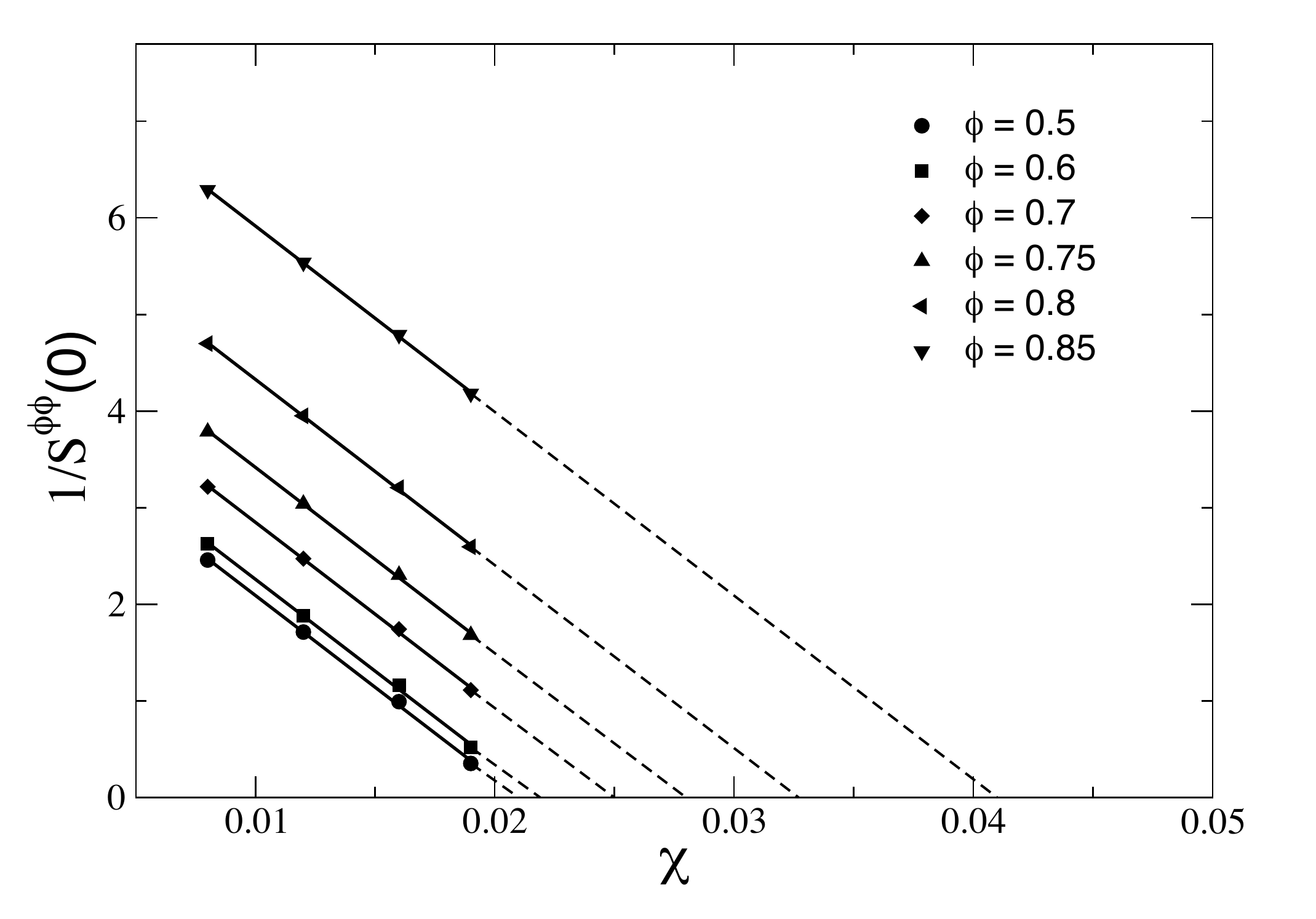}
\\
\includegraphics[scale=0.43]{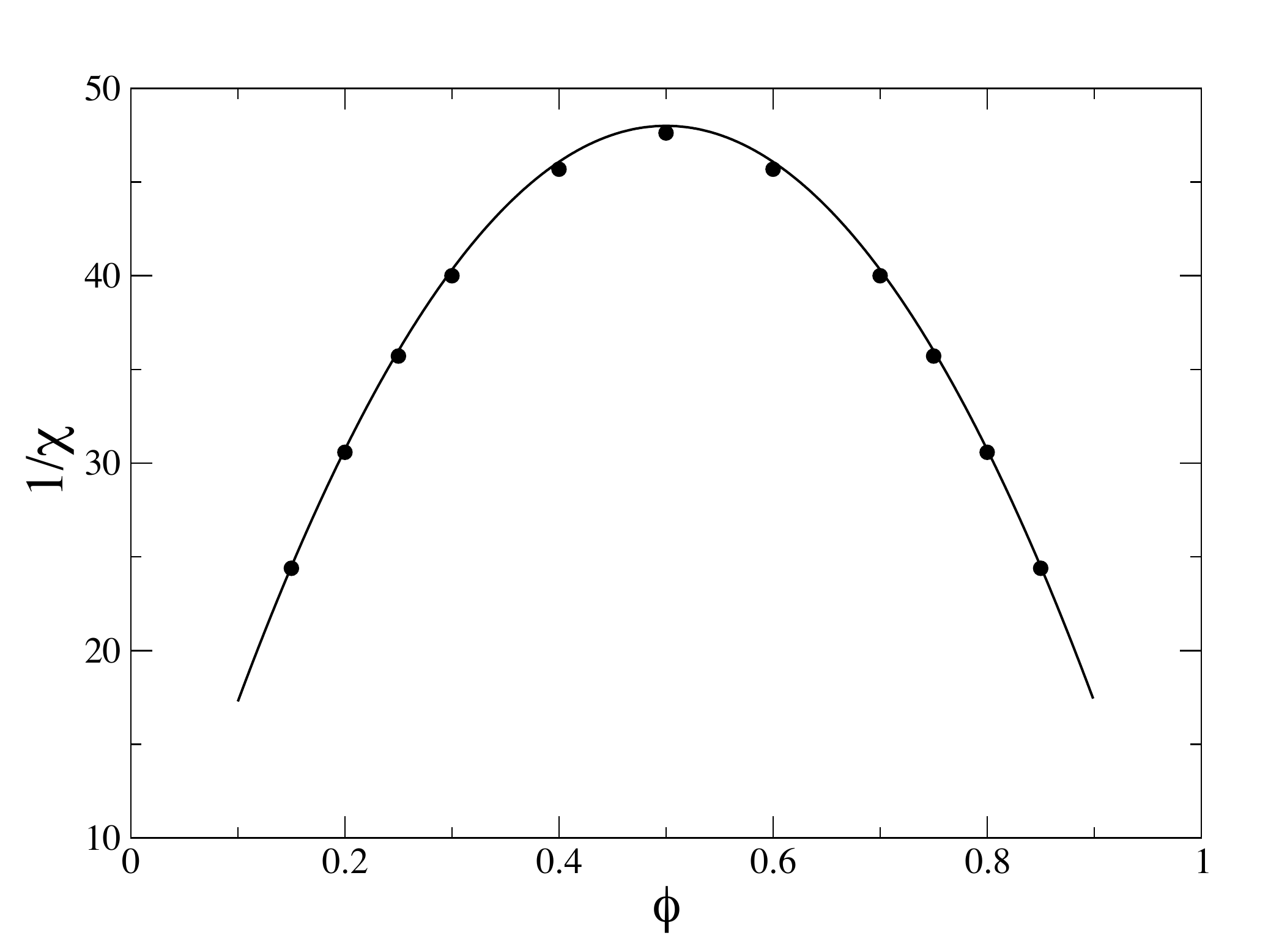} \\
\end{tabular}
\caption{\small{Top: Inverse concentration fluctuation structure factor, $S^{\phi \phi}(0)$ plotted against the interaction parameter, $\chi$ for different values of $\phi$, for the mixture hhPP/PE. The solid line depicts a linear fit to the data and the dashed line shows the extrapolation to the spinodal. Bottom: Phase diagram for the coarse-grained mixture obtained from the above extrapolation to the spinodal, $\chi_s$. The solid curve depicts the Florry-Huggins analytical expression.}}
\label{FG:PHASE}
\end{figure}

Finally, in Figure \ref{FG:PHASE} a plot of the inverse structure factor, $S^{\phi \phi}(0)$, vs. $\chi$ at each value of $\phi$ shows the linear behavior from which the spinodal, $\chi_s$, may be extrapolated and used to sketch the phase diagram of the system. In the bottom panel of Figure \ref {FG:PHASE} the spinodal curve is compared to the predicted Flory-Huggins model, $\chi_s=[2 N_A
\phi]^{-1}+ [2 N_B (1-\phi)]^{-1}$, which was used in Equation \ref{EQ:XI}. The spinodal curve from our simulation exhibits a characteristic parabolic shape consistent with mean-field theory, where $\xi_{\phi}\sim (1-\chi/\chi_s)^{-\nu}$, $\nu=1/2$. In the immediate region of the critical temperature, mean-field theory breaks down, and Ising-type critical behavior is expected. For this narrow temperature region, the linear extrapolation in Figure \ref{FG:PHASE} would be invalid and the spinodal will exhibit a flatter peak.\cite{SARIBAN} For long polymer chains, the temperature region for which mean field theory becomes invalid is very small, since the Ginzburg number, which scales inversely with N, is small.\cite{BINDR} As seen in the upper panel of Figure \ref{FG:PHASE} most of the simulations performed are well within the temperature region described by mean-field theory. Although the linear extrapolation becomes less quantitative near the horizontal axis, the mean-field approximation is consistent with our data.

\section{Corrections to the Debye Intramolecular Form Factor}
\label{SX:DEBYE}
Upon examination of Figure \ref{FG:BLNS}, it appears that there is slightly better agreement between UA MD simulations when compared to our analytical results using the Pad\'e approximation than with the full Debye form (as indicated by the dashed lines). Since the Pad\'e form is an approximation, this improvement is likely due to a cancellation of errors. The Debye formula is exact for ideal Gaussian chains; however, Wittmer, \emph{et. al}, have recently shown that dense polymer melts exhibit deviations from ideal Gaussian behavior because of long-range correlations arising from the repulsive interaction of chain segments.\cite{Wittmer} These deviations become more significant for polymers confined between walls in ultrathin films.\cite{Cavallo} In this section we investigate the implementation of higher order corrections to the Gaussian approximation on the effective pair potential by evaluating Equation \ref{EQ:HCCK} numerically with a corrected from of the intramolecular form factor. 

\begin{figure*}[t!]
\centering
\begin{tabular}{cc}
\begin{minipage}{3in}
\includegraphics[scale=0.3]{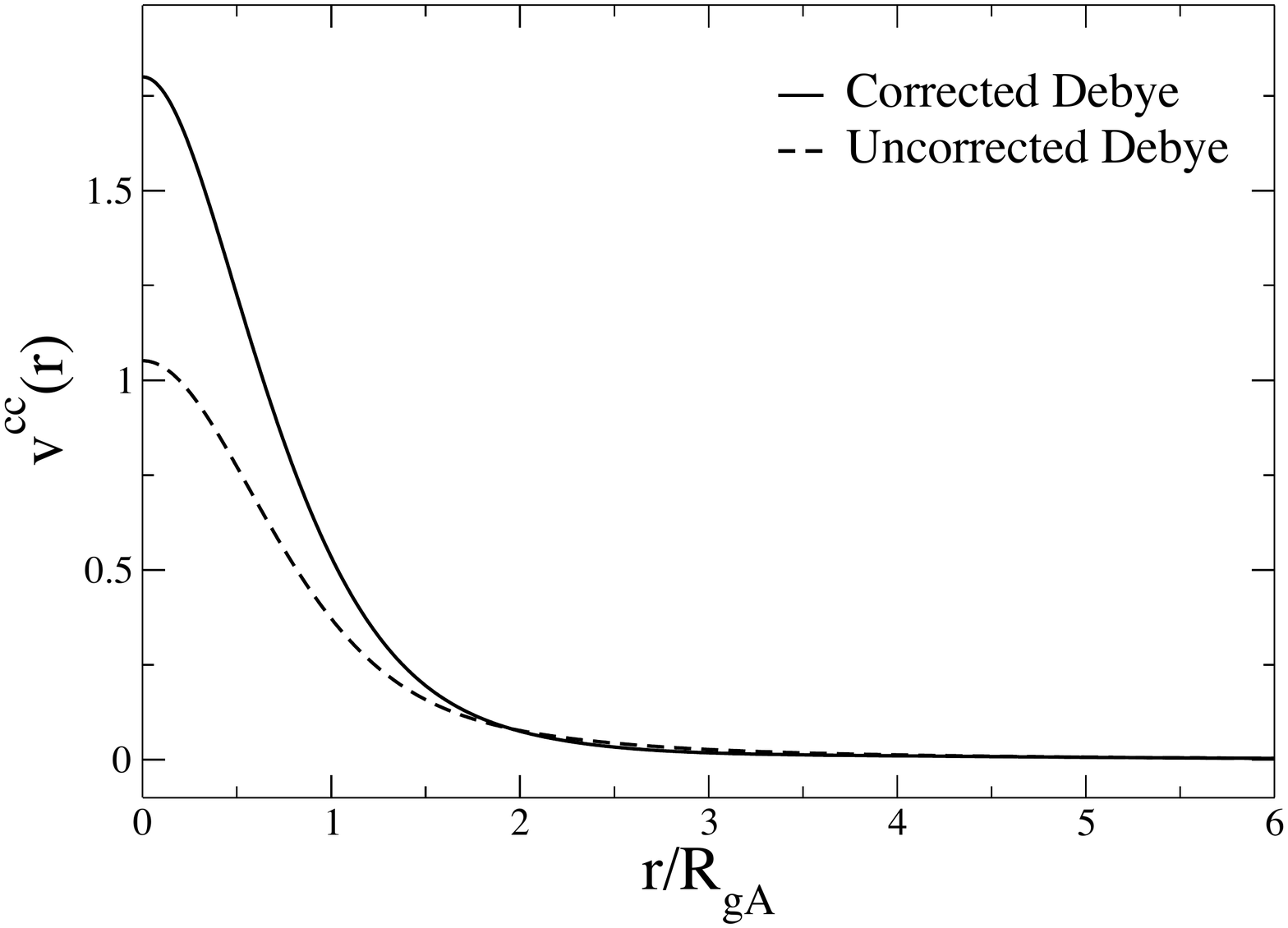}
\end{minipage}
&
\begin{minipage}{3in}
\includegraphics[scale=0.3]{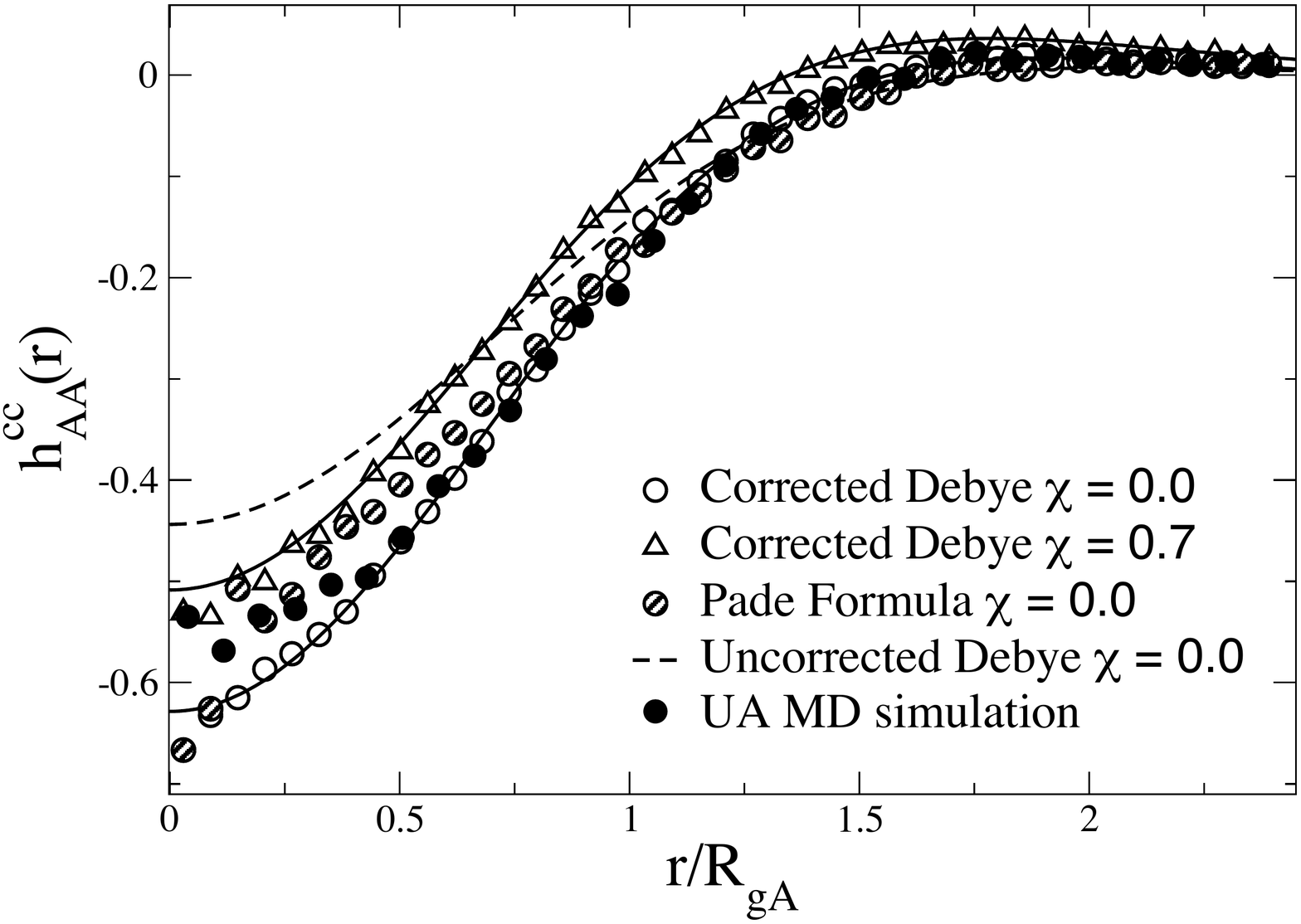}
\end{minipage} \\
\begin{minipage}{3in}
\includegraphics[scale=0.3]{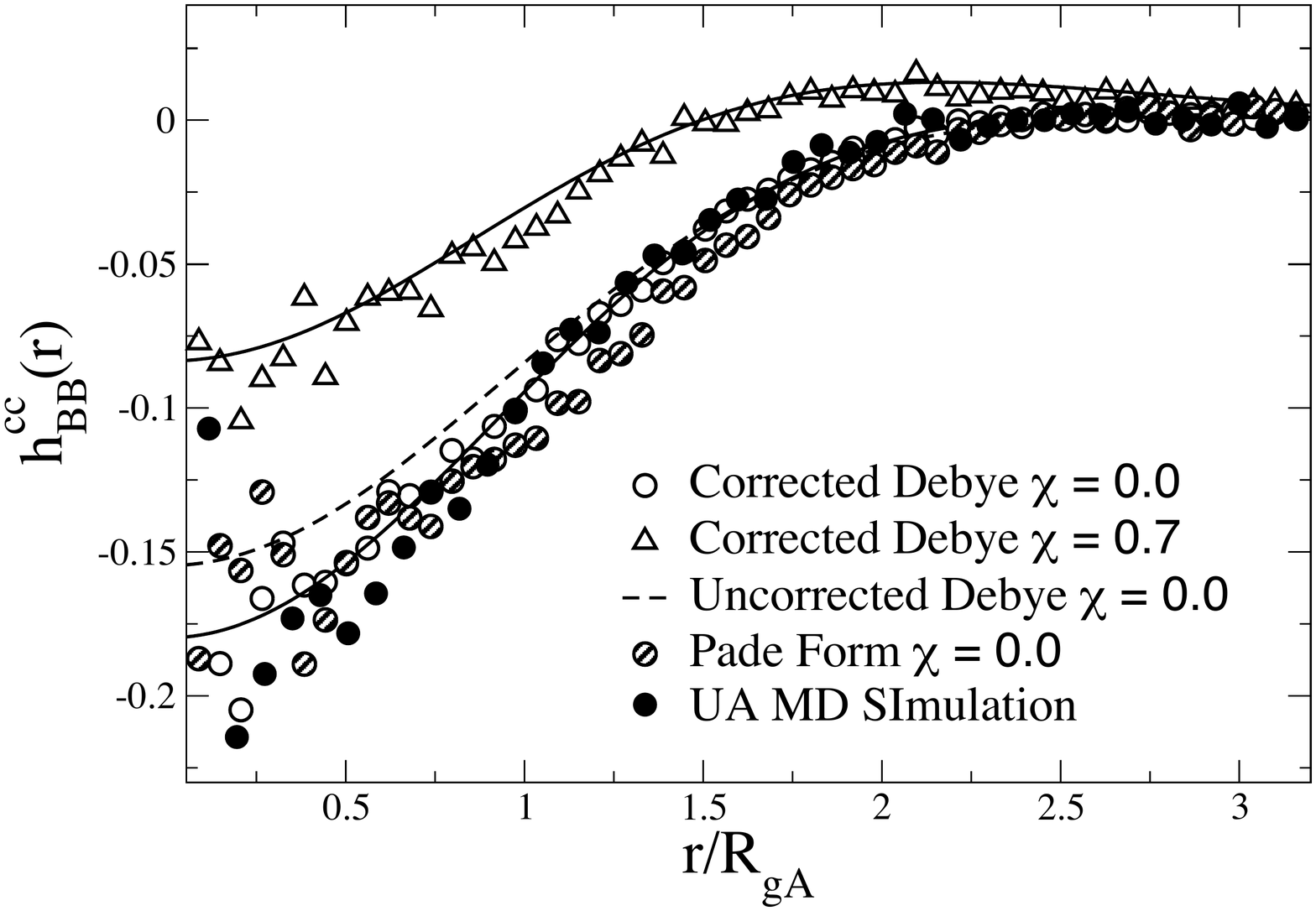}
\end{minipage}
&
\begin{minipage}{3in}
\includegraphics[scale=0.3]{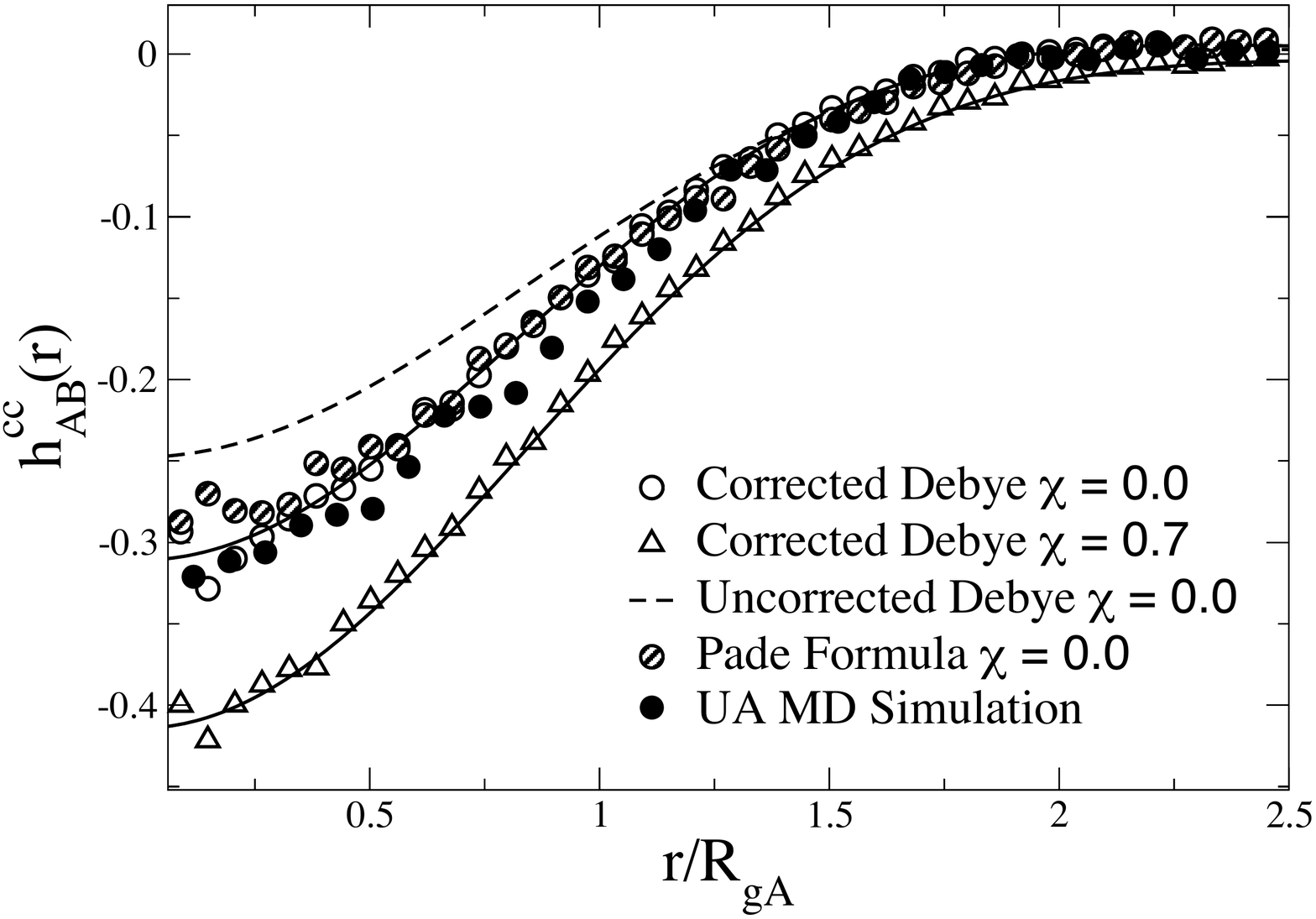}
\end{minipage}
\end{tabular}
\caption{\small{Top Left: The effective pair potential between A type chains for an athermal mixture of hhPP/PE with $\phi = 0.5$ when corrections to the Debye formula are included (solid line). The dashed line indicates the potential obtained using the uncorrected Debye formula. Top Right: The AA component of the correlation function calculated from mesoscale simulations using the corrected Debye formula at $\chi=0.0$ (open circles) and $\chi=0.7$ (open triangles). The solid line represents theoretical predictions, and the dashed line indicates predictions using the Debye formula. Filled circles represent UA MD simulations. Bottom Left: BB component and Bottom Right: AB component of the correlation function for the same mixture. The pcfs obtained using the Pad\'e approximation are shown to nearly superimpose with the corrected Debye form. (Partially shaded circles).}}
\label{FG:DebyeCorrect}
\end{figure*}

In the infinite chain limit ($N \rightarrow \infty$) it has been proposed that corrections to the Debye formula in the intermediate wave vector range depends only on the monomer density, such that\cite{Wittmer, Beckrich} 
\begin{equation}
\frac{1}{\omega^{mm}(k)}=\frac{1}{\omega^{mm}_{Debye}(k)}+\frac{1}{32}\frac{k^3}{\rho}
\label{EQ:CORR}
\end{equation}
Although approximate for finite chain lengths, Equation \ref{EQ:CORR} was input into Equation \ref{EQ:HCCK} to obtain a corrected form of the pair potential which is shown in Figure \ref{FG:DebyeCorrect} (top left) for a 50:50 mixture of hhPP/PE. The resulting correlation functions, displayed in Figure \ref{FG:DebyeCorrect}, show that the corrected Debye formula agrees very well with UA MD simulations for this sample, indicating that the disagreement between mesoscale simulations using the Debye formula and UA simulations on intermediate lengthscales is due to non-Gaussian behavior of real chains as the Flory ideality hypothesis breaks down. On the local scale, however, the corrected Debye and the UA-MD simulations tend to disagree. This is not relevant for systems with long chains, such as the hhPP:PE mixture, but it becomes important for short chains, e.g. mixtures of PIB chains, where the behavior at short distance becomes unphysical.
In conclusion, while in the current publication we limit our investigation to just this correction term for the hhPP:PE mixture, further study is necessary to investigate if the observed improvement is a common feature of long-chain mixtures, independent of their monomeric structures. The pcfs obtained using the Pad\'e approximation (Figure \ref{FG:BLNS}) are also shown in Figure \ref{FG:DebyeCorrect} and compare well with the corrected Debye results. 

\section{Applications to Miscible LCST Blends}
\label{SX:LCST}
While most polymer blends are immiscible and tend to demix at experimentally relevant temperatures, some systems are known to be miscible having a lower critical solution temperature (LCST). In this section we demonstrate the extension of our approach to model LCST blends where the effective $\chi$ parameter may be negative over most of the temperature range of interest. It is worth noticing that while the hhPP/PIB blend is miscible, the iPP/PIB blend is immiscible, indicating that subtle changes in the specific polyolefin architecture may give rise to a completely different phase diagram. The temperature dependence of the $\chi$ parameter for the miscible hhPP/PIB blend is reported in Table \ref{Table1}. The $\chi$ parameter in the literature is defined on a monomer basis and must be divided by the number of united atom sites per monomer (4.8 for hhPP/PIB) to be consistent with the UA site description used here. We performed mesoscale simulations for various temperatures of a mixture of 50:50 hhPP/PIB ($\chi_s=0.021$). 

\begin{figure*}[t!]
\centering
\begin{tabular}{cc}
\begin{minipage}{3in}
\includegraphics[scale=0.3]{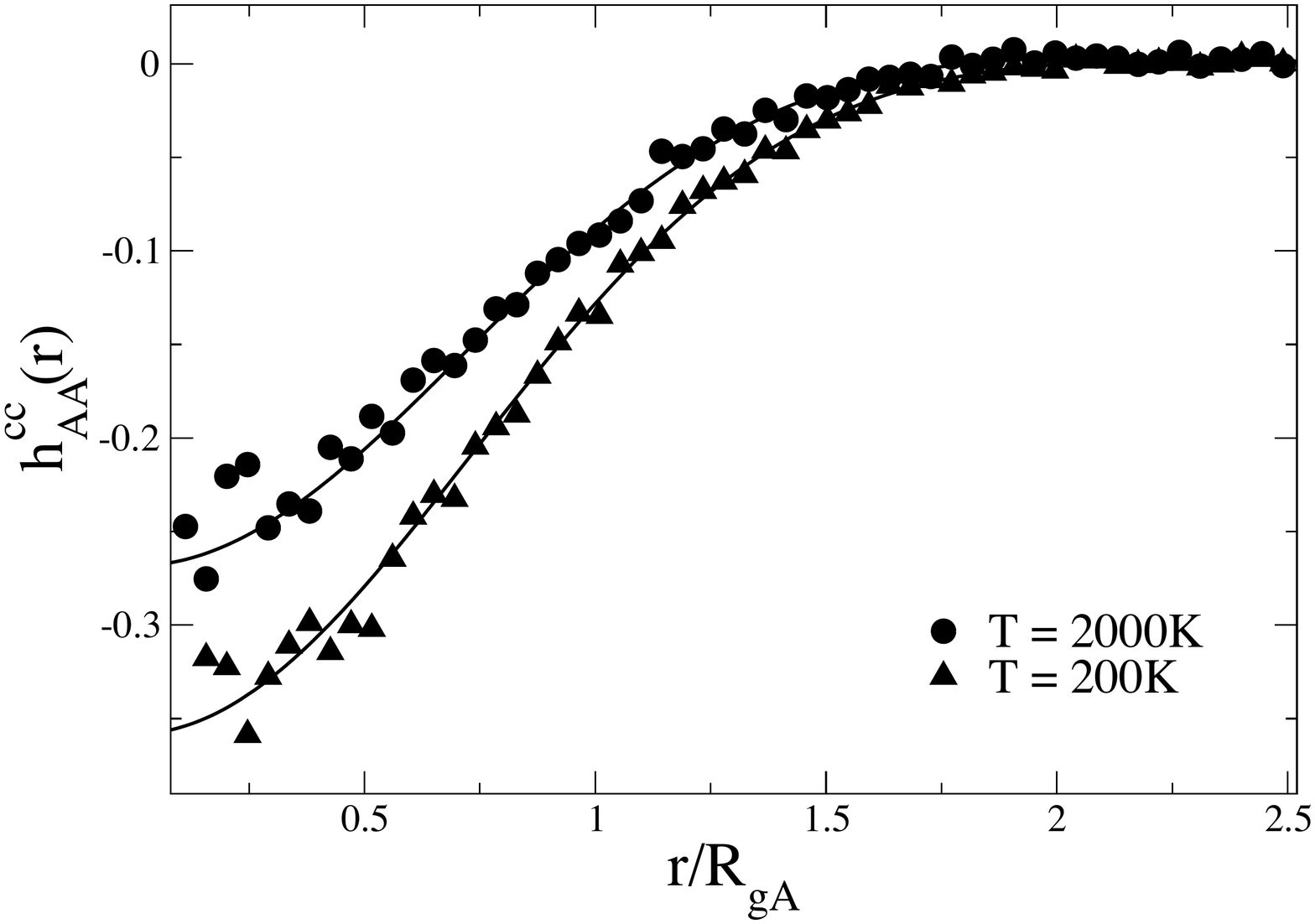}
\end{minipage}
&
\begin{minipage}{3in}
\includegraphics[scale=0.3]{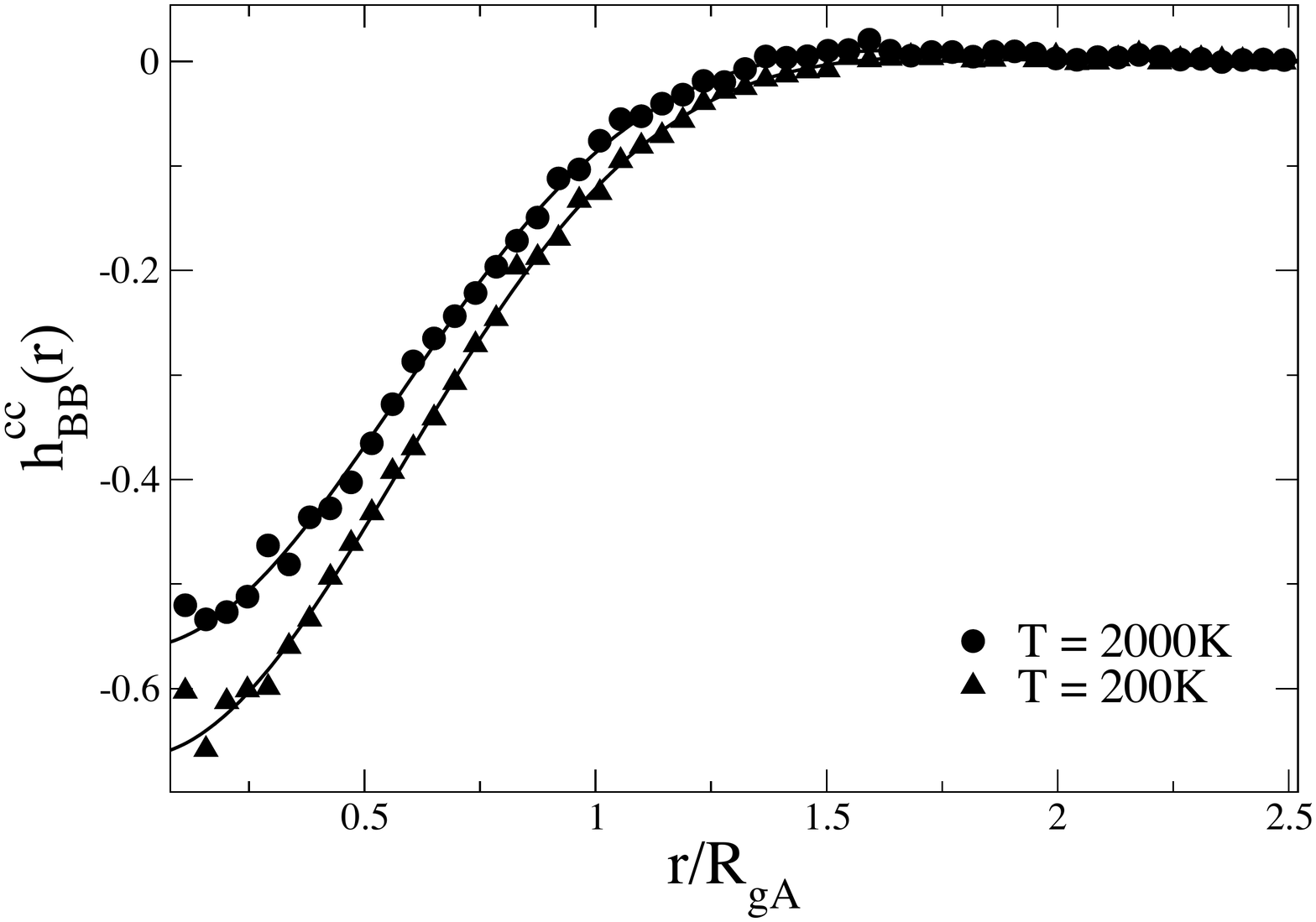}
\end{minipage} \\
\begin{minipage}{3in}
\includegraphics[scale=0.3]{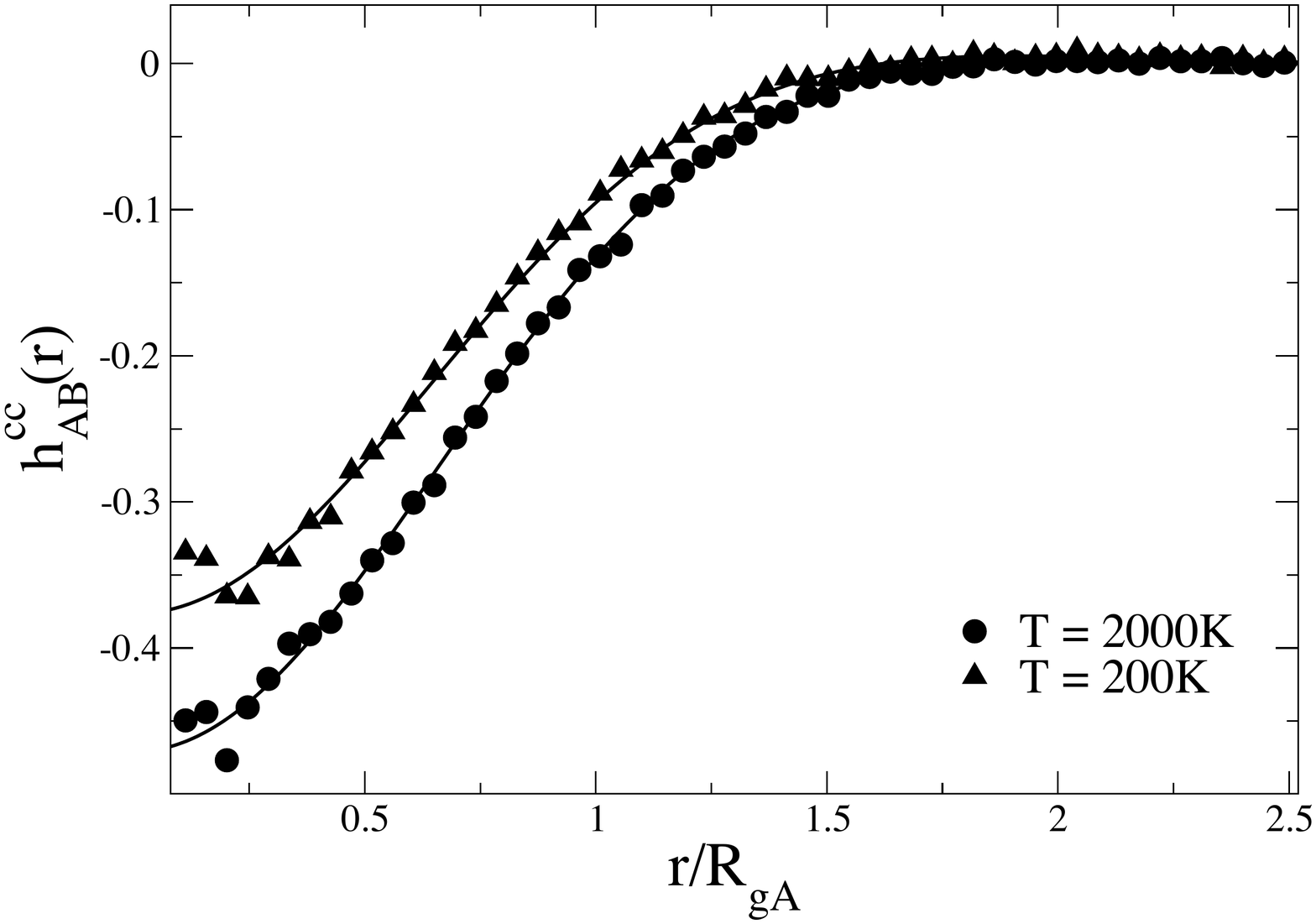}
\end{minipage}
&
\begin{minipage}{3in}
\includegraphics[scale=0.3]{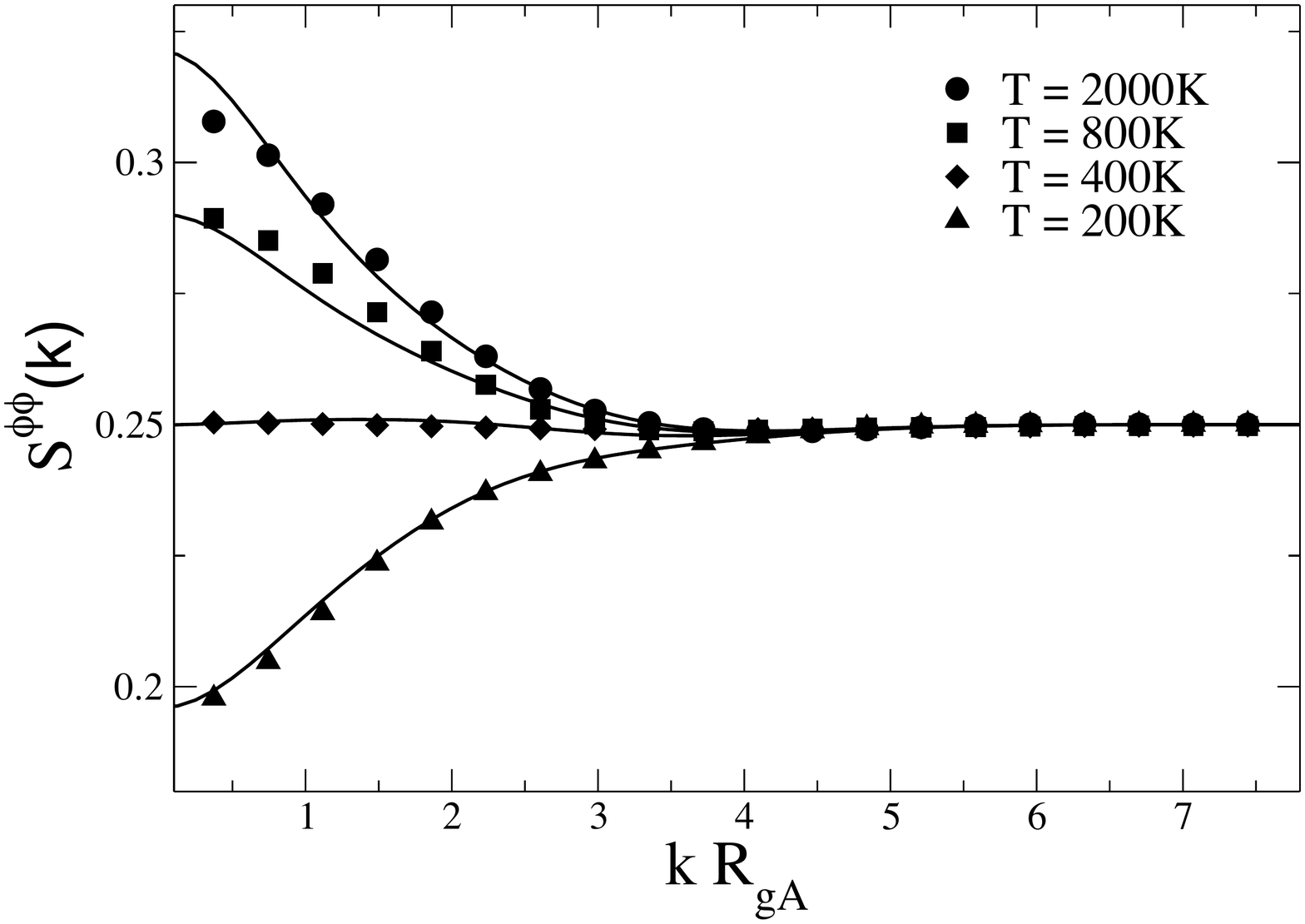}
\end{minipage}
\end{tabular}
\caption{\small{Top Left: AA component of the correlation function for the miscible blend, hhPP/PIB ($\phi =0.5$) at $T=2000K$ (circles) and $T=200K$ (triagles). Theoretical predictions are indicated as solid lines. Top Right: BB component and Bottom Left: AB component of the same mixture. Bottom Right: The concentration fluctuation structure factor for hhPP/PIB obtained from mesoscale simulation (symbols) and from theory (solid line) at various temperatures.}}
\label{FG:HHPP_PIB}
\end{figure*}

The resulting correlation functions determined for two temperatures, 2000K and 200K, from mesoscale simulations are shown in Figure \ref{FG:HHPP_PIB}. When compared with Figure \ref{FG:THERM} it is evident that the pcfs for the hhPP/PIB blend exhibit an opposite trend with temperature. These differences are clearly evident in the concentration fluctuation structure factor, which was calculated from these pcfs at various temperatures and is shown in the bottom right of Figure \ref{FG:HHPP_PIB}. As depicted in the low wave vector behavior of $S^{\phi \phi}(k)$, fluctuations in the concentration become smaller as the temperature is decreased, and the system becomes more stable. These results indicate that our procedure of mapping polymer blends as soft-colloids and performing mesoscopic simulations using an effective pair potential can be applied to miscible LCST blends given that the temperature dependence of the $\chi$ parameter is known. 

\section{Summary}
\label{SX:SUMM}
In this paper, we present the implementation of our
analytical coarse-grained description for polymer mixtures in 
mesoscopic modeling through computer simulation.  Using the analytical
representations for the intermolecular total pair correlation
functions at the center-of-mass level and the hypernetted-chain
closure, we derive the effective pair interaction potentials which are
the required input to the coarse-grained simulations.  The simulations
are then carried out and the coarse-grained liquid structure, as given
by the intermolecular pair correlation function, is extracted from the
trajectories.  In the athermal regime, results are compared with our theoretical predictions and
data obtained from united atom molecular dynamics simulations.  In the thermal regime, mesoscopic simulations capture the relevant trends for demixing of polymers in the miscible regime approaching the spinodal. These results are used to calculate static structure factors which are related to the increasing concentration fluctuations of the mixture. By extrapolation to the low wave vector limit, we are able to determine the phase diagram of the coarse-grained mixture which is consistent with mean-field theory predictions. The consistency of all representations supports the theoretical foundation
of our development.

While other methods exist to obtain equilibrium properties of blends, an analytical potential is desirable for many of these approaches. For, example, the Gaussian equivalent representation (GER) has been implemented using a purely repulsive Gaussian potential for a system of interacting particles to obtain a field-theoretic representation of the partition function, and used to compute structural and thermodynamic quantities of interest.\cite{BaeurleEPL2006, BaeurleJCP2006} Possible future applications of the analytical potential derived in this publication could include using it in low-cost approximation methods, such as the zeroth-order GER model formalism.

Although the current publication is focused on determining the equilibrium structure of blends under various conditions, the proposed procedure of obtaining the effective potential and performing mesoscale simulations should be useful in determining the non-equilibrium and dynamic behavior of these systems as well, where time must be treated explicitly. Future work in this direction should include examining the phase transition that occurs when the thermodynamic conditions of the system are suddenly changed. In the context of a multiscale modeling procedure, mesoscale simulations, such as those performed here, may be coupled to a more detailed description in order to combine local and global information over multiple length and time scales.

\section{Acknowledgements}
This research was supported by the National Science Foundation through grant DMR-0804145.
Computational resources were provided by LONI through the TeraGrid project supported by NSF.

\end{document}